\documentclass[a4paper,11pt]{article}
\usepackage{jheppub} 
\usepackage[T1]{fontenc} 
\usepackage{amsmath}
\usepackage{empheq}
\usepackage{graphicx,subfigure}
\usepackage{epsf}
\usepackage{amssymb}
\usepackage{float}
\usepackage{slashed}
\usepackage{xcolor}

\DeclareMathOperator{\csch}{csch}

\def\[{\left[}
\def\]{\right]}
\def\({\left(}
\def\){\right)}

\def\del{{\partial}}

\newcommand{\be}{\begin{equation}}
\newcommand{\ee}{\end{equation}}
\newcommand{\bea}{\begin{eqnarray}}
\newcommand{\eea}{\end{eqnarray}}

\newcommand{\lan}{\langle}
\newcommand{\ran}{\rangle}
\renewcommand{\ln}{\log}

\DeclareMathOperator{\Tr}{Tr}

\newcommand{\bpsi}{\bar{\psi}}

\newcommand{\bu}{\bar{u}}
\newcommand{\bv}{\bar{v}}
\newcommand{\tT}{\text{T}}
\newcommand{\ttb}{$T\bar{T}$}
\newcommand{\ttbd}{$T\bar{T}$-deformation}
\def \be {\begin{equation}}
\def \ee {\end{equation}}
\def \bea {\begin{eqnarray}}
\def \eea {\end{eqnarray}}
\def \beal#1 {\begin{align}#1\end{align}}

\everymath{\displaystyle}

 \title{\boldmath Perturbative renormalization of the $T \bar{T}$-deformed free massive Dirac fermion}

\author[1]{Anshuman Dey,\note{anshuman.dey@mail.huji.ac.il}} 
\author[2]{Aryeh Fortinsky\note{aryeh.fortinsky@mail.huji.ac.il}} 
\affiliation{The Racah Institute of Physics, The Hebrew University of Jerusalem, \\ Jerusalem 91904, Israel}

\abstract{In this paper we explicitly carry out the perturbative renormalization of the $T\bar{T}$-deformed free massive Dirac fermion in two dimensions up to second order in the coupling constant. This is done by computing the two-to-two $S$-matrix using the LSZ reduction formula and canceling out the divergences by introducing counterterms. We demonstrate that the renormalized Lagrangian is unambiguously determined by demanding that it gives the correct $S$-matrix of a $T\bar{T}$-deformed integrable field theory.  Remarkably, the renormalized Lagrangian is qualitatively very different from its classical counterpart.}

\begin{document} 
\maketitle
\flushbottom

\section{Introduction} 
Quantum field theory (QFT) is one of the most elegant and accurate frameworks which describes nature to a very high accuracy. QFTs admit deformations by operators which trigger a flow, known as the renormalization group (RG) flow, as we look into the theory at different scales \cite{Wilson}, \cite{Polchinski}. The fixed points of the RG-flow trajectory define conformal field theories (CFT) which are crucial in understanding aspects of condensed matter physics and  string theory. However, it is equally significant to understand the RG flow away from the fixed points by deforming the QFT with a relevant or irrelevant operator. A relevant deformation stimulates the flow at lower energies (IR) while an irrelevant deformation drives the flow at higher energies (UV). Exploring the later is more difficult as it requires infinitely many counterterms in computing physical quantities in such a theory. However, a special kind of irrelevant deformation, known as the the $T\bar{T}$-deformation, has been introduced recently in two dimensions \cite{Zamolodchikov}, \cite{Smirnov} and has received wide attention in the past few years as the theory is solvable.  
\\\\
The $T\bar{T}$-deformation is generated by the irrelevant operator $\det (T_{\mu\nu})$, the determinant of the energy-momentum tensor $T_{\mu\nu}$ of the theory, and thus can be considered in any generic QFT. At the classical level, the Lagrangian of the deformed theory itself looks very interesting and was obtained in \cite{Andrea}, \cite{Bonelli_2018}. For example, the Lagrangian of a \ttb-deformed free massless boson is equivalent to the Nambu-Goto action for a string in three spacetime dimensions in the static gauge \cite{Andrea}. Although the deformation makes the theory non-local, often complicated and non-renormalizable, there are several remarkable features that make the deformed theory so compelling. One such feature is that the energy spectrum of the deformed theory can be derived non-perturbatively and in a compact form \cite{Zamolodchikov}. However, the spectrum appears to be sensitive to the sign of the \ttb-coupling. For one of the signs, the highly excited states in the spectrum carry complex energies where the deformed theory becomes non-unitary. For this particular sign of the coupling, a holographic dual was proposed in \cite{McGough} and was further investigated in \cite{Kraus} - \cite{Caputa}. For the other sign of the coupling, the spectrum is real and the deformed theory is unitary. Moreover, at high energies, the density of states exhibit Hagedorn behaviour like the two dimensional little string theory (LST). In the context of holography, it was argued that certain single-trace \ttb-deformation of two dimensional CFTs corresponds to a  two dimensional vacuum of LST \cite{Giveon1}, see \cite{Giveon2}-\cite{Barbon} for related exciting works in this direction.
Another interesting aspect of the two dimensional \ttb-deformed theories is the partition function - one can compute the partition function of the deformed theory on  a torus or a cylinder or a disk and each of them satisfy a linear diffusion-type differential equation \cite{Cardy1}. In particular, \ttb-deformation of a CFT was considered in \cite{Datta} and the torus partition function of the deformed CFT was obtained to be modular invariant despite the fact that the deformed theory is not conformal. See \cite{Aharony}, \cite{Jiang} for further exciting results concerning the partition functions in the \ttb-deformed CFTs.
\\\\
Although the $T\bar{T}$-deformation can be defined in any two dimensional QFT, a special interest has been taken in the study of integrable quantum systems \cite{Smirnov}, \cite{Delfino}-\cite{Caselle}. Integrable QFTs contain infinite number of  conserved charges. The reason for this interest is that a $T\bar{T}$ deformation preserves the integrable structure of the theory \cite{Smirnov}. In \cite{Smirnov}, it was also argued that the $T\bar{T}$-deformation modifies the $S$-matrix only by a CDD phase factor. 
\\\\
Being an irrelevant deformation, the Lagrangian of the $T\bar{T}$-deformed theory is apparently non-renormalizable as it involves an infinite number of counterterms. However, it is the integrable structure of the theory that enables one to obtain a renormalized Lagrangian in such a naively non-renormalizable theory. The $T\bar{T}$-deformed integrable theory provides an infinite number of constraints which  uniquely fix the all counterterms that appear in the theory.  In \cite{Rosenhaus}, the authors considered a $T\bar{T}$-deformed free scalar field theory and showed how to derive the renormalized Lagrangian perturbatively by demanding that it should produce the correct $S$-matrix. Obtaining a renormalized Lagrangian unambiguously allows one to compute all physical quantities of interest. This motivates us to consider the $T\bar{T}$-deformed free massive Dirac fermion in two dimensions and to compute the renormalized Lagrangian of such a theory perturbatively. As we will see, the renormalized Lagrangian exists and is qualitatively very different from the classical Lagrangian. In particular, while the classical Lagrangian has only one scale, namely the bare coupling constant, the renormalized Lagrangian to second order reveals three different scales in the theory.
\\\\
In this paper, the perturbative renormalization of the \ttb-deformed free massive Dirac fermion in two dimensions will be explicitly performed. This will be done using the LSZ reduction formula to compute the two-to-two $S$-matrix and introducing counterterms to cancel divergences. The paper is organized as follows: In section \ref{integrability} the basics of \ttb-deformations as well as some related  aspects of integrability will be reviewed. In section \ref{classicalLag} the \ttb-deformed free massive Dirac fermion in two dimensions will be considered. In section \ref{renormalizedLag} the renormalized Lagrangian to second order in the \ttb-coupling constant will be computed. Finally, in section \ref{Discussion} the main results and future directions will be discussed. 

\section{{Integrability and \ttbd s}}
\label{integrability}
In this section several aspects of integrability, two-dimensional quantum field theories and \ttb-deformations are discussed.
\vspace{2mm}
\\
In two dimensional integrable theories the $S$-matrix of any scattering process factorizes into two-to-two $S$-matrices and there is no particle production \cite{ZZ}-\cite{Bombardelli}.
Consider a two-to-two scattering process of  particles with identical mass $m$ in two dimensions. The 2-momentum of the $i^{th}$ particle can be parameterized by its rapidity, $\theta_i$\footnote{Rapidity or the parameter of  velocity is defined as, $\theta_i=\tanh^{-1}(v_i)$, where $v_i$ is the velocity of the $i$-th particle and the speed of light is set to $c=1$.}, as,
\begin{equation}
	p_i^{\mu}=
	\left(\begin{array}{lr}
		m\cosh\theta_i\\
		m\sinh\theta_i	
	\end{array}\right)\ .
	\label{twomomenta}
\end{equation}
 Identify $p_i^{0}=m  \cosh\theta_i\equiv E_i(\theta_i)$ as the energy of the $i^{th}$ particle and $p_i^{1}=m \sinh\theta_i\equiv P_i(\theta_i)$ as the momentum of the $i^{th}$ particle. 
\\
Two dimensional kinematics of the two-to-two scattering processes have the unique property that the incoming momenta of the particles equal the outgoing momenta of the particles. This fact and the Lorentz invariance implies that the $S$-matrix, denoted by $S(\theta)$, will only be a function of the difference of rapidities, $\theta=\theta_1-\theta_2$.
\vspace{2mm}
\\
In what follows the \ttb-deformation of an integrable field theory will be considered.  If a \ttb-deformation is performed on an integrable field theory, the deformed theory is integrable as well \cite{Smirnov}. As we mentioned earlier,  there is no particle production in a scattering process in an integrable theory and hence, unitarity demands
\bea
|S(\theta)|^2=1\ .
\label{unitarity}
\eea
On the other hand, the crossing symmetry of the $S$-matrix\footnote{It is the symmetry of the $S$-matrix under the interchange of the $s$ and $u$-channels as will be shown later.} implies,
 \bea
  S(\theta)=S(i\pi-\theta)\ . 
  \label{crossing}
  \eea
  The solution to (\ref{unitarity}) and (\ref{crossing}) is simple and given by the CDD factor,
\bea
S_{\alpha}(\theta)=\frac{\sinh\theta-i\sin\alpha}{\sinh\theta+i\sin\alpha} \ ,
\label{CDD}
\eea
where $\alpha$ is a real parameter related to the coupling constant. The product of $S_{\alpha}(\theta)$ over $\alpha$, $\prod_{\alpha}S_{\alpha}(\theta)$, is also a solution to (\ref{unitarity}) and (\ref{crossing}).
\\
\\
An alternative solution for the CDD factor was recently considered by Smirnov and Zamolodchikov \cite{Smirnov}, which admits the following representation
 \bea
S_{\alpha}(\theta)= e^{i\alpha \sinh\theta} \ .
\label{CDDSZ}
\eea
The authors considered \ttb-deformed theories in two dimensions\footnote{To be precise, the authors in \cite{Smirnov} considered integrable quantum field theories (IQFT) and deformed them by generic scalars $X_s$ such that the deformations preserve integrability. Being IQFT, these theories have an infinite number of conserved currents ($T_{s+1}(z), \Theta_{s-1}(z)$) and ($\bar{T}_{s+1}(z), \bar{\Theta}_{s-1}(z)$), where the index $s$ represents the spin of the corresponding fields thus labeling the currents. The scalars $X_s$ are defined in terms of these local currents. For example, $X_1$ is precisely the composite operator \ttb  \hspace{1mm}considered in our paper.} where the deformation produces a one-parameter family of Lagrangians obeying the $T\bar{T}$ flow equation,
\bea
\frac{\partial \mathcal{L}(\lambda)}{\partial \lambda}=-4\Big(T^{\lambda}(z)\bar{T}^{\lambda}(z)-\theta^{\lambda}(z)\bar\theta^{\lambda}(z)\Big)\ ,
\label{floweqn}
\eea
 $\lambda$ being the \ttb-coupling constant, $T^{\lambda}=T_{zz}^{\lambda}$, $\bar{T}^{\lambda}=T_{\bar{z} \bar{z}}^{\lambda}$ and $\theta^{\lambda}=\bar\theta^{\lambda}=T_{z\bar{z}}^{\lambda}$ are the components of the energy momentum tensor of the deformed theory and $\mathcal{L}(\lambda=0)$  is the Lagrangian of the undeformed theory. 
It was argued that the deformed theory is integrable and the $S$-matrix of this theory can be expressed in a factorizable form, $\hat{S'}(\theta)S(\theta)$, where $S(\theta)$ is the CDD factor determined by unitarity and crossing-symmetry of the $S$-matrix,
\bea
S(\theta)=e^{i\lambda m^2 \sinh\theta}\ .
\label{CDDSZ1}
\eea 
The factor $\hat{S'}(\theta)$ signifies the presence of mass degeneracies in the spectrum and it satisfies the Yang-Baxter equation \cite{Bombardelli}. This typically fixes the “flavor” structure of the $S$-matrix. Simply, the \ttb-deformation (\ref{floweqn}) corresponds to multiplying the $S$-matrix by the factor (\ref{CDDSZ1}).
\vspace{2mm}
\\
However, observe that (\ref{CDDSZ1}) grows exponentially at large imaginary momenta.
This behaviour is inconsistent with the analytic behaviour of $S$-matrices in a local QFT.  
 Nevertheless, rather than just throwing out these theories, one can try to understand such theories as QFTs coupled to gravity \cite{Gorbenko} - \cite{Conti2}.  
 In this paper, however, we will be restricted to low-energy regime of the \ttb-deformed theories. In this regime these are quantum field theories and can be studied perturbatively in the \ttb-coupling $\lambda$ around $\lambda=0$ \cite{Kraus}, \cite{Rosenhaus}, \cite{He1}-\cite{Dey} \footnote{For non-perturbative studies of the \ttb-deformed theories, see \cite{Cardy2}, \cite{Haruna}.}.
 
\section{The \ttb-deformed free massive Dirac fermion}
\label{classicalLag}
Consider the Euclidean action for the free massive  Dirac fermion with mass $m$,
\be
I_0=\int dx_1 dx_2\, \mathcal{L}_{0} =\int dx_1 dx_2\, \big(i \bpsi \gamma^{\mu}\del_{\mu} \psi -m \bpsi \psi \big)~.
\label{freeaction}
\ee
Performing a Wick rotation to the Euclidean action $x_1=i t$ and $x_2=x$, yields the Lorentzian action,
\be
-I_0=-\int dx_1 dx_2\, \mathcal{L}_{0} =i\int dt dx\, \big(-i \bpsi \gamma^{\mu}\del_{\mu} \psi +m \bpsi \psi \big)\ ,
\label{freeactionLor}
\ee
where the $\gamma^{\mu}$s are the two dimensional gamma matrices satisfying the Clifford algebra,
\be
\{\gamma^{\mu},\gamma^{\nu}\}=2\ \eta^{\mu\nu} \ \mathbb{I}
\label{Cliffordalgebra}
\ee
and $\eta_{\mu\nu}$ is the two dimensional Minkowski metric. We represent the gamma matrices in terms of the Pauli matrices: $\gamma_0=\sigma_x$ and $\gamma_1=-i\sigma_y$.
\\\\
The equation of motion for the fermionic fields $\psi$ and $\bpsi$ are given by,
\bea
i \gamma^{\mu}\del_{\mu}\psi-m\psi = 0\:\:\:\:\:\:\text{and}\:\:\:\:\:\:
i \del_{\mu}\bpsi \gamma^{\mu}+m\bpsi =0.
\label{Diraceq}
\eea
To solve the Dirac equation (\ref{Diraceq}), make the ansatz, $\psi(x)=u(k)e^{-i k .  x}$.  Plugging the ansatz
into the Dirac equation and using the normalization $u^{\dagger}(k)u(k)=1$, one finds the normalized positive energy plane wave solution,
\begin{align}
u(k)=\left(\begin{array}{c}\sqrt{E_{+}+k^1}\\ \frac{m}{\sqrt{E_{+}+k^1}}\end{array}\right) \hspace{3mm}
= \hspace{2mm}\left(\begin{array}{c}
	\sqrt{k^0+k^1}\\\sqrt{k^0-k^1}
\end{array}\right)
\label{+vesoln}
\end{align}
where, $E_{+}=k^0=\sqrt{(k^1)^2+m^2}$.
\\\\
Similarly, using $\psi(x)=v(k)e^{i k .  x}$ one can obtain the negative energy plane wave solution,
\begin{align}
	v(k)=\left(\begin{array}{c}-\sqrt{E_{-}+k^1}\\ \frac{m}{\sqrt{E_{-}+k^1}}\end{array}\right) \hspace{3mm}
	= \hspace{2mm}\left(\begin{array}{c}
		-\sqrt{k^0+k^1}\\\sqrt{k^0-k^1}
	\end{array}\right)
	\label{-vesoln}
\end{align}
where, $E_{-}=k^0=-\sqrt{(k^1)^2+m^2}$.
\\\\
Hence, the solution to the wave equation for the free massive Dirac fermion is,
\bea
	\psi(x)&=&\int \frac{d\vec{k}}{2\pi}\frac{1}{\sqrt{2E}}\left(a(\vec{k})u(\vec{k})e^{-ik\cdot x}+b^{\dagger}(\vec{k})v(\vec{k})e^{ik\cdot x}\right)\ ,
\eea
where $a^{\dagger}(\vec{k})$ and $a(\vec{k})$ are the fermion creation and annihilation operators respectively, while $b^{\dagger}(\vec{k})$ and $b(\vec{k})$ are the anti-fermion creation and annihilation operators respectively. The creation and annihilation operators satisfy the Clifford algebra,
\bea
\{a(\vec{k_1}), a^{\dagger}(\vec{k_2})\}=2\pi \delta(\vec{k_1}-\vec{k_2})\:\:\:\:\:\:\text{and}\:\:\:\:\:\:
\{b(\vec{k_1}), b^{\dagger}(\vec{k_2})\}=2\pi \delta(\vec{k_1}-\vec{k_2})\ , 
\eea
while all other anti-commutators vanish. The Dirac adjoint of a spinor $\psi$ is defined as, $\bpsi=\psi^{\dagger}\gamma^0$.
\\\\
Consider the $\bar{T}T$-deformation of the free massive Dirac fermion in two dimensional Euclidean spacetime,
\be
I=\int d^{\,2}x\, \mathcal{L}(\lambda)=\int dx_1 dx_2\, \big(i \bpsi \gamma^{\mu}\del_{\mu} \psi -m \bpsi \psi \big) + \lambda \int dx_1 dx_2\, \mathcal{O}_{T\bar{T}}\ ,
\label{deformedaction}
\ee
where $\mathcal{O}_{T\bar{T}}$ is the local $T\bar{T}$-operator given by the determinant of the energy-momentum tensor,
\be
\mathcal{O}_{T\bar{T}}=\det (T^{(\lambda)})=\frac{1}{2} \epsilon^{\mu\nu}\epsilon^{\rho\sigma}T_{\mu\nu}^{(\lambda)}T_{\rho\sigma}^{(\lambda)}
=\frac{1}{2}\Big[\big(T_{\mu}^{ \ \mu(\lambda)}\big)^2-T_{\mu\nu}^{(\lambda)}T^{\mu\nu(\lambda)}\Big]\ ,
\ee
$\epsilon^{\mu\nu}$  the two-dimensional Levi-Civita tensor and $T_{\mu\nu}^{(\lambda)}$ the energy-momentum tensor of the finite-$\lambda$ theory.
\\
The canonical energy-momentum tensor of the undeformed theory is given by,
\be
T_{\mu\nu (c)}^{(0)}=X_{\mu\nu}-\delta_{\mu\nu}(\Tr X -m \bpsi\psi)\ ,
\label{canemtensor}
\ee
where
\bea 
X_{\mu\nu}=\frac{i}{2}  \big(\bpsi\gamma_{\mu}\del_{\nu}\psi
-\del_{\nu}\bpsi \gamma_{\mu}\psi\big)\ .
\label{Xmunu}
\eea
The above canonical energy-momentum tensor can be symmetrized using the Belinfante technique, yielding
\bea
T_{\mu\nu}^{(0)}&=&\tilde{X}_{\mu\nu}-\delta_{\mu\nu}(\Tr X -m \bpsi\psi)\ ,
\label{emtensor}
\eea
where
\bea
\tilde{X}_{\mu\nu}&=&\frac{i}{2}  \Big(\bpsi\gamma_{(\mu}\del_{\nu)}\psi
-\del_{(\mu}\bpsi \gamma_{\nu)}\psi\Big) \nonumber\\
&=&\frac{i}{4}  \big(\bpsi\gamma_{\mu}\del_{\nu}\psi+\bpsi\gamma_{\nu}\del_{\mu}\psi
-\del_{\mu}\bpsi \gamma_{\nu}\psi-\del_{\nu}\bpsi \gamma_{\mu}\psi\big)\ .
\label{Xtmunu}
\eea
The \ttb-deformed Lagrangian of the free massive Dirac fermion can be obtained by solving the $T\bar{T}$ flow equation \cite{Bonelli_2018},
\bea
\frac{\partial \mathcal{L}(\lambda)}{\partial{\lambda}}=\mathcal{O}_{T\bar{T}}\ ,
\label{floweqn1}
\eea
with the initial condition $\mathcal{L}(\lambda=0)=\mathcal{L}_0$. 
\\
\\
Solving (\ref{floweqn1}) perturbatively in the \ttb-coupling $\lambda$, one finds the \ttb-deformed Lagrangian for the free massive Dirac fermion \cite{Bonelli_2018},
\bea
\mathcal{L}(\lambda)&=& \big(i \bpsi \gamma^{\mu}\del_{\mu} \psi -m \bpsi \psi \big)-\frac{\lambda}{2}\Big(\tilde{X}_{\mu\nu}\tilde{X}^{\mu\nu}-(\tilde{X}_{\mu}^{ \ \mu})^2+2m\bpsi\psi  \tilde{X}_{\mu}^{ \ \mu}-2m^2(\bpsi\psi)^2\Big)\nonumber\\
&&+\frac{\lambda^2}{2}m\bpsi\psi\Big(\tilde{X}_{\mu\nu}\tilde{X}^{\mu\nu}-(\tilde{X}_{\mu}^{ \ \mu})^2\Big)\ ,
\label{deformedLag1}
\eea
where $\tilde{X}_{\mu\nu}$ is given by (\ref{Xtmunu}). It is noteworthy to mention that the \ttb-deformed Lagrangian (\ref{deformedLag1}) is exact in $\lambda$, all the higher order terms in $\lambda$ vanish identically due to the Grassmann nature of the fermion fields.\footnote{Although, terms proportional to $\tilde{X}^4$ can be present at third order in the \ttb-coupling $\lambda$, the authors of \cite{Bonelli_2018} claimed that the $\mathcal{O}(\lambda^3)$ term vanishes by using Fierz identities. It is also possible to verify this claim directly by plugging the expression into MATHEMATICA.}
\\\\
Therefore, the \ttb-deformed Lorentzian action of the free massive Dirac fermion is,
\bea
-I=-\int dx_1 dx_2\, \mathcal{L} =i&\int& dt dx\, \Big[\big(-i \bpsi \gamma^{\mu}\del_{\mu} \psi +m \bpsi \psi \big)+\frac{\lambda}{2}\Big(\tilde{X}_{\mu\nu}\tilde{X}^{\mu\nu}-(\tilde{X}_{\mu}^{ \ \mu})^2\nonumber\\
&+&2m\bpsi\psi  \tilde{X}_{\mu}^{ \ \mu}-2m^2(\bpsi\psi)^2\Big)
-\frac{\lambda^2}{2}m\bpsi\psi\Big(\tilde{X}_{\mu\nu}\tilde{X}^{\mu\nu}-(\tilde{X}_{\mu}^{ \ \mu})^2\Big) \Big] \nonumber\\
\label{deformedactionLor}
\eea
In what follows two-to-two scattering in the \ttb-deformed free massive Dirac fermion theory will be considered. The $S$-matrix, $S(\theta)$, will be computed perturbatively to second order in the \ttb-coupling $\lambda$. Finally the $S$-matrix will be compared with (\ref{CDDSZ1}) and the renormalized Lagrangian will be constructed.

\section{Renormalization of the \ttb-deformed free massive Dirac fermion}
\label{renormalizedLag}
In  section \ref{classicalLag}, the classical Lagrangian of the \ttb-deformed free massive Dirac fermion was stated. In this section the renormalized Lagrangian will be computed to second order.
\vspace{2mm}
\\
In order to compute the renormalized Lagrangian similar methods to those found in \cite{Rosenhaus} will be used. First, the $S$-matrix will be computed using the classical Lagrangian giving
rise to UV divergences. Next, counterterms will be added to the Lagrangian in order to
cancel the divergences and ensure that the final $S$-matrix is given by (\ref{CDDSZ1}).

\subsection{The $S$-matrix}
\label{subsecSmatrix}
Consider the two-to-two scattering of a fermion and anti-fermion in the \ttb-deformed free massive Dirac theory\footnote{One can also consider the other possible two-to-two scattering, namely, the fermion-fermion scattering. However, the resulting renormalized Lagrangian would be the same as it does not depend on the particular scattering process. Because of this fact only the fermion anti-fermion scattering process is considered.}, 
\bea
f_1+\bar{f_2}\rightarrow f_3+\bar{f_4}\nonumber
\eea
where $f_1$ represents the incoming fermion with momentum $p_1$, $\bar{f_2}$ the incoming anti-fermion with momentum $p_2$, $f_3$ the outgoing fermion with momentum $p_3$ and  $\bar{f_4}$ the outgoing anti-fermion with momentum $p_4$.
\\
Recall, due to the two dimensional kinematics of two-to-two scattering the momenta of the incoming particles equal the  momenta of the outgoing particles. When considering fermion anti-fermion scattering the $0^{th}$-order $S$-matrix is just the identity. Hence, the momentum of the incoming fermion must equal the momentum of the outgoing fermion and the momentum of the incoming anti-fermion must equal the momentum of the outgoing anti-fermion\footnote{If  scattering in a scalar theory is considered \cite{Rosenhaus}, one can choose $p_1=p_4$ and $p_2=p_3$ also. However, it is easy to see that if one works with fermions, we must have $p_1=p_3$ and $p_2=p_4$ to have a non-zero scattering matrix.}. In particular,
\bea
p_1=p_3 \ \ \ \ \ \ \ \ \text{and} \ \ \ \ \ \ \ \ \ p_2=p_4 .
\label{inoutmomenta}
\eea
By (\ref{twomomenta}) and (\ref{inoutmomenta}) the Mandelstam variables in the $(+,-)$ signature take the form,
\bea
s&=&(p_1+p_2)^2=2m^2(1+\cosh\theta)\ , \nonumber \\
t&=&(p_1-p_3)^2=0 \ , \nonumber\\
u&=&(p_1-p_4)^2=2m^2(1-\cosh\theta)=4m^2-s\ .
\label{Mandelstam}
\eea
Observe that the Mandelstam variables can be related to each other by the transformations, 
\bea 
u=s|_{\theta\rightarrow i\pi-\theta} \ \ \ \ \  \text{and} \ \ \ \ \ \ t=s|_{\theta\rightarrow i\pi}. 
\label{crossingsymmetry}
\eea
Finally due to (\ref{inoutmomenta}), for the two-to-two scattering in an integrable theory the $S$-matrix, $S(\theta)$, is defined as, 
\bea
{}_{out} \lan p_3, p_4| p_1, p_2\ran_{in}&=& (2\pi)^2  \delta(p_1 - p_3) \delta(p_2 - p_4) 2E(p_1) 2 E(p_2) S(\theta)\ ,
\eea
where the zeroth order $S$-matrix in the \ttb-coupling is $S^{(0)}(\theta) = 1$. 
The $S$-matrix $S(\theta)$ is related to the scattering amplitude $\mathcal{A}$ by \cite{Rosenhaus},
\be 
S(\theta)= \frac{\mathcal{A}}{ 4 m^2\sinh\theta}~.
\label{AtoS}
\ee
However, before computing the $S$-matrix, express the deformed action (\ref{deformedactionLor}) in a  simpler form. By the field redefinitions, 
\bea
\psi_{a}\rightarrow \psi_a+ \lambda \frac{m}{2} \psi_{a}(\bpsi\psi)\:\:\:\:\:\text{and}\:\:\:\:\:
\bpsi_{a}\rightarrow \bpsi_a+ \lambda \frac{m}{2}(\bpsi\psi) \bpsi_{a}\ ,
\label{newfields}
\eea
where $a=\{1,2\}$ are the spinor indices, one can write the action (\ref{deformedactionLor}) as,
\bea
-I=i\int dt dx\, &\Big[&\big(-i \bpsi \gamma^{\mu}\del_{\mu} \psi +m \bpsi \psi \big)+\frac{\lambda}{2}\Big(\tilde{X}_{\mu\nu}\tilde{X}^{\mu\nu}-(\tilde{X}_{\mu}^{\mu})^2\Big)\nonumber\\
&+&\frac{\lambda^2}{2}m\bpsi\psi\Big(\tilde{X}_{\mu\nu}\tilde{X}^{\mu\nu}-(\tilde{X}_{\mu}^{\mu})^2\Big)+\frac{7}{4}\lambda^2 m^2(\bar\psi\psi)^2 \ \tilde{X}_{\mu}^{\mu}-\frac{7}{4}\lambda^2 m^3 (\bar\psi\psi)^3
\Big].\nonumber\\
\label{deformedactionLor2}
\eea 
For details see Appendix \ref{AppendixA}.
\\
\\
Observe that the new action no longer contains linear terms of the form $\bpsi\psi \tilde{X}_{\mu}^{ \ \mu}$ and $(\bpsi\psi)^2$, making it simpler to compute the one-loop bubble diagrams by decreasing the total number of terms. But, at second order, two new terms $(\bpsi\psi)^2 \tilde{X}_{\mu}^{ \ \mu}$ and $(\bpsi\psi)^3$ are generated which contribute to the tadpole diagrams at one-loop. However, these new terms give vanishing contributions as will be shown later.
\vspace{2mm}
\\
The two dimensional Dirac spinors $u(p_i)$, $\bar u(p_i)$, $v(p_i)$ and $\bar v(p_i)$ will arise in the computation of the $S$-matrix. Using (\ref{+vesoln}), (\ref{-vesoln}) and  the fact that $p_3^{\mu}=p_1^{\mu}=(m \cosh \theta_1,  \ m\sinh \theta_1)$ and $p_4^{\mu}=p_2^{\mu}=(m \cosh \theta_2,  \ m\sinh \theta_2)$ the two dimensional Dirac spinors can be written as,

\begin{align}
u(p_1)=\sqrt{m}\left(\begin{array}{c}
		\sqrt{\cosh\theta_1+\sinh\theta_1}\\\sqrt{\cosh\theta_1-\sinh\theta_1}
	\end{array}\right)=u(p_3) \ , \hspace{4mm}
	u(p_2)=\sqrt{m}\left(\begin{array}{c}
		\sqrt{\cosh\theta_2+\sinh\theta_2}\\\sqrt{\cosh\theta_2-\sinh\theta_2}
	\end{array}\right)=u(p_4) \nonumber\\
v(p_1)=\sqrt{m}\left(\begin{array}{c}
		-\sqrt{\cosh\theta_1+\sinh\theta_1}\\\sqrt{\cosh\theta_1-\sinh\theta_1}
	\end{array}\right)=v(p_3)  \ , \hspace{4mm}
	v(p_2)=\sqrt{m}\left(\begin{array}{c}
		-\sqrt{\cosh\theta_2+\sinh\theta_2}\\\sqrt{\cosh\theta_2-\sinh\theta_2}
	\end{array}\right)=v(p_4)\ .\nonumber\\
	\label{uv}
\end{align}
The Dirac adjoints  $\bar u(p_i)=u(p_i)^{\dagger} \gamma^0$ and $\bar v(p_i)=v(p_i)^{\dagger} \gamma^0$ can be written as,
\begin{align}
	\bar u(p_1)=\sqrt{m}\left(\begin{array}{c}
		\sqrt{\cosh\theta_1-\sinh\theta_1} \hspace{10mm} \sqrt{\cosh\theta_1+\sinh\theta_1}
	\end{array}\right)=\bar u(p_3)\ , \nonumber\\
	\bar u(p_2)=\sqrt{m}\left(\begin{array}{c}
		\sqrt{\cosh\theta_2-\sinh\theta_2} \hspace{10mm} \sqrt{\cosh\theta_2+\sinh\theta_2}
	\end{array}\right)=\bar u(p_4)\ , \nonumber\\
	\bar v(p_1)=\sqrt{m}\left(\begin{array}{c}
		\sqrt{\cosh\theta_1-\sinh\theta_1} \hspace{10mm} -\sqrt{\cosh\theta_1+\sinh\theta_1}
	\end{array}\right)=\bar v(p_3)\ , \nonumber\\
	\bar v(p_2)=\sqrt{m}\left(\begin{array}{c}
		\sqrt{\cosh\theta_2-\sinh\theta_2} \hspace{10mm} -\sqrt{\cosh\theta_2+\sinh\theta_2}
	\end{array}\right)=\bar v(p_4)\ .
	\label{ubarvbar}
\end{align}
It is finally time to compute the $S$-matrix perturbatively up to second order in the \ttb-coupling $\lambda$ using the redefined Lagrangian,
\bea
\mathcal{L}(\lambda)=\mathcal{L}_0+\mathcal{L}_1(\lambda)+\mathcal{L}_2(\lambda)\ ,
\label{redefinedLag}
\eea
where,
\bea
\mathcal{L}_0&=&-i \bpsi \gamma^{\mu}\del_{\mu} \psi +m \bpsi \psi\ , \nonumber\\
\mathcal{L}_1(\lambda)&=&\frac{\lambda}{2}\Big(\tilde{X}_{\mu\nu}\tilde{X}^{\mu\nu}-(\tilde{X}_{\mu}^{\ \mu})^2\Big)\ ,\nonumber\\
\mathcal{L}_2(\lambda)&=&\frac{\lambda^2}{2}m\bpsi\psi\Big(\tilde{X}_{\mu\nu}\tilde{X}^{\mu\nu}-(\tilde{X}_{\mu}^{ \ \mu})^2\Big)+\frac{7}{4}\lambda^2 m^2(\bpsi\psi)^2 \tilde{X}_{\mu}^{ \ \mu}-\frac{7}{4}\lambda^2 m^3(\bpsi\psi)^3 . 
\label{redefinedL0L1L2}
\eea

\vspace{1mm}
\subsubsection{First order $S$-matrix}
\begin{figure}[h!]
	\centering \noindent
	\includegraphics[width=8cm]{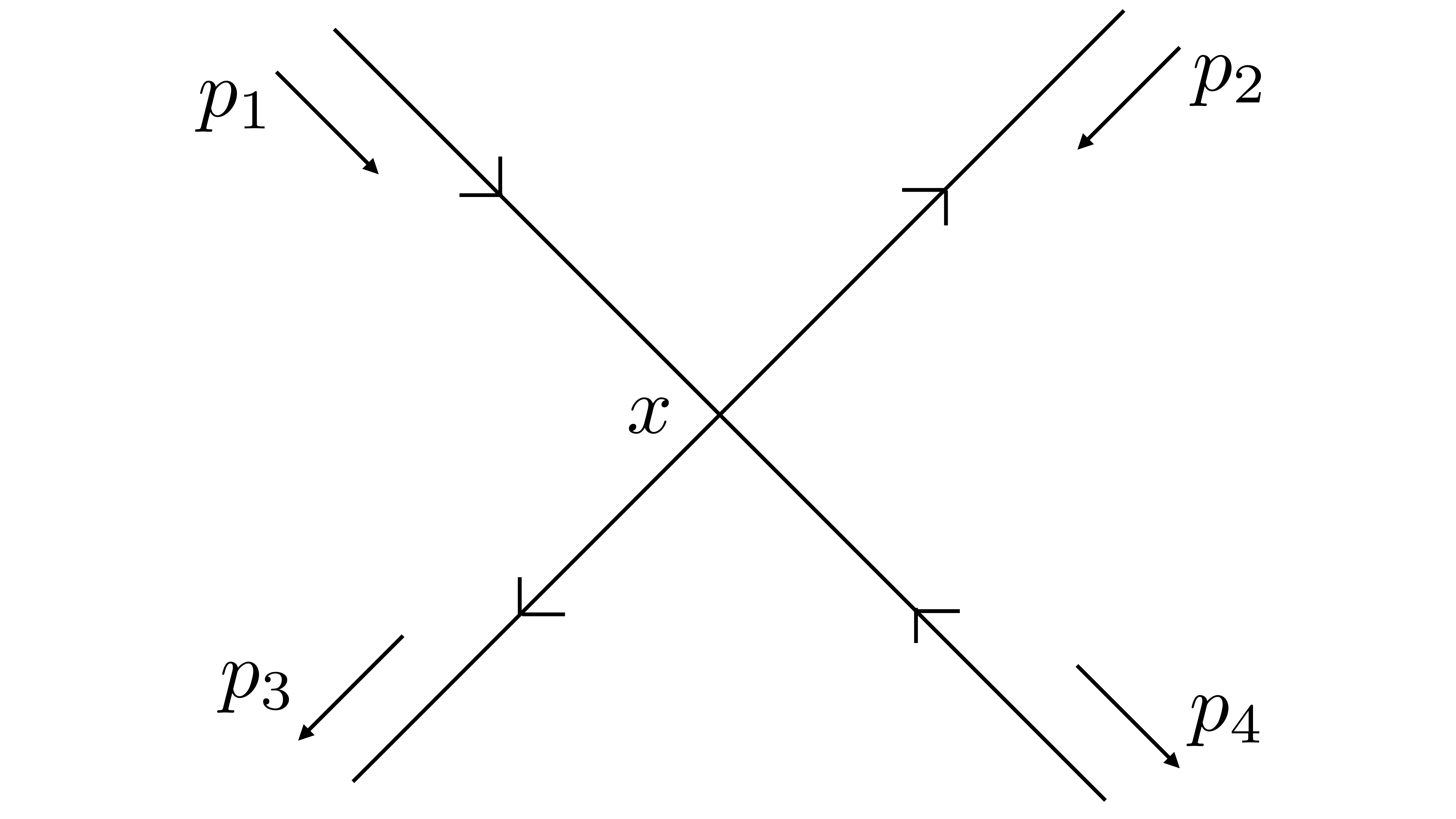}
	\caption{Tree-level diagram that contributes to the $S$-matrix at first order. 
	}
	\label{treelevel diagram}
\end{figure}
At first order in the \ttb-coupling $\lambda$, the $S$-matrix only gets a contribution from the tree-level diagrams of the form found in figure \ref{treelevel diagram}, where the vertex corresponds to the quartic couplings, $\tilde{X}_{\mu\nu} \tilde{X}^{\mu\nu}$ or $(\tilde{X}_{\mu}^{\ \mu})^2$. The total contribution to the tree-level amplitude of a fermion anti-fermion scatering process is,
\bea
\mathcal{A}^{(1)}=i \frac{\lambda}{2}\lan 0 | b(p_4) a(p_3) \tT\big[\tilde{X}_{\mu\nu}\tilde{X}^{\mu\nu}-(\tilde{X}_{\mu}^{\ \mu})^2\big] a^{\dagger}(p_1) b^{\dagger}(p_2)|0 \ran.
\label{treeamplitude}
\eea
In order to evaluate (\ref{treeamplitude}) compute the general contribution to the amplitude of four external fields with arbitrary indices for the fermion anti-fermion scattering process,
\bea
\mathcal{A}^{(1)}_{\bpsi_a\psi_b\bpsi_c\psi_d}&=&\lan 0 | b(p_4) a(p_3) \tT\big[\bpsi_a\psi_b\bpsi_c\psi_d\big] a^{\dagger}(p_1) b^{\dagger}(p_2)|0 \ran\nonumber\\ &=&\bu_a(p_3)v_b(p_4)\bv_c(p_2)u_d(p_1)+\bv_a(p_2)u_b(p_1)\bu_c(p_3)v_d(p_4)
-\bv_a(p_2)v_b(p_4)\bu_c(p_3)u_d(p_1)\nonumber\\
&&-\bu_a(p_3)u_b(p_1)\bv_c(p_2)v_d(p_4)
\label{generaltree}
\eea
where the four possible Wick contractions were performed, $u_a$, $v_a$, $\bar{u}_a$ and $\bar{v}_a$ are the Dirac spinors given by (\ref{uv}) and (\ref{ubarvbar}) and $a, b, c, d=\{1,2\}$ are the spinor indices.
\vspace{2mm}
\\
 The total first order amplitude can be computed by plugging each vertex into (\ref{generaltree}) and adding the results together. 
By (\ref{Xtmunu}), the quartic couplings $\tilde{X}_{\mu\nu}\tilde{X}^{\mu\nu}$ and $(\tilde{X}_{\mu}^{\ \mu})^2$ can be expressed as,
\begin{align}
	\tilde{X}_{\mu\nu}\tilde{X}^{\mu\nu}
	&=-\frac{1}{4}  \Big(\bpsi\gamma_{(\mu}\del_{\nu)}\psi \bpsi\gamma^{(\mu}\del^{\nu)}\psi
	-2\bpsi\gamma_{(\mu}\del_{\nu)}\psi \del^{(\mu}\bpsi \gamma^{\nu)}\psi
	+\del_{(\mu}\bpsi \gamma_{\nu)}\psi \del^{(\mu}\bpsi \gamma^{\nu)}\psi\Big) 
	\label{TmunuSquare}\\ 
	(\tilde{X}_{\mu}^{\ \mu})^2
	&=-\frac{1}{4}  \Big(\bpsi\gamma^{\mu}\del_{\mu}\psi \ \bpsi\gamma^{\nu}\del_{\nu}\psi
	-2\bpsi\gamma^{\mu}\del_{\mu}\psi \ \del_{\nu}\bpsi \gamma^{\nu}\psi
	+\del_{\mu}\bpsi \gamma^{\mu}\psi \ \del_{\nu}\bpsi \gamma^{\nu}\psi\Big) 
	\label{TmumuSquare}.
\end{align}
Using (\ref{generaltree}) the tree-level contribution from each of these terms in (\ref{TmunuSquare}) can be computed and thus the contribution from $\tilde{X}_{\mu\nu}\tilde{X}^{\mu\nu}$ to the tree-level amplitude can be determined.
For example, the first term in (\ref{TmunuSquare}), $\bpsi\gamma_{(\mu}\del_{\nu)}\psi \bpsi\gamma^{(\mu}\del^{\nu)}\psi$, contributes,
\bea
&&\mathcal{A}_{\bpsi\gamma_{(\mu}\del_{\nu)}\psi \bpsi\gamma^{(\mu}\del^{\nu)}\psi}^{(1)}\nonumber\\
&=&\lan 0 | b(p_4) a(p_3) \tT\big[\bpsi\gamma_{(\mu}\del_{\nu)}\psi \bpsi\gamma^{(\mu}\del^{\nu)}\psi\big] a^{\dagger}(p_1) b^{\dagger}(p_2)|0 \ran \nonumber\\
 &=&\gamma_{ab(\mu}\gamma_{cd}^{(\mu}\lan 0 | b(p_4) a(p_3) \tT\big[\bpsi_a \del_{\nu)}\psi_b\bpsi_c\del^{\nu)}\psi_d\big] a^{\dagger}(p_1) b^{\dagger}(p_2)|0 \ran\nonumber\\
&=&\gamma_{ab(\mu}\gamma_{cd}^{(\mu} \Big[\bu_a(p_3)v_b(p_4)\bv_c(p_2)u_d(p_1) i p_{4\nu)} (-i)p_1^{\nu)}
+\bv_a(p_2)u_b(p_1)\bu_c(p_3)v_d(p_4)(-i)p_{1\nu)} i p_4^{\nu)}\nonumber\\
&&-\bv_a(p_2)v_b(p_4)\bu_c(p_3)u_d(p_1)i p_{4\nu)} (-i)p_1^{\nu)}
-\bu_a(p_3)u_b(p_1)\bv_c(p_2)v_d(p_4)(-i)p_{1\nu)} i p_4^{\nu)}\Big]\nonumber\\
&=& \bu(p_1)\cdot \gamma_{(\mu}p_{2\nu)} \cdot v(p_2) \
\bv(p_2)\cdot \gamma^{(\mu}p_1^{\nu)}\cdot  u(p_1)
+\bv(p_2)\cdot \gamma_{(\mu}p_{1\nu)} \cdot u(p_1) \
\bu(p_1)\cdot \gamma^{(\mu}p_2^{\nu)}\cdot  v(p_2)\nonumber\\
&&-\bv(p_2)\cdot \gamma_{(\mu}p_{2\nu)} \cdot v(p_2) \
\bu(p_1)\cdot \gamma^{(\mu}p_1^{\nu)}\cdot  u(p_1)
-\bu(p_1)\cdot \gamma_{(\mu}p_{1\nu)} \cdot u(p_1) \
\bv(p_2)\cdot \gamma^{(\mu}p_2^{\nu)}\cdot  v(p_2)\nonumber\\
&=& -2m^4\big(1+3\cosh \theta +2 \cosh 2\theta\big).
\label{tree1}
\eea
where on the third line the general amplitude, (\ref{generaltree}), was used, a factor of $\pm i$ appears whenever a derivative operator acts on $\psi$ or $\bpsi$ \footnote{$+i$ appears if the derivative operator acting on $\psi(x)$ gives an outgoing momentum $p_3$ or $p_4$, while $-i$ appears if the derivative operator produces an ingoing momentum $p_1$ or $p_2$ upon acting on $\psi(x)$.}. (\ref{inoutmomenta}) was used to get the fourth equality and on the final line the mathematical expressions for the Dirac spinors (\ref{uv}) and (\ref{ubarvbar}) were plugged in.
\vspace{2mm}
\\
Similarly, the contributions from the other two terms in (\ref{TmunuSquare}) can be computed. The computation yields,
\bea
\mathcal{A}_{\bpsi\gamma_{(\mu}\del_{\nu)}\psi \del^{(\mu}\bpsi \gamma^{\nu)}\psi}^{(1)}&=&\lan 0 | b(p_4) a(p_3) \tT\big[\bpsi\gamma_{(\mu}\del_{\nu)}\psi \del^{(\mu}\bpsi \gamma^{\nu)}\psi\big] a^{\dagger}(p_1) b^{\dagger}(p_2)|0 \ran \nonumber\\
&=& -2m^4\big(1-\cosh \theta -2 \cosh 2\theta\big) 
\label{tree2}
\eea
\bea
\mathcal{A}_{\del_{(\mu}\bpsi \gamma_{\nu)}\psi \del^{(\mu}\bpsi \gamma^{\nu)}\psi}^{(1)}&=&\lan 0 | b(p_4) a(p_3) \tT\big[\del_{(\mu}\bpsi \gamma_{\nu)}\psi \del^{(\mu}\bpsi \gamma^{\nu)}\psi\big] a^{\dagger}(p_1) b^{\dagger}(p_2)|0 \ran \nonumber\\
&=& -2m^4\big(1+3\cosh \theta +2 \cosh 2\theta\big)\ .
\label{tree3}
\eea
Adding together (\ref{tree1}), (\ref{tree2}) and (\ref{tree3}) according to (\ref{TmunuSquare}) gives the contribution of the quartic coupling $\tilde{X}_{\mu\nu}\tilde{X}^{\mu\nu}$,
\bea
\mathcal{A}_{\tilde{X}_{\mu\nu}\tilde{X}^{\mu\nu}}^{(1)}&=&- \  \frac{1}{4}\Big(\mathcal{A}_{\bpsi\gamma_{(\mu}\del_{\nu)}\psi \bpsi\gamma^{(\mu}\del^{\nu)}\psi}^{(1)}-2\mathcal{A}_{\bpsi\gamma_{(\mu}\del_{\nu)}\psi \del^{(\mu}\bpsi \gamma^{\nu)}\psi}^{(1)}+\mathcal{A}_{\del_{(\mu}\bpsi \gamma_{\nu)}\psi \del^{(\mu}\bpsi \gamma^{\nu)}\psi}^{(1)}\Big)\nonumber\\
&=&4 m^4  (\cosh\theta+\cosh2\theta).
\label{treeamplitudeTmunuSq}
\eea
One can compute the contributions from the three terms found in (\ref{TmumuSquare}) to evaluate the contribution from $(\tilde{X}_{\mu}^{ \ \mu})^2$ to the tree level amplitude in a similar manner,
\bea
\mathcal{A}_{\bpsi\gamma^{\mu}\del_{\mu}\psi \bpsi\gamma^{\nu}\del_{\nu}\psi}^{(1)}=\lan 0 | b(p_4) a(p_3) \tT\big[\bpsi\gamma^{\mu}\del_{\mu}\psi \bpsi\gamma^{\nu}\del_{\nu}\psi\big] a^{\dagger}(p_1) b^{\dagger}(p_2)|0 \ran &=& -4m^4\big(1+\cosh \theta\big) \nonumber\\
\mathcal{A}_{\bpsi\gamma^{\mu}\del_{\mu}\psi \del_{\nu}\bpsi \gamma^{\nu}\psi}^{(1)}=\lan 0 | b(p_4) a(p_3) \tT\big[\bpsi\gamma^{\mu}\del_{\mu}\psi \del_{\nu}\bpsi \gamma^{\nu}\psi\big] a^{\dagger}(p_1) b^{\dagger}(p_2)|0 \ran &=& 4m^4\big(1+\cosh \theta \big) \nonumber\\
\mathcal{A}_{\del_{\mu}\bpsi \gamma^{\mu}\psi\del_{\nu}\bpsi \gamma^{\nu}\psi}^{(1)}=\lan 0 | b(p_4) a(p_3) \tT\big[\del_{\mu}\bpsi \gamma^{\mu}\psi\del_{\nu}\bpsi \gamma^{\nu}\psi\big] a^{\dagger}(p_1) b^{\dagger}(p_2)|0 \ran &=& -4m^4\big(1+\cosh \theta\big). \nonumber\\
\label{tree4_5_6}
\eea
Adding the terms in (\ref{tree4_5_6}) according to (\ref{TmumuSquare}) yields the contribution from the quartic coupling $(\tilde{X}_{\mu}^{\ \mu})^2$,
\bea
\mathcal{A}_{(\tilde{X}_{\mu}^{\ \mu})^2}^{(1)}&=&-  \frac{1}{4}\Big(\mathcal{A}_{\bpsi\gamma^{\mu}\del_{\mu}\psi \bpsi\gamma^{\nu}\del_{\nu}\psi}^{(1)}-2\mathcal{A}_{\bpsi\gamma^{\mu}\del_{\mu}\psi \del_{\nu}\bpsi \gamma^{\nu}\psi}^{(1)}+\mathcal{A}_{\del_{\mu}\bpsi \gamma^{\mu}\psi\del_{\nu}\bpsi \gamma^{\nu}\psi}^{(1)}\Big)\nonumber\\
&=&4 m^4  (1+\cosh\theta).
\label{treeamplitudeTmumuSq}
\eea
Substituting (\ref{treeamplitudeTmunuSq}) and (\ref{treeamplitudeTmumuSq}) into (\ref{treeamplitude}) gives the total tree-level amplitude,
\bea
\mathcal{A}^{(1)}&=&i\frac{\lambda}{2}\Big[\mathcal{A}_{\tilde{X}_{\mu\nu}\tilde{X}^{\mu\nu}}^{(1)}-\mathcal{A}_{(\tilde{X}_{\mu}^{\ \mu})^2}^{(1)}\Big]
=4 i \lambda m^4  \sinh ^2\theta. 
\label{treeamplitudefinal}
\eea
Therefore, by (\ref{AtoS}) the first order $S$-matrix is,
\bea
S^{(1)}(\theta)=\frac{\mathcal{A}^{(1)}}{4 m^2 \sinh \theta}= i \lambda m^2  \sinh\theta \ .
\label{treeSmatrix}
\eea
This result exactly matches with what one would expect from (\ref{CDDSZ1}) at linear order in the \ttb-coupling $\lambda$.
\subsubsection{Second order $S$-matrix}
At second order the $S$-matrix gets contribution from one-loop diagrams. At one-loop, two types of diagrams can contribute to the second order amplitude: the first one is the tadpole diagram while the second one is the bubble diagram. Both of these diagrams will be computed in this section.
\subsubsection*{Contribution from tadpole diagrams}
The second order Lagrangian of the \ttb-deformed free massive Dirac fermion can be written as,
\bea
i\mathcal{L}_2(\lambda)&=&i\Big[\frac{\lambda^2}{2}m\bpsi\psi\Big(\tilde{X}_{\mu\nu}\tilde{X}^{\mu\nu}-(\tilde{X}_{\mu}^{ \ \mu})^2\Big)+\frac{7}{4}\lambda^2 m^2(\bpsi\psi)^2 \tilde{X}_{\mu}^{ \ \mu}-\frac{7}{4}\lambda^2 m^3(\bpsi\psi)^3 \Big]
\label{L2L0}
\eea
where,
\bea 
\bpsi\psi \tilde{X}_{\mu\nu}\tilde{X}^{\mu\nu}
&=&-\frac{1}{4}  \bpsi\psi \Big(\bpsi\gamma_{(\mu}\del_{\nu)}\psi \bpsi\gamma^{(\mu}\del^{\nu)}\psi
-2\bpsi\gamma_{(\mu}\del_{\nu)}\psi \del^{(\mu}\bpsi \gamma^{\nu)}\psi
+\del_{(\mu}\bpsi \gamma_{\nu)}\psi \del^{(\mu}\bpsi \gamma^{\nu)}\psi\Big)\ , \nonumber\\
\label{psibarpsiTmunuSquare}
\eea
\bea
\hspace{-8mm}\bpsi\psi(\tilde{X}_{\mu}^{ \ \mu})^2
&=&-\frac{1}{4}  \bpsi\psi \Big(\bpsi\gamma^{\mu}\del_{\mu}\psi \ \bpsi\gamma^{\nu}\del_{\nu}\psi
-2\bpsi\gamma^{\mu}\del_{\mu}\psi \ \del_{\nu}\bpsi \gamma^{\nu}\psi
+\del_{\mu}\bpsi \gamma^{\mu}\psi \ \del_{\nu}\bpsi \gamma^{\nu}\psi\Big)\ , 
\label{psibarpsiTmumuSquare}
\eea
\bea
\hspace{-7.1cm}(\bpsi\psi)^2 \tilde{X}_{\mu}^{\ \mu}
&=&\frac{i}{2} (\bpsi\psi)^2   \big(\bpsi\gamma^{\mu}\del_{\mu}\psi
-\del_{\mu}\bpsi \gamma^{\mu}\psi\big) \ .
\label{psibarpsiSqTmumu}
\eea
Since the interaction vertices contain six fields, two internal fields must be contracted resulting in loops.
\begin{figure}[t!]
	\centering \noindent
	\includegraphics[width=8cm]{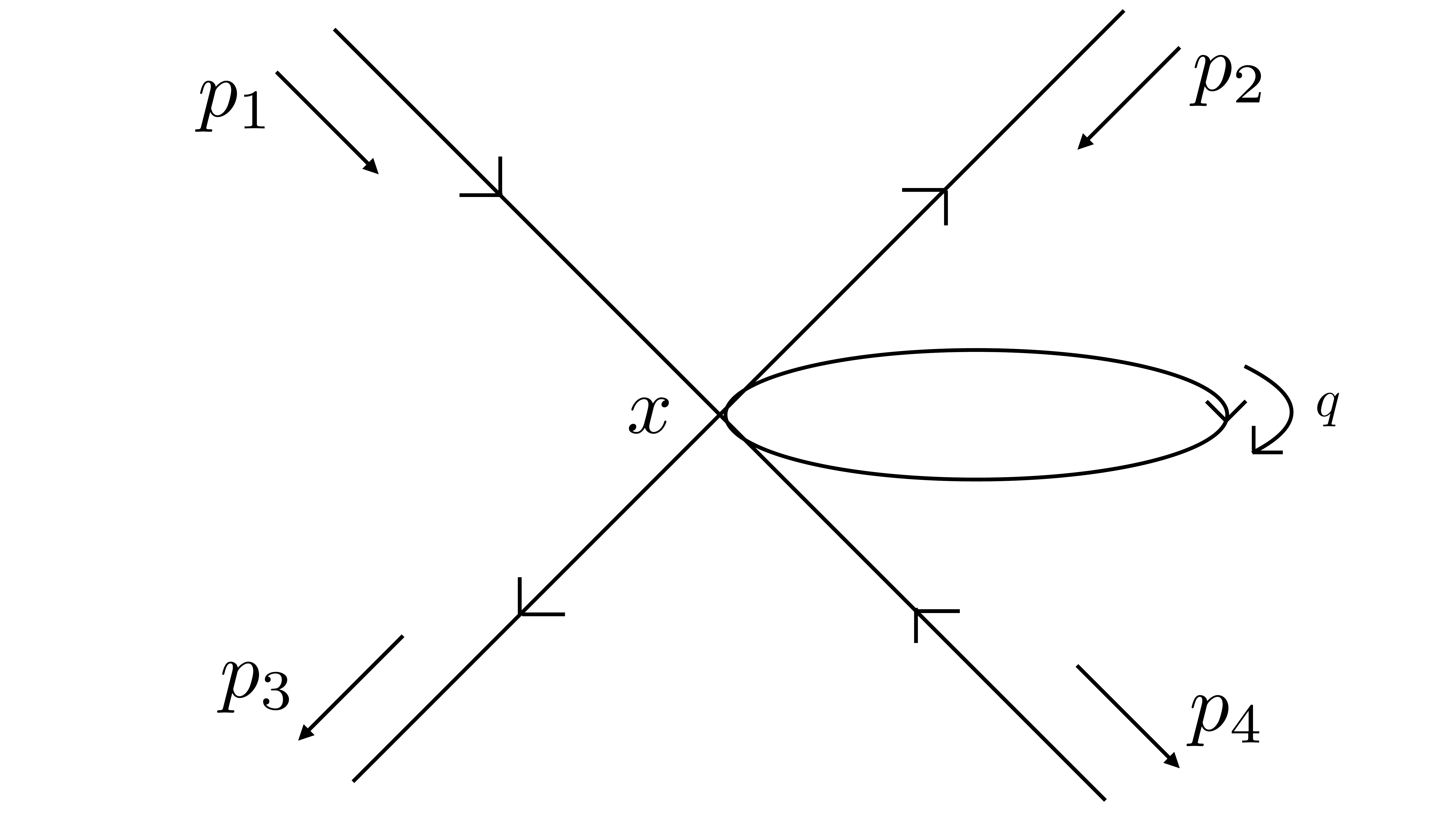}
	\caption{Tadpole diagram that contributes to the $S$-matrix at second order.}
	\label{tadpole diagram}
\end{figure}
The sextic couplings give rise to the one-loop tadpole diagrams shown in figure \ref{tadpole diagram} and contribute to the second order $S$-matrix. The corresponding amplitude is given by,
\bea
\mathcal{A}^{\text{tad}}&=&i\frac{\lambda^2}{2}\lan 0 | b(p_4) a(p_3) \tT\Big[m\bpsi\psi\Big(\tilde{X}_{\mu\nu}\tilde{X}^{\mu\nu}-(\tilde{X}_{\mu}^{ \ \mu})^2\Big)+\frac{7}{2}m^2(\bpsi\psi)^2 \tilde{X}_{\mu}^{ \ \mu}\nonumber\\
&& \hspace{4.3cm}-\frac{7}{2} m^3(\bpsi\psi)^3 \Big] a^{\dagger}(p_1) b^{\dagger}(p_2)|0 \ran.
\label{tadpoleamplitude}
\eea
Just as in the first order case one can evaluate (\ref{tadpoleamplitude}) by first deriving the general amplitude of a general sextic coupling,
\bea
\mathcal{A}_{\bpsi_a\psi_b\bpsi_c\psi_d\bpsi_e\psi_f}^{\text{tad}}&=&\lan 0 | b(p_4) a(p_3) \tT\Big[\bpsi_a\psi_b\bpsi_c\psi_d\bpsi_e\psi_f\Big] a^{\dagger}(p_1) b^{\dagger}(p_2)|0 \ran\nonumber\\
 &=&-\left\lan\psi_b(x)\bpsi_a(x)\right\ran\Big[\bu_c(p_3)v_d(p_4)\bv_e(p_2)u_f(p_1)
+\bv_c(p_2)u_d(p_1)\bu_e(p_3)v_f(p_4)\nonumber\\ 
&&\hspace{2.5cm}-\bv_c(p_2)v_d(p_4)\bu_e(p_3)u_f(p_1)
-\bu_c(p_3)u_d(p_1)\bv_e(p_2)v_f(p_4)\Big]\nonumber\\
&&+\left\lan\psi_d(x)\bpsi_a(x)\right\ran\Big[\bu_c(p_3)v_b(p_4)\bv_e(p_2)u_f(p_1)+\bv_c(p_2)u_b(p_1)\bu_e(p_3)v_f(p_4)\nonumber\\
&&\hspace{2.5cm}-\bv_c(p_2)v_b(p_4)\bu_e(p_3)u_f(p_1)-\bu_c(p_3)u_b(p_1)\bv_e(p_2)v_f(p_4)\Big]\nonumber\\
&&-\left\lan\psi_f(x)\bpsi_a(x)\right\ran\Big[\bu_c(p_3)v_b(p_4)\bv_e(p_2)u_d(p_1)
+\bv_c(p_2)u_b(p_1)\bu_e(p_3)v_d(p_4)\nonumber\\
&&\hspace{2.5cm}-\bv_c(p_2)v_b(p_4)\bu_e(p_3)u_d(p_1)-\bu_c(p_3)u_b(p_1)\bv_e(p_2)v_d(p_4)\Big]\nonumber\\ 
&&+\left\lan\psi_b(x)\bpsi_c(x)\right\ran\Big[\bu_a(p_3)v_d(p_4)\bv_e(p_2)u_f(p_1)
+\bv_a(p_2)u_d(p_1)\bu_e(p_3)v_f(p_4)\nonumber\\
&&\hspace{2.5cm}-\bv_a(p_2)v_d(p_4)\bu_e(p_3)u_f(p_1)-\bu_a(p_3)u_d(p_1)\bv_e(p_2)v_f(p_4)\Big]\nonumber\\
&&-\left\lan\psi_b(x)\bpsi_e(x)\right\ran\Big[\bu_a(p_3)v_d(p_4)\bv_c(p_2)u_f(p_1)+\bv_a(p_2)u_d(p_1)\bu_c(p_3)v_f(p_4)\nonumber\\
&&\hspace{2.5cm}-\bv_a(p_2)v_d(p_4)\bu_c(p_3)u_f(p_1)-\bu_a(p_3)u_d(p_1)\bv_c(p_2)v_f(p_4)\Big]\nonumber\\
&&-\left\lan\psi_d(x)\bpsi_c(x)\right\ran\Big[\bu_a(p_3)v_b(p_4)\bv_e(p_2)u_f(p_1)
+\bv_a(p_2)u_b(p_1)\bu_e(p_3)v_f(p_4)\nonumber\\
&&\hspace{2.5cm}-\bv_a(p_2)v_b(p_4)\bu_e(p_3)u_f(p_1)-\bu_a(p_3)u_b(p_1)\bv_e(p_2)v_f(p_4)\Big]\nonumber\\ 
&&+\left\lan\psi_f(x)\bpsi_c(x)\right\ran\Big[\bu_a(p_3)v_b(p_4)\bv_e(p_2)u_d(p_1)
+\bv_a(p_2)u_b(p_1)\bu_e(p_3)v_d(p_4)\nonumber\\
&&\hspace{2.5cm}-\bv_a(p_2)v_b(p_4)\bu_e(p_3)u_d(p_1)-\bu_a(p_3)u_b(p_1)\bv_e(p_2)v_d(p_4)\Big]\nonumber\\
&&+\left\lan\psi_d(x)\bpsi_e(x)\right\ran\Big[\bu_a(p_3)v_b(p_4)\bv_c(p_2)u_f(p_1)+\bv_a(p_2)u_b(p_1)\bu_c(p_3)v_f(p_4)\nonumber\\
&&\hspace{2.5cm}-\bv_a(p_2)v_b(p_4)\bu_c(p_3)u_f(p_1)-\bu_a(p_3)u_b(p_1)\bv_c(p_2)v_f(p_4)\Big]\nonumber\\
&&-\left\lan\psi_f(x)\bpsi_e(x)\right\ran\Big[\bu_a(p_3)v_b(p_4)\bv_c(p_2)u_d(p_1)
+\bv_a(p_2)u_b(p_1)\bu_c(p_3)v_d(p_4)\nonumber\\
&&\hspace{2.5cm}-\bv_a(p_2)v_b(p_4)\bu_c(p_3)u_d(p_1)-\bu_a(p_3)u_b(p_1)\bv_c(p_2)v_d(p_4)\Big]\nonumber\\
\label{generaltadpole}
\eea
where all possible Wick contractions of fermionic fields were performed. The internal propagators give rise to loops whose values are derived in Appendix \ref{AppendixB} and given by, 
\begin{align}
\langle \psi_a(x)\bar{\psi}_b(x)\rangle&=N_0\delta_{ab} ,\label{tadpoleprop}\\
\langle \partial^\mu\psi_a(x)\bar{\psi}_b(x)\rangle&=	-\langle \psi_a(x)\partial^\mu\bar{\psi}_b(x)\rangle=N_1\gamma^{\mu}_{ab} ,\label{prop2}\\
\langle \partial^\mu\psi_a(x)\partial^\nu\bar{\psi}_b(x)\rangle&=imN_1\eta^{\mu\nu}\delta_{ab}\label{prop3}\,	
\end{align}
where,
\bea
N_0&=&-\frac{m}{4\pi}\log(\frac{m^2}{\Lambda^2})\ ,
\label{N0}\\
 N_1&=&\frac{i\Lambda^2}{8\pi}\Big[1+\frac{m^2}{\Lambda^2}\log(\frac{m^2}{\Lambda^2})\Big] \label{N1}\ ,
\eea
and $\Lambda$ is the UV cut-off.
\vspace{2mm}
\\
Using the general expression for the amplitude of a sextic coupling, (\ref{generaltadpole}), one can compute the contribution of $(\bpsi\psi)^3$ to the amplitude,
\bea
\mathcal{A}_{(\bpsi\psi)^3}^{\text{tad}}&=&
\lan 0 | b(p_4) a(p_3) \tT\big[(\bpsi\psi)^3\big] a^{\dagger}(p_1) b^{\dagger}(p_2)|0 \ran\nonumber\\ &=&\delta_{ab}\delta_{cd}\delta_{ef}\lan 0 | b(p_4) a(p_3) \tT\big[\bpsi_a\psi_b\bpsi_c\psi_d\bpsi_e\psi_f\big] a^{\dagger}(p_1) b^{\dagger}(p_2)|0 \ran=0
\label{tadpoleamplitudepsibarpsiCube}
\eea
where (\ref{tadpoleprop}), (\ref{uv}) and (\ref{ubarvbar}) were used to obtain the above result.
\vspace{2mm}
\\
The contribution to the amplitude due to the coupling $\bpsi \psi \tilde{X}_{\mu\nu}\tilde{X}^{\mu\nu}$ can be evaluated by computing the contributions from each term of  (\ref{psibarpsiTmunuSquare}).
The first term in (\ref{psibarpsiTmunuSquare}) contributes,
\bea
&&\mathcal{A}_{\bpsi \psi \bpsi \gamma_{(\mu}\del_{\nu)}\psi \bpsi\gamma^{(\mu}\del^{\nu)}\psi}^{\text{tad}}\nonumber\\
=&&\lan 0 | b(p_4) a(p_3) \tT\Big[\bpsi \psi \ \bpsi \gamma_{(\mu}\del_{\nu)}\psi \bpsi\gamma^{(\mu}\del^{\nu)}\psi\Big] a^{\dagger}(p_1) b^{\dagger}(p_2)|0 \ran \nonumber\\
=&&\delta_{ab}\gamma_{cd(\mu}\gamma_{ef}^{(\mu}\lan 0 | b(p_4) a(p_3) \tT\Big[\bpsi_a\psi_b \bpsi_c \del_{\nu)}\psi_d\bpsi_e\del^{\nu)}\psi_f\Big] a^{\dagger}(p_1) b^{\dagger}(p_2)|0 \ran\nonumber\\
=&-&2N_0\Big[\bu(p_3)\cdot i\gamma_{(\mu}p_{4\nu)}\cdot v(p_4) \ \bv(p_2)\cdot (-i)\gamma^{(\mu}p_1^{\nu)}\cdot u(p_1)
+\bv(p_2)\cdot (-i)\gamma_{(\mu}p_{1\nu)}\cdot u(p_1) \ \bu(p_3)\cdot i\gamma^{(\mu}p_4^{\nu)}\cdot v(p_4)\nonumber\\ 
&&-\bv(p_2)\cdot i\gamma_{(\mu}p_{4\nu)}\cdot v(p_4) \ \bu(p_3)\cdot (-i)\gamma^{(\mu}p_1^{\nu)}\cdot u(p_1)
-\bu(p_3)\cdot (-i)\gamma_{(\mu}p_{1\nu)}\cdot u(p_1) \ \bv(p_2)\cdot i\gamma^{(\mu}p_4^{\nu)}\cdot v(p_4)\Big]\nonumber\\
&+&N_1\Big[\bu(p_3)\cdot \gamma_{(\mu}\gamma_{\nu)}\cdot v(p_4) \ \bv(p_2)\cdot (-i)\gamma^{(\mu}p_1^{\nu)}\cdot u(p_1)
+\bv(p_2)\cdot \gamma_{(\mu}\gamma_{\nu)}\cdot u(p_1) \ \bu(p_3)\cdot i\gamma^{(\mu}p_4^{\nu)}\cdot v(p_4)\nonumber\\ 
&&-\bv(p_2)\cdot \gamma_{(\mu}\gamma_{\nu)}\cdot v(p_4) \ \bu(p_3)\cdot (-i)\gamma^{(\mu}p_1^{\nu)}\cdot u(p_1)
-\bu(p_3)\cdot \gamma_{(\mu}\gamma_{\nu)}\cdot u(p_1) \ \bv(p_2)\cdot i\gamma^{(\mu}p_4^{\nu)}\cdot v(p_4)\Big]\nonumber\\
&-&N_1\Big[\bv(p_2)\cdot \gamma_{(\mu}\gamma_{\nu)}\cdot v(p_4) \ \bu(p_3)\cdot (-i)\gamma_{(\mu}p_{1\nu)}\cdot u(p_1)
+\bu(p_3)\cdot \gamma_{(\mu}\gamma_{\nu)}\cdot u(p_1) \ \bv(p_2)\cdot i\gamma_{(\mu}p_{4\nu)}\cdot v(p_4)\nonumber\\ 
&&-\bu(p_3)\cdot \gamma_{(\mu}\gamma_{\nu)}\cdot v(p_4) \ \bv(p_2)\cdot (-i)\gamma_{(\mu}p_{1\nu)}\cdot u(p_1)
-\bv(p_2)\cdot \gamma_{(\mu}\gamma_{\nu)}\cdot u(p_1) \ \bu(p_3)\cdot i\gamma_{(\mu}p_{4\nu)}\cdot v(p_4)\Big]\nonumber\\
&+&N_0\Big[\bu(p_3)\cdot i\gamma_{(\mu}p_{4\nu)}\cdot v(p_4) \ \bv(p_2)\cdot (-i)\gamma^{(\mu}p_1^{\nu)}\cdot u(p_1)
+\bv(p_2)\cdot (-i)\gamma_{(\mu}p_{1\nu)}\cdot u(p_1) \ \bu(p_3)\cdot i\gamma^{(\mu}p_4^{\nu)}\cdot v(p_4)\nonumber\\ 
&&-\bv(p_2)\cdot i\gamma_{(\mu}p_{4\nu)}\cdot v(p_4) \ \bu(p_3)\cdot (-i)\gamma^{(\mu}p_1^{\nu)}\cdot u(p_1)
-\bu(p_3)\cdot (-i)\gamma_{(\mu}p_{1\nu)}\cdot u(p_1) \ \bv(p_2)\cdot i\gamma^{(\mu}p_4^{\nu)}\cdot v(p_4)\Big]\nonumber\\
&-&N_0\Big[\bv(p_2)\cdot i\gamma_{(\mu}p_{4\nu)}\cdot v(p_4) \ \bu(p_3)\cdot (-i)\gamma^{(\mu}p_1^{\nu)}\cdot u(p_1)
+\bu(p_3)\cdot (-i)\gamma_{(\mu}p_{1\nu)}\cdot u(p_1) \ \bv(p_2)\cdot i\gamma^{(\mu}p_4^{\nu)}\cdot v(p_4)\nonumber\\ 
&&-\bu(p_3)\cdot i\gamma_{(\mu}p_{4\nu)}\cdot v(p_4) \ \bv(p_2)\cdot (-i)\gamma^{(\mu}p_1^{\nu)}\cdot u(p_1)
-\bv(p_2)\cdot (-i)\gamma_{(\mu}p_{1\nu)}\cdot u(p_1) \ \bu(p_3)\cdot i\gamma^{(\mu}p_4^{\nu)}\cdot v(p_4)\Big]\nonumber\\
&-&N_1 \ \Tr\big(\gamma_{(\mu}\gamma_{\nu)}\big)\Big[\bu(p_3)\cdot  v(p_4) \ \bv(p_2)\cdot (-i)\gamma^{(\mu}p_1^{\nu)}\cdot u(p_1)
+\bv(p_2)\cdot u(p_1) \ \bu(p_3)\cdot i\gamma^{(\mu}p_4^{\nu)}\cdot v(p_4)\nonumber\\ 
&&-\bv(p_2)\cdot  v(p_4) \ \bu(p_3)\cdot (-i)\gamma^{(\mu}p_1^{\nu)}\cdot u(p_1)
-\bu(p_3)\cdot u(p_1) \ \bv(p_2)\cdot i\gamma^{(\mu}p_4^{\nu)}\cdot v(p_4)\Big]\nonumber
\eea
\bea
&+&N_1\Big[\bu(p_3)\cdot  v(p_4) \ \bv(p_2)\cdot (-i)\gamma^{(\mu}\gamma^{\nu)}\gamma_{(\mu}p_{1\nu)}\cdot u(p_1)
+\bv(p_2) \cdot u(p_1) \ \bu(p_3)\cdot i \gamma^{(\mu}\gamma^{\nu)}\gamma_{(\mu}p_{4\nu)}\cdot v(p_4)\nonumber\\ 
&&-\bv(p_2)\cdot  v(p_4) \ \bu(p_3)\cdot (-i) \gamma^{(\mu}\gamma^{\nu)}\gamma_{(\mu}p_{1\nu)}\cdot u(p_1)
-\bu(p_3)\cdot  u(p_1) \ \bv(p_2)\cdot i \gamma^{(\mu}\gamma^{\nu)}\gamma_{(\mu}p_{4\nu)}\cdot v(p_4)\Big]\nonumber\\
&+&N_1\Big[\bu(p_3)\cdot  v(p_4) \ \bv(p_2)\cdot (-i)\gamma_{(\mu}\gamma_{\nu)}\gamma^{(\mu}p_1^{\nu)}\cdot u(p_1)
+\bv(p_2) \cdot u(p_1) \ \bu(p_3)\cdot i \gamma_{(\mu}\gamma_{\nu)}\gamma^{(\mu}p_4^{\nu)}\cdot v(p_4)\nonumber\\ 
&&-\bv(p_2)\cdot  v(p_4) \ \bu(p_3)\cdot (-i) \gamma_{(\mu}\gamma_{\nu)}\gamma^{(\mu}p_1^{\nu)}\cdot u(p_1)
-\bu(p_3)\cdot  u(p_1) \ \bv(p_2)\cdot i \gamma_{(\mu}\gamma_{\nu)}\gamma^{(\mu}p_4^{\nu)}\cdot v(p_4)\Big]\nonumber\\
&-&N_1\Tr\big(\gamma^{(\mu}\gamma^{\nu)}\big)\Big[\bu(p_3)\cdot  v(p_4) \ \bv(p_2)\cdot (-i)\gamma_{(\mu}p_{1\nu)}\cdot u(p_1)
+\bv(p_2) \cdot u(p_1) \ \bu(p_3)\cdot i \gamma_{(\mu}p_{4\nu)}\cdot v(p_4)\nonumber\\ 
&&-\bv(p_2)\cdot  v(p_4) \ \bu(p_3)\cdot (-i) \gamma_{(\mu}p_{1\nu)}\cdot u(p_1)
-\bu(p_3)\cdot  u(p_1) \ \bv(p_2)\cdot i \gamma_{(\mu}p_{4\nu)}\cdot v(p_4)\Big] \,
\label{psibarpsiTmunuSquare1_step}
\eea
where $N_0$ and $N_1$ are given by (\ref{N0}) and (\ref{N1}) respectively.  Evaluating (\ref{psibarpsiTmunuSquare1_step}) one sees that, 
\bea
\mathcal{A}_{\bpsi \psi \bpsi \gamma_{(\mu}\del_{\nu)}\psi \bpsi\gamma^{(\mu}\del^{\nu)}\psi}^{\text{tad}}=0.
\label{psibarpsiTmunuSquare1}
\eea
Notice that the last term in (\ref{psibarpsiTmunuSquare}) is  the complex conjugate of the first term and will hence give a vanishing contribution to the amplitude as well,
\bea
\mathcal{A}_{\bpsi \psi \del_{(\mu}\bpsi\gamma_{\nu)}\psi \del^{(\mu}\bpsi\gamma^{\nu)}\psi}^{\text{tad}}=\lan 0 | b(p_4) a(p_3) \tT\Big[\bpsi \psi \del_{(\mu}\bpsi\gamma_{\nu)}\psi \del^{(\mu}\bpsi\gamma^{\nu)}\psi\Big] a^{\dagger}(p_1) b^{\dagger}(p_2)|0 \ran=0.
\label{psibarpsiTmunuSquare3}
\eea
The second term in (\ref{psibarpsiTmunuSquare}) gives a non-zero contribution to the amplitude and can be computed in a similar way,
\bea
\mathcal{A}_{\bpsi \psi \bpsi\gamma_{(\mu}\del_{\nu)}\psi \del^{(\mu}\bpsi\gamma^{\nu)}\psi}^{\text{tad}}&=&\lan 0 | b(p_4) a(p_3) \tT\Big[\bpsi \psi \bpsi\gamma_{(\mu}\del_{\nu)}\psi \del^{(\mu}\bpsi\gamma^{\nu)}\psi\Big] a^{\dagger}(p_1) b^{\dagger}(p_2)|0 \ran\nonumber\\
&=&-\frac{m^3\Lambda^2}{\pi}\Big[1+4\big(2-\cosh\theta\big)\frac{m^2}{\Lambda^2}\log\frac{m^2}{\Lambda^2}\Big] \cosh^2\frac{\theta}{2}.
\label{psibarpsiTmunuSquare2}
\eea
Combining (\ref{psibarpsiTmunuSquare1}), (\ref{psibarpsiTmunuSquare3}) and (\ref{psibarpsiTmunuSquare2}) according to (\ref{psibarpsiTmunuSquare}) the total contribution from the sextic coupling $\bpsi\psi \tilde{X}_{\mu\nu}\tilde{X}^{\mu\nu}$ is,
\bea
\mathcal{A}_{\bpsi\psi \tilde{X}_{\mu\nu}\tilde{X}^{\mu\nu}}^{\text{tad}}&=&- \  \frac{1}{4}\Big[\mathcal{A}_{\bpsi\psi \bpsi\gamma_{(\mu}\del_{\nu)}\psi \bpsi\gamma^{(\mu}\del^{\nu)}\psi}^{\text{tad}}-2\mathcal{A}_{\bpsi\psi \bpsi\gamma_{(\mu}\del_{\nu)}\psi \del^{(\mu}\bpsi \gamma^{\nu)}\psi}^{\text{tad}}+\mathcal{A}_{\bpsi\psi \del_{(\mu}\bpsi \gamma_{\nu)}\psi \del^{(\mu}\bpsi \gamma^{\nu)}\psi}^{\text{tad}}\Big]\nonumber\\
&=&-\frac{m^3\Lambda^2}{2\pi}\Big[1+4\big(2-\cosh\theta\big)\frac{m^2}{\Lambda^2}\log\frac{m^2}{\Lambda^2}\Big] \cosh^2\frac{\theta}{2}.
\label{tadpoleamplitudepsibpsiTmunuSq}
\eea
Following the same steps as in (\ref{psibarpsiTmunuSquare1_step}) one can compute the contribution from the sextic coupling $\bpsi\psi(\tilde{X}_{\mu}^{\ \mu})^2$ by determining the contribution from each term in (\ref{psibarpsiTmumuSquare}),
\bea
\mathcal{A}_{\bpsi\psi\bpsi\gamma^{\mu}\del_{\mu}\psi \bpsi\gamma^{\nu}\del_{\nu}\psi}^{\text{tad}}=\lan 0 | b(p_4) a(p_3) \tT\big[\bpsi\psi\bpsi\gamma^{\mu}\del_{\mu}\psi \bpsi\gamma^{\nu}\del_{\nu}\psi\big] a^{\dagger}(p_1) b^{\dagger}(p_2)|0 \ran &=& 0\nonumber\\
\mathcal{A}_{\bpsi\psi\bpsi\gamma^{\mu}\del_{\mu}\psi \del_{\nu}\bpsi \gamma^{\nu}\psi}^{\text{tad}}=\lan 0 | b(p_4) a(p_3) \tT\big[\bpsi\psi\bpsi\gamma^{\mu}\del_{\mu}\psi \del_{\nu}\bpsi \gamma^{\nu}\psi\big] a^{\dagger}(p_1) b^{\dagger}(p_2)|0 \ran &=& \frac{m^3\Lambda^2}{\pi}\big(1+\cosh \theta \big) \nonumber\\
\mathcal{A}_{\bpsi\psi\del_{\mu}\bpsi \gamma^{\mu}\psi\del_{\nu}\bpsi \gamma^{\nu}\psi}^{\text{tad}}=\lan 0 | b(p_4) a(p_3) \tT\big[\bpsi\psi\del_{\mu}\bpsi \gamma^{\mu}\psi\del_{\nu}\bpsi \gamma^{\nu}\psi\big] a^{\dagger}(p_1) b^{\dagger}(p_2)|0 \ran &=& 0.
\label{tadpole4_5_6}
\eea
Combining the individual amplitudes according to (\ref{psibarpsiTmumuSquare}) the total contribution from the sextic coupling $\bpsi\psi (\tilde{X}_{\mu}^{\ \mu})^2$ is,
\bea
\mathcal{A}_{\bpsi\psi (\tilde{X}_{\mu}^{\ \mu})^2}^{\text{tad}}&=&- \  \frac{1}{4}\Big[\mathcal{A}_{\bpsi\psi\bpsi\gamma^{\mu}\del_{\mu}\psi \bpsi\gamma^{\nu}\del_{\nu}\psi}^{\text{tad}}-2\mathcal{A}_{\bpsi\psi\bpsi\gamma^{\mu}\del_{\mu}\psi \del_{\nu}\bpsi \gamma^{\nu}\psi}^{\text{tad}}+\mathcal{A}_{\bpsi\psi\del_{\mu}\bpsi \gamma^{\mu}\psi\del_{\nu}\bpsi \gamma^{\nu}\psi}^{\text{tad}}\Big]\nonumber\\
&=&\frac{m^3\Lambda^2}{2\pi}\big(1+\cosh \theta \big)
\label{tadpoleamplitudepsibpsiTmumuSq}
\eea
Similarly, evaluating the contribution to the amplitude from each term in $(\bpsi\psi)^2 \tilde{X}_{\mu}^{\ \mu}$, (\ref{psibarpsiSqTmumu}), yields a vanishing contribution,
\bea
\mathcal{A}_{(\bpsi\psi)^2 \tilde{X}_{\mu}^{\ \mu}}^{\text{tad}}&=&\  \frac{i}{2}\Big[\mathcal{A}_{(\bpsi\psi)^2   \bpsi\gamma^{\mu}\del_{\mu}\psi}^{\text{tad}}-\mathcal{A}_{(\bpsi\psi)^2\del_{\mu}\bpsi \gamma^{\mu}\psi}^{\text{tad}}\Big]=0
\label{tadpoleamplitudepsibpsiSqTmumu}
\eea
Finally, substituting (\ref{tadpoleamplitudepsibarpsiCube}), (\ref{tadpoleamplitudepsibpsiTmunuSq}), (\ref{tadpoleamplitudepsibpsiTmumuSq}) and (\ref{tadpoleamplitudepsibpsiSqTmumu}) into (\ref{tadpoleamplitude}) yields the contribution to the amplitude from the tadpole diagrams,
\bea
\mathcal{A}^{\text{tad}}&=&i\frac{\lambda^2}{2}\Big[m\Big(\mathcal{A}_{\bpsi\psi \tilde{X}_{\mu\nu}\tilde{X}^{\mu\nu}}^{\text{tad}}-\mathcal{A}_{\bpsi\psi (\tilde{X}_{\mu}^{\ \mu})^2}^{\text{tad}}\Big)+\frac{7}{2}m^2\mathcal{A}_{(\bpsi\psi)^2 \tilde{X}_{\mu}^{\ \mu}}^{\text{tad}}-\frac{7}{2}m^3\mathcal{A}_{(\bpsi\psi)^3}^{\text{tad}} \Big]\nonumber\\
&=&-i\frac{\lambda^2 m^6 }{4\pi}\cosh^2\frac{\theta}{2}\Big[3\frac{\Lambda^2}{m^2}+4\big(2-\cosh\theta\big)\log\frac{m^2}{\Lambda^2}\Big] \ .
\label{tadpoleamplitudefinal}
\eea
Converting from amplitude to $S$-matrix using (\ref{AtoS}) yields the tadpole contribution to the $S$-matrix,
\bea
S^{\text{tad}}(\theta)=\frac{\mathcal{A}^{\text{tad}}}{4 m^2 \sinh \theta}= -i\frac{\lambda^2 m^4 }{32\pi}\coth\frac{\theta}{2}\Big[3\frac{\Lambda^2}{m^2}+4\big(2-\cosh\theta\big)\log\frac{m^2}{\Lambda^2}\Big] \ .
\label{tadpoleSmatrix}
\eea
\
\subsubsection*{Contribution from bubble diagrams}
\label{Bubble_Contribution}
\begin{figure}[h!]
	\centering
	\subfigure[$s$-channel]{
		\includegraphics[width=0.31\columnwidth,height=0.21\columnwidth]{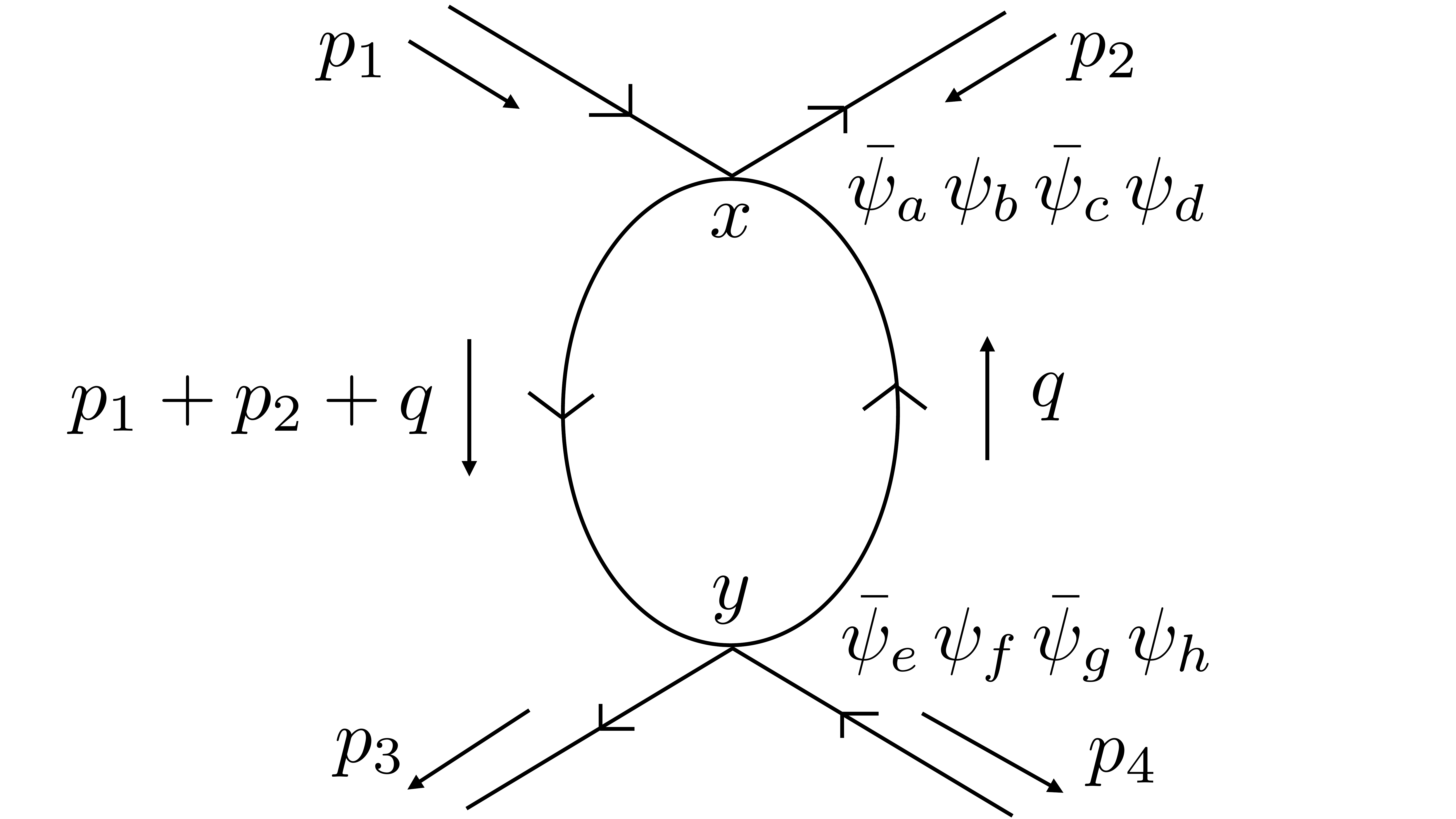}
		\label{bubble diagram_s} } 
	\subfigure[$t$-channel]{
		\includegraphics[width=0.31\columnwidth,height=0.21\columnwidth]{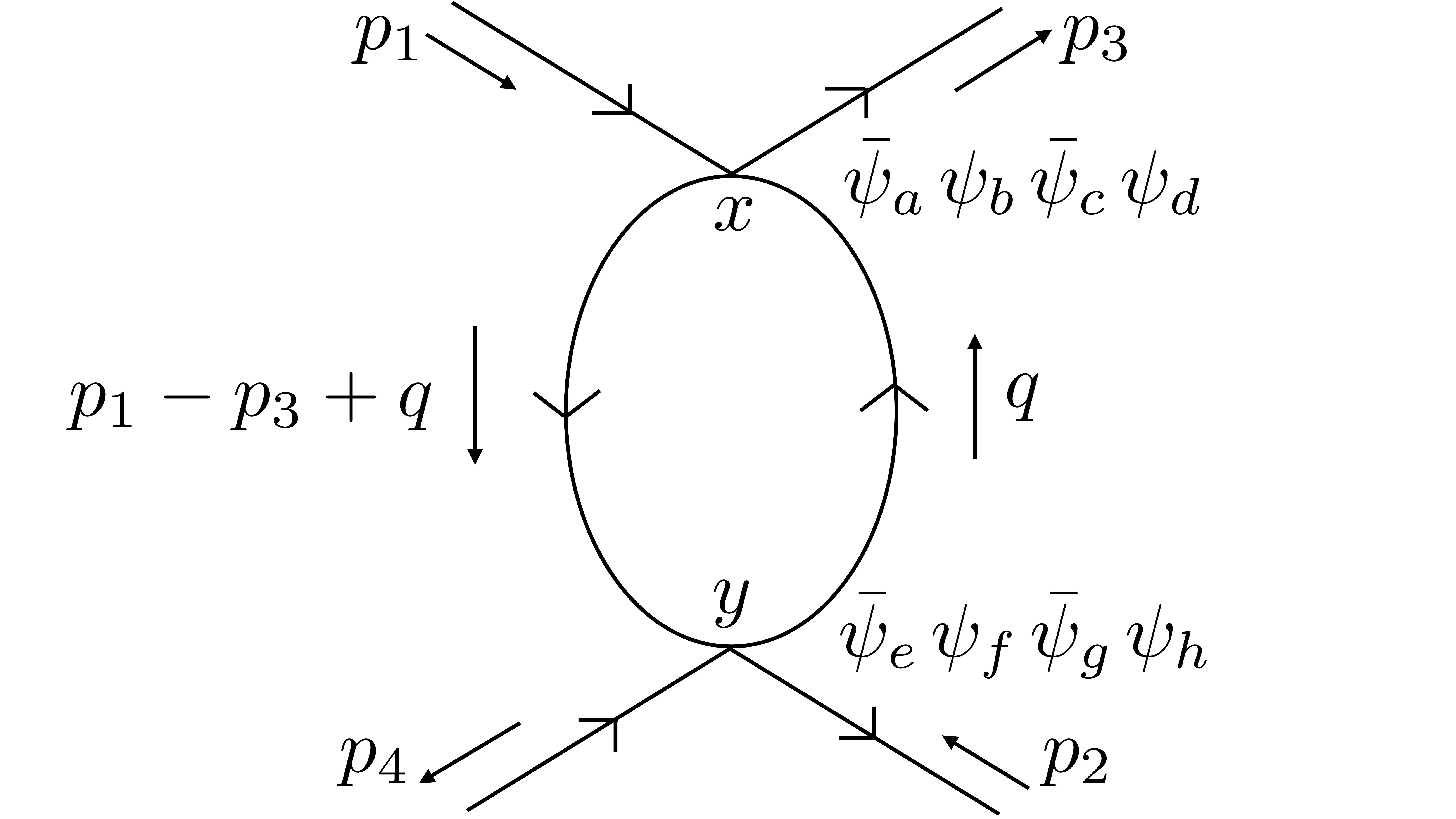}
		\label{bubble diagram_t} }
	\subfigure[$u$-channel]{
		\includegraphics[width=0.31\columnwidth,height=0.21\columnwidth]{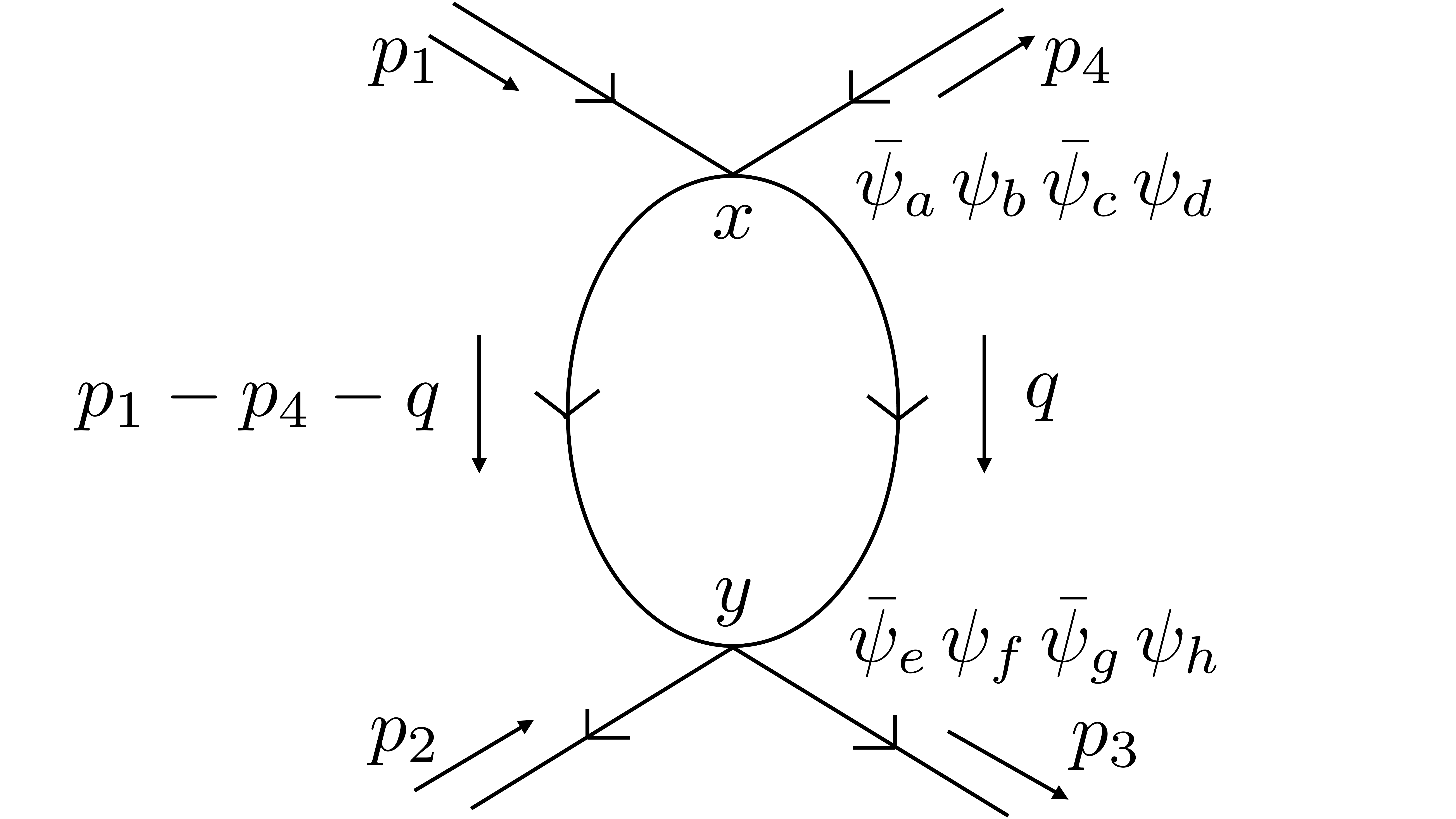}
		\label{bubble diagram_u} }
	\caption{\small The $s$, $t$ and $u$-channels of a typical bubble diagram that contribute to the $S$-matrix
		at second order.}
	\label{bubble diagram}
\end{figure}

The final and most complicated contribution to the $S$-matrix is the contribution that arises from the first order term in the Lagrangian squared, 
\bea
\frac{1}{2!}\left(i\mathcal{L}_1(\lambda)\right)^2=\frac{1}{2!}\Big(\frac{i\lambda}{2}\Big)^2\Big(\tilde{X}_{\mu\nu}\tilde{X}^{\mu\nu}-(\tilde{X}_{\mu}^{\mu})^2\Big)^2
\label{L2NL0}.
\eea
These interaction vertices give rise to one-loop bubble diagrams of the form figure \ref{bubble diagram}.
\\\\
From (\ref{L2NL0}) there are three different kinds of bubble diagrams: 
\begin{enumerate}
	\item[\textbf{(a)}]  both vertices contain $\tilde{X}_{\mu\nu}\tilde{X}^{\mu\nu}$ interactions, 
	\item[\textbf{(b)}]  one vertex contains interaction $\tilde{X}_{\rho\lambda}\tilde{X}^{\rho\lambda}$  while the other contains $(\tilde{X}_{\mu}^{\ \mu})^2$ and 
	\item[\textbf{(c)}]  both vertices contain $(\tilde{X}_{\mu}^{\ \mu})^2$ interactions.
\end{enumerate}
For the bubble diagram case the amplitudes will be split up into $s$, $t$ and $u$-channel contributions to the amplitude, as shown in figure (\ref{bubble diagram}). As done in the previous cases, begin by writing down the general expressions for the amplitude where both vertices $x$ and $y$ contain arbitrary non-derivative quartic couplings.  These expressions will be the building blocks for the computation of the amplitude from the bubble diagrams.
\vspace{2mm}
\\
In order to compute the general contributions of the $s$, $t$ and $u$-channels to the amplitude consider the case when the vertex $x$ involves a quartic coupling $\bpsi_a(x)\psi_b(x)\bpsi_c(x)\psi_d(x)$ and vertex $y$ contains another quartic coupling $\bpsi_e(y)\psi_f(y)\bpsi_g(y)\psi_h(y)$.
\\\\
 The general contribution of the $s$-channel (figure \ref{bubble diagram_s}) to the amplitude is given by,
\bea
\mathcal{B}_{abcdefgh}^{(s)}&=&\lan 0 | b(p_4) a(p_3) \tT\Big[\bpsi_e(y)\psi_f(y)\bpsi_g(y)\psi_h(y)\bpsi_a(x)\psi_b(x)\bpsi_c(x)\psi_d(x)\Big] a^{\dagger}(p_1) b^{\dagger}(p_2)|0 \ran^{(s)}\nonumber\\
&=&-\bv_c(p_2)u_d(p_1)\int\frac{d^2q}{(2\pi)^2}\Big(\bu_e(p_3)v_f(p_4)G_{ha}(\xi+q)G_{bg}(q)-\bu_e(p_3)v_h(p_4)G_{fa}(\xi+q)G_{bg}(q)\nonumber\\
&-&\bu_{g}(p_3)v_f(p_4)G_{ha}(\xi+q)G_{be}(q)+\bu_g(p_3)v_h(p_4)G_{fa}(\xi+q)G_{be}(q)\Big)
\nonumber\\
&-&\bv_a(p_2)u_b(p_1)\int\frac{d^2q}{(2\pi)^2}\Big(\bu_e(p_3)v_f(p_4)G_{hc}(\xi+q)G_{dg}(q)-\bu_e(p_3)v_h(p_4)G_{fc}(\xi+q)G_{dg}(q)\nonumber\\
&-&\bu_{g}(p_3)v_f(p_4)G_{hc}(\xi+q)G_{de}(q)+\bu_g(p_3)v_h(p_4)G_{fc}(\xi+q)G_{de}(q)\Big)
\nonumber\\
&+&\bv_a(p_2)u_d(p_1)\int\frac{d^2q}{(2\pi)^2}\Big(\bu_e(p_3)v_f(p_4)G_{hc}(\xi+q)G_{bg}(q)-\bu_e(p_3)v_h(p_4)G_{fc}(\xi+q)G_{bg}(q)\nonumber\\
&-&\bu_{g}(p_3)v_f(p_4)G_{hc}(\xi+q)G_{be}(q)+\bu_g(p_3)v_h(p_4)G_{fc}(\xi+q)G_{be}(q)\Big)
\nonumber\\
&+&\bv_c(p_2)u_b(p_1)\int\frac{d^2q}{(2\pi)^2}\Big(\bu_e(p_3)v_f(p_4)G_{ha}(\xi+q)G_{dg}(q)-\bu_e(p_3)v_h(p_4)G_{fa}(\xi+q)G_{dg}(q)\nonumber\\
&-&\bu_{g}(p_3)v_f(p_4)G_{ha}(\xi+q)G_{de}(q)+\bu_g(p_3)v_h(p_4)G_{fa}(\xi+q)G_{de}(q)\Big)\ ,
\label{schannelbubblegeneral}
\eea
where $\xi^2=(p_1+p_2)^2=s$ and $G_{ab}(q)=\frac{i(\gamma\cdot q+m)_{ab}}{q^2-m^2+i\epsilon}$ is the fermionic propagator in the free theory.
\\\\
Next consider the contribution of the $t$-channel (figure \ref{bubble diagram_t}) to the amplitude. Since $p_1=p_3$, the general $t$-channel contribution to the amplitude is given by,
\bea
\mathcal{B}_{abcdefgh}^{(t)}&=&\lan 0 | b(p_4) a(p_3) \tT\Big[\bpsi_e(y)\psi_f(y)\bpsi_g(y)\psi_h(y)\bpsi_a(x)\psi_b(x)\bpsi_c(x)\psi_d(x)\Big] a^{\dagger}(p_1) b^{\dagger}(p_2)|0 \ran^{(t)}\nonumber\\
&=&\bu_c(p_3)u_d(p_1)\int\frac{d^2q}{(2\pi)^2}\Big(\bv_g(p_2)v_h(p_4)G_{fa}(q)G_{be}(q)-\bv_g(p_2)v_f(p_4)G_{ha}(q)G_{be}(q)\nonumber\\
&-&\bv_{e}(p_2)v_h(p_4)G_{fa}(q)G_{bg}(q)+\bv_e(p_2)v_f(p_4)G_{ha}(q)G_{bg}(q)\Big)
\nonumber\\
&+&\bu_a(p_3)u_b(p_1)\int\frac{d^2q}{(2\pi)^2}\Big(\bv_g(p_2)v_h(p_4)G_{fc}(q)G_{de}(q)-\bv_g(p_2)v_f(p_4)G_{hc}(q)G_{de}(q)\nonumber\\
&-&\bv_{e}(p_2)v_h(p_4)G_{fc}(q)G_{dg}(q)+\bv_e(p_2)v_f(p_4)G_{hc}(q)G_{dg}(q)\Big)
\nonumber\eea\bea
&-&\bu_a(p_3)u_d(p_1)\int\frac{d^2q}{(2\pi)^2}\Big(\bv_g(p_2)v_h(p_4)G_{fc}(q)G_{be}(q)-\bv_g(p_2)v_f(p_4)G_{hc}(q)G_{be}(q)\nonumber\\
&-&\bv_{e}(p_2)v_h(p_4)G_{fc}(q)G_{bg}(q)+\bv_e(p_2)v_f(p_4)G_{hc}(q)G_{bg}(q)\Big)
\nonumber\\
&-&\bu_c(p_3)u_b(p_1)\int\frac{d^2q}{(2\pi)^2}\Big(\bv_g(p_2)v_h(p_4)G_{fa}(q)G_{de}(q)-\bv_g(p_2)v_f(p_4)G_{ha}(q)G_{de}(q)\nonumber\\
&-&\bv_{e}(p_2)v_h(p_4)G_{fa}(q)G_{dg}(q)+\bv_e(p_2)v_f(p_4)G_{ha}(q)G_{dg}(q)\Big)\ .
\label{tchannelbubblegeneral}
\eea
\\
Finally consider the contribution of the $u$-channel (figure \ref{bubble diagram_u}) to the amplitude. The general $u$-channel contribution to the amplitude is given by,
\bea
\mathcal{B}_{abcdefgh}^{(u)}&=&\lan 0 | b(p_4) a(p_3) \tT\Big[\bpsi_e(y)\psi_f(y)\bpsi_g(y)\psi_h(y)\bpsi_a(x)\psi_b(x)\bpsi_c(x)\psi_d(x)\Big] a^{\dagger}(p_1) b^{\dagger}(p_2)|0 \ran^{(u)}\nonumber\\
&=&\Big(u_d(p_1)v_b(p_4)\bu_e(p_3)\bv_g(p_2)-u_d(p_1)v_b(p_4)\bu_g(p_3)\bv_e(p_2)-u_b(p_1)v_d(p_4)\bu_e(p_3)\bv_g(p_2)\nonumber\\
&+&u_b(p_1)v_d(p_4)\bu_g(p_3)\bv_e(p_2)\Big)\int\frac{d^2q}{(2\pi)^2}\Big(G_{hc}(\zeta-q)G_{fa}(q)-G_{ha}(\zeta-q)G_{fc}(q)\Big)\ ,\nonumber\\
\label{uchannelbubblegeneral}
\eea
where $\zeta^2=(p_1-p_4)^2=(p_1-p_2)^2=u$ as $p_2=p_4$.
\\
\\
Notice that a general $s$-channel amplitude (\ref{schannelbubblegeneral}) involves evaluation of the one-loop integral,
\bea
I_{abcd}^{(s)}(\xi)&=&\int  \frac{d^2 q}{(2\pi)^2}G_{ab}(\xi+q)G_{cd}(q)=\int  \frac{d^2 q}{(2\pi)^2} \frac{i\big(\gamma\cdot (\xi+q)+m\big)_{ab}}{(\xi+q)^2-m^2} \frac{i\big(\gamma\cdot q+m\big)_{cd}}{q^2-m^2} \nonumber\\
&=&-\int  \frac{d^2 q}{(2\pi)^2}\frac{\gamma^{\mu}_{ab}\gamma^{\nu}_{cd}(\xi+q)_{\mu}q_{\nu}+m\delta_{ab}\gamma^{\mu}_{cd}q_{\mu}+m\gamma^{\mu}_{ab}(\xi+q)_{\mu}\delta_{cd}+m^2\delta_{ab}\delta_{cd}}
{\big[(\xi+q)^2-m^2\big](q^2-m^2)}\nonumber\\
&=&-L^{(s)}(\xi)\Big(m\slashed{\xi}_{ab}\delta_{cd}+m^2\delta_{ab}\delta_{cd}\Big)-\Big(\slashed{\xi}_{ab}\slashed{L}^{(s)}_{cd}(\xi)+m\delta_{ab}\slashed{L}^{(s)}_{cd}(\xi)+m\slashed{L}^{(s)}_{ab}(\xi)\delta_{cd}\Big)\nonumber\\
&&-\gamma^{\mu}_{ab}\gamma^{\nu}_{cd}L^{(s)}_{\mu\nu}(\xi)\ ,
\label{I_abcd_s}
\eea
where $\ \ \ \xi^{\mu}=(p_1+p_2)^{\mu}$, $ \ \ \slashed{L}^{(s)}_{ab}=\gamma^{\mu}_{ab}L^{(s)}_{\mu}$ , $  \ \ \slashed{\xi}_{ab}=\gamma^{\mu}_{ab}\xi_{\mu}\ \ $   and
\bea
L^{(s)}&=&\int  \frac{d^2 q}{(2\pi)^2} \frac{1}{\big[(\xi+q)^2-m^2\big]\big(q^2-m^2\big)}=-\frac{\pi+i\theta}{4\pi m^2 \sinh\theta} \ , \nonumber\\
L^{(s)}_{\mu}&=&\int  \frac{d^2 q}{(2\pi)^2} \frac{q_{\mu}}{\big[(\xi+q)^2-m^2\big]\big(q^2-m^2\big)}=\frac{\pi+i\theta}{8\pi m^2 \sinh\theta} \ \xi_{\mu} \ , \nonumber\\
L^{(s)}_{\mu\nu}&=&\int  \frac{d^2 q}{(2\pi)^2} \frac{q_{\mu}q_{\nu}}{\big[(\xi+q)^2-m^2\big]\big(q^2-m^2\big)}\nonumber\\
&=&-\frac{i}{8\pi}\bigg[1+\log \frac{\Lambda^2}{m^2}+\big(i\pi-\theta\big)\tanh\frac{\theta}{2}\bigg]\eta_{\mu\nu}+\frac{i}{16\pi m^2\cosh^2\frac{\theta}{2}}\bigg[1+\big(i\pi-\theta\big)\coth\theta\bigg]\xi_{\mu}\xi_{\nu} \ .\nonumber\\
 \label{L_s_trial}
\eea
The above integrals are derived in detail in Appendix \ref{AppendixC}.
\\
\\
Similar yet simpler integrals are involved in the evaluation of the general contribution to the amplitude of $t$-channel (\ref{tchannelbubblegeneral}). In the $t$-channel case the integrals are much simpler as both the internal propagators carry momentum $q$ since $p_1=p_3$ and take the form,
\bea
I_{abcd}^{(t)}&=&\int  \frac{d^2 q}{(2\pi)^2}G_{ab}(q)G_{cd}(q)\nonumber\\
&=&-m^2\delta_{ab}\delta_{cd}L^{(t)}-\Big(m\delta_{ab}\slashed{L}^{(t)}_{cd}+m\slashed{L}^{(t)}_{ab}\delta_{cd}\Big)-\gamma^{\mu}_{ab}\gamma^{\nu}_{cd}L^{(t)}_{\mu\nu}\ ,
\label{I_abcd_t}
\eea
where,
\bea
L^{(t)}=\int  \frac{d^2 q}{(2\pi)^2} \frac{1}{\big(q^2-m^2\big)^2}\ ,  \ \ \ 
L^{(t)}_{\mu}=\int  \frac{d^2 q}{(2\pi)^2} \frac{q_{\mu}}{\big(q^2-m^2\big)^2}, \ \ \
L^{(t)}_{\mu\nu}=\int  \frac{d^2 q}{(2\pi)^2} \frac{q_{\mu}q_{\nu}}{\big(q^2-m^2\big)^2}\nonumber.\\
\label{L_t_trial}
\eea
However, there is no need to explicitly compute $L^{(t)}$, $L^{(t)}_{\mu}$ and $L^{(t)}_{\mu\nu}$ as their values can easily be obtained from their $s$-channel counterparts (\ref{L_s_trial}). In section \ref{subsecSmatrix}, it was shown that the $t$-channel corresponds to $\theta\rightarrow i\pi$. Thus, one can simply substitute $\theta\rightarrow i\pi$ and $\xi\rightarrow 0$ into (\ref{L_s_trial}) to obtain the corresponding $t$-channel loop integrals $L^{(t)}$, $L^{(t)}_{\mu}$ and $L^{(t)}_{\mu\nu}$. Hence,
\bea
L^{(t)}=\frac{i}{4\pi m^2} \ , \ \ \ \ \ \ \
L^{(t)}_{\mu}=0 \ , \ \ \ \ \ \ \ 
L^{(t)}_{\mu\nu}=\frac{i}{8\pi}\Big(1-\log\frac{\Lambda^2}{m^2}\Big)\eta_{\mu\nu}.
\eea
In the $u$-channel case (\ref{uchannelbubblegeneral}) one must compute the following loop-integral similar to the $s$-channel case,
\bea
I_{abcd}^{(u)}(\zeta)&=&\int  \frac{d^2 q}{(2\pi)^2}G_{ab}(\zeta-q)G_{cd}(q)\nonumber\\
&=&-L^{(u)}(\zeta)\Big(m\slashed{\zeta}_{ab}\delta_{cd}+m^2\delta_{ab}\delta_{cd}\Big)-\Big(\slashed{\zeta}_{ab}\slashed{L}_{cd}^{(u)}(\zeta)+m\delta_{ab}\slashed{L}_{cd}^{(u)}(\zeta)-m\slashed{L}_{ab}^{(u)}(\zeta)\delta_{cd}\Big)\nonumber\\
&&+\gamma^{\mu}_{ab}\gamma^{\nu}_{cd}L_{\mu\nu}^{(u)}(\zeta)\ ,
\label{I_abcd_u_trial}
\eea
where $\zeta^{\mu}=(p_1-p_4)^{\mu}=(p_1-p_2)^{\mu} \ \ \ $ and 
\bea
L^{(u)}&=&\int  \frac{d^2 q}{(2\pi)^2} \frac{1}{\big[(\zeta-q)^2-m^2\big]\big(q^2-m^2\big)}\ , \nonumber\\
L^{(u)}_{\mu}&=&\int  \frac{d^2 q}{(2\pi)^2} \frac{q_{\mu}}{\big[(\zeta-q)^2-m^2\big]\big(q^2-m^2\big)}\ , \nonumber\\
L^{(u)}_{\mu\nu}&=&\int  \frac{d^2 q}{(2\pi)^2} \frac{q_{\mu}q_{\nu}}{\big[(\zeta-q)^2-m^2\big]\big(q^2-m^2\big)}\ .
\label{L_u_trial}
\eea
Similarly to the $t$-channel case, it was shown in section \ref{subsecSmatrix} that the $u$-channel corresponds to $\theta\rightarrow i\pi-\theta$. Hence, naively one can evaluate the $u$-channel loop-integrals $L^{(u)}$, $L^{(u)}_{\mu}$ and $L^{(u)}_{\mu\nu}$ directly from their $s$-channel counterparts (\ref{L_s_trial}) by replacing $\theta\rightarrow i\pi-\theta$.\footnote{Replacing $\theta$ by $i\pi-\theta$ is equivalent to replacing $\xi^2$ by $\zeta^2$.} However, notice that in the u-channel loop integrals (\ref{L_u_trial}) the functional dependence of the integrands on $\zeta$ is of the form $\zeta-q$, while in the $s$-channel loop integrals (\ref{L_s_trial}) the corresponding functional dependence is $\xi+q$.
Therefore, since the functional dependence of the integrands on $\zeta$ is $\zeta-q$, an extra minus sign must be included whenever the integrand is odd in $q$,
 \bea
 L^{(u)}&=&i\frac{\theta}{4\pi m^2\sinh\theta}\ ,\nonumber\\
 L^{(u)}_{\mu}&=&i\frac{\theta}{8\pi m^2\sinh\theta}\zeta_{\mu}\ ,\nonumber\\
 L^{(u)}_{\mu\nu}&=&-\frac{i}{8\pi}\bigg[1+\log \frac{\Lambda^2}{m^2}-\theta\coth\frac{\theta}{2}\bigg]\eta_{\mu\nu}-\frac{i}{16\pi m^2\sinh^2\frac{\theta}{2}}\big(1-\theta\coth\theta\big)\zeta_{\mu}\zeta_{\nu}\ .
 \eea
The derivations of the above integrals are given explicitly in  Appendix \ref{AppendixC}.
\vspace{2mm}
\\
When the vertices have derivative interactions the amplitudes involve more complicated one-loop integrals,
\bea
(I_{\mu_1\cdots\mu_n})_{abcd}(\xi)&=&\int  \frac{d^2 q}{(2\pi)^2}\prod_{i=1}^{n} q_{\mu_i} G_{ab}(\xi+q)G_{cd}(q)\ ,
\eea
which requires the computation of integrals of the form,
\bea
	L_{\mu_1\cdots\mu_{n+2}}=\int \frac{d^2q}{2\pi^2} \frac{\prod_{i=1}^{n+2} q_{\mu_i}}{\big[(\xi+q)^2-m^2\big](q^2-m^2)}\label{L_i}
\eea
where $n\in\{0,...,4\}$.
A detailed discussion on the evaluation of these one-loop integrals can be found in Appendix \ref{AppendixC}.
At this point one can plug the one-loop integral expressions and vertices into MATHEMATICA to obtain the contribution to the amplitudes from the $s$, $t$ and $u$-channels.
\vspace{2mm}
\\
It is important, however, to mention that in the $S$-matrix computation for a \ttb-deformed scalar \cite{Rosenhaus}, one does not need to calculate amplitudes for all the channels explicitly to compute the total amplitude. Rather, one can use the following trick: evaluate the s-channel amplitude explicitly, then consider the appropriate limits to obtain the $t$ and $u$-channel amplitudes. However, such a procedure can not be performed when there are non-trivial polarization vectors that depend on the momenta or rapidities of the external particles. 
To compute the total amplitude of the \ttb-deformed free massive Dirac fermionic theory one must take into account the external polarization vectors described by the Dirac spinors $u_a$, $v_a$, $\bar{u}_a$ and $\bar{v}_a$ which depend on the rapidities $\theta_i$ of the particles. Thus the naive substitution $\theta\rightarrow i\pi$ or $\theta\rightarrow i\pi-\theta$ into the final $s$-channel amplitude would produce incorrect results for the amplitude from the other channels and hence they must be explicitly computed.
\subsubsection*{(a) Both vertices contain interaction $\tilde{X}_{\mu\nu}\tilde{X}^{\mu\nu}$:}
\begin{figure}[h!]
	\centering
	\subfigure[$s$-channel]{
		\includegraphics[width=0.31\columnwidth,height=0.21\columnwidth]{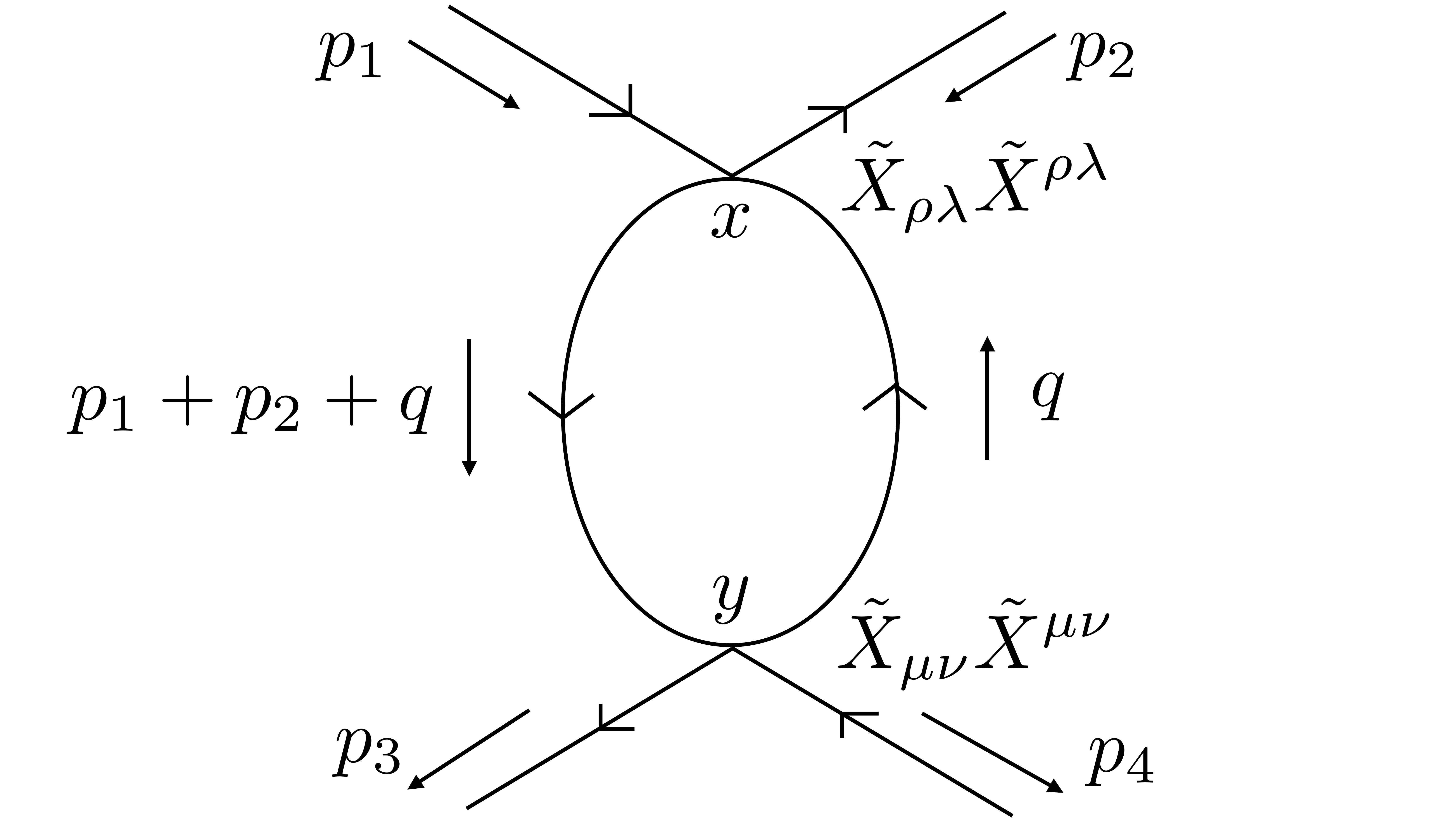}
		\label{bubblediagram_twoderv_s} } 
	\subfigure[$t$-channel]{
		\includegraphics[width=0.31\columnwidth,height=0.21\columnwidth]{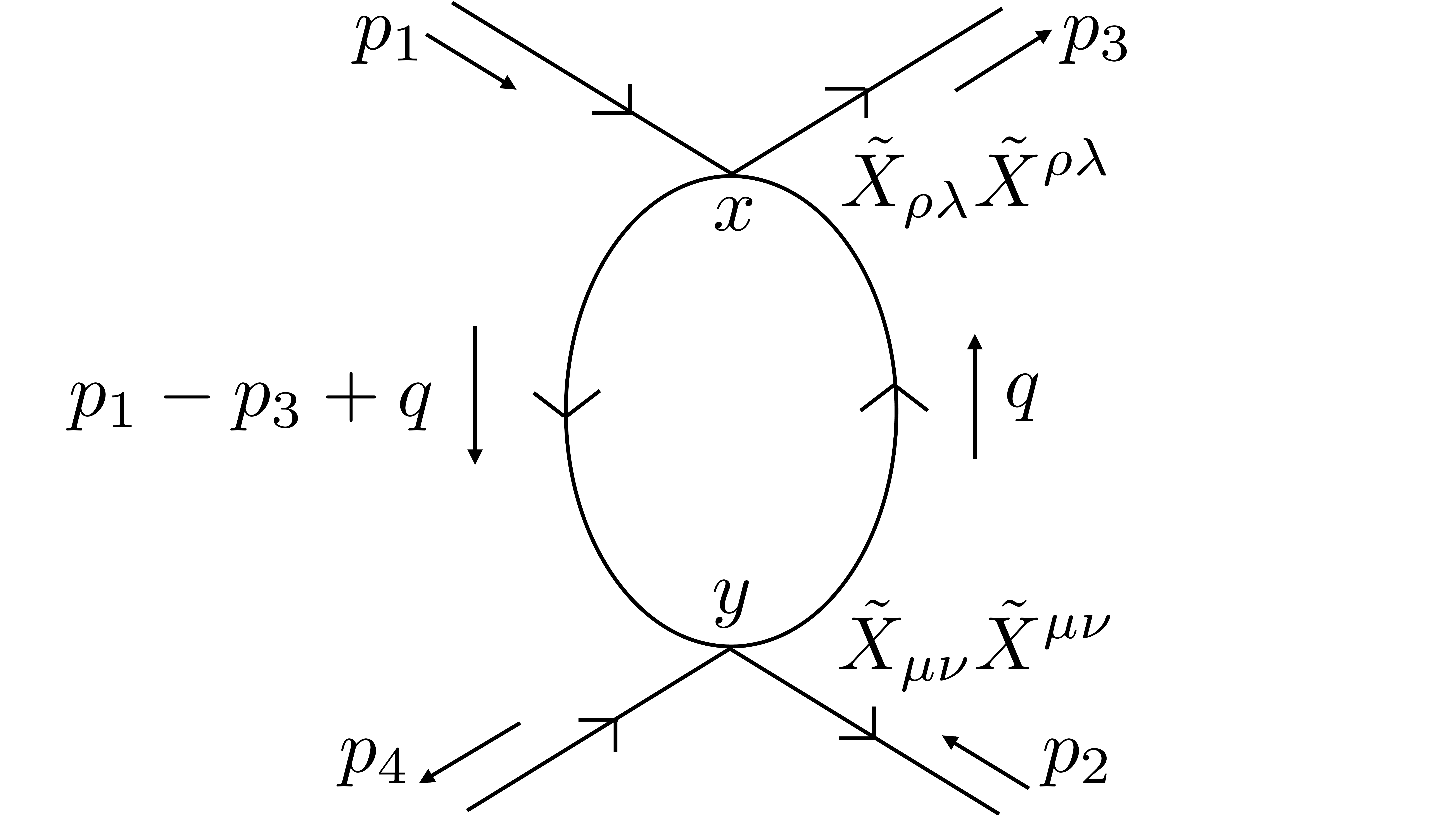}
		\label{bubblediagram_twoderv_t} }
	\subfigure[$u$-channel]{
		\includegraphics[width=0.31\columnwidth,height=0.21\columnwidth]{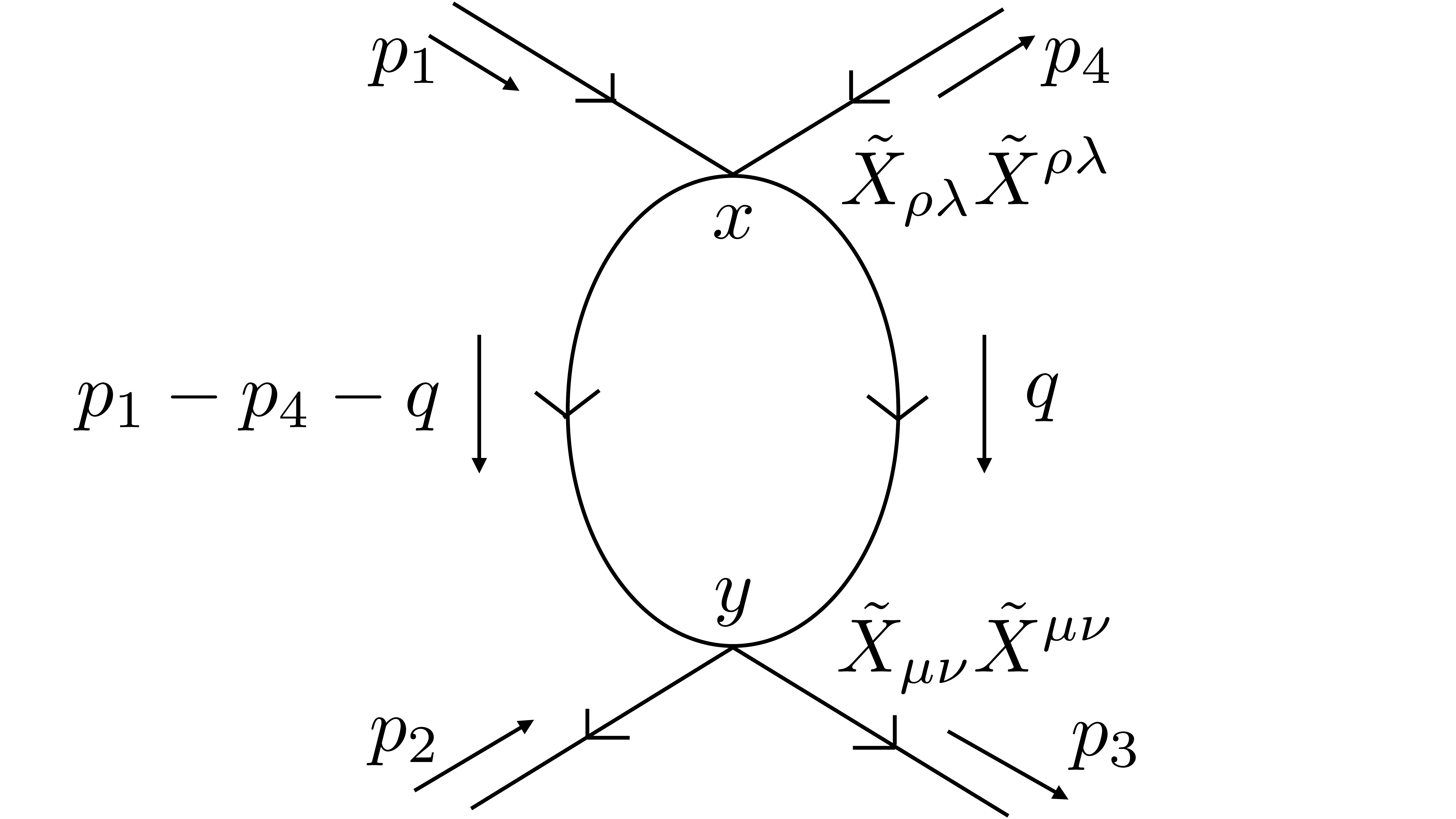}
		\label{bubblediagram_twoderv_u} }
	\caption{\small The $s$, $t$ and $u$-channel amplitudes when both interaction vertices are $\tilde{X}_{\mu\nu}\tilde{X}^{\mu\nu}$.}
	\label{bubblediagram_twoderv}
\end{figure}

First consider the bubble diagrams shown in figure \ref{bubblediagram_twoderv} where both vertices contain quartic couplings $\tilde{X}_{\mu\nu}\tilde{X}^{\mu\nu}$.
The amplitude due to these kinds of bubble diagrams is given by,
\bea 
	\mathcal{A}_{\text{a}}&=& \frac{1}{2!} \big(i\frac{\lambda}{2}\big)^2\lan 0 | b(p_4) a(p_3) \tT\Big[\tilde{X}_{\mu\nu}(y)\tilde{X}^{\mu\nu}(y) \tilde{X}_{\rho\lambda}(x)\tilde{X}^{\rho\lambda}(x)\Big] a^{\dagger}(p_1) b^{\dagger}(p_2)|0 \ran\nonumber\\
&=&\frac{1}{2!} \big(i\frac{\lambda}{2}\big)^2\frac{1}{16}\lan 0 | b(p_4) a(p_3) \tT\Big[
\bpsi(y)\gamma_{(\mu}\del_{\nu)}\psi(y)\bpsi(y)\gamma^{(\mu}\del^{\nu)}\psi(y) \ \ \bpsi(x)\gamma_{(\rho}\del_{\lambda)}\psi(x) \bpsi(x)\gamma^{(\rho}\del^{\lambda)}\psi(x)\nonumber\\
&-&2\bpsi(y)\gamma_{(\mu}\del_{\nu)}\psi(y)\bpsi(y)\gamma^{(\mu}\del^{\nu)}\psi(y) \ \ \bpsi(x)\gamma_{(\rho}\del_{\lambda)}\psi(x) \del^{(\rho}\bpsi(x) \gamma^{\lambda)}\psi(x)\nonumber\\
&+&\bpsi(y)\gamma_{(\mu}\del_{\nu)}\psi(y)\bpsi(y)\gamma^{(\mu}\del^{\nu)}\psi(y) \ \
\del_{(\rho}\bpsi(x) \gamma_{\lambda)}\psi(x) \del^{(\rho}\bpsi(x) \gamma^{\lambda)}\psi(x)\nonumber\\
&-&2\bpsi(y)\gamma_{(\mu}\del_{\nu)}\psi(y) \del^{(\mu}\bpsi(y) \gamma^{\nu)}\psi(y) \ \ \bpsi(x)\gamma_{(\rho}\del_{\lambda)}\psi(x) \bpsi(x)\gamma^{(\rho}\del^{\lambda)}\psi(x)\nonumber\\
&+&4\bpsi(y)\gamma_{(\mu}\del_{\nu)}\psi(y) \del^{(\mu}\bpsi(y) \gamma^{\nu)}\psi(y) \ \ \bpsi(x)\gamma_{(\rho}\del_{\lambda)}\psi(x) \del^{(\rho}\bpsi(x) \gamma^{\lambda)}\psi(x)\nonumber\\
&-&2\bpsi(y)\gamma_{(\mu}\del_{\nu)}\psi(y) \del^{(\mu}\bpsi(y) \gamma^{\nu)}\psi(y) \ \  \del_{(\rho}\bpsi(x) \gamma_{\lambda)}\psi(x) \del^{(\rho}\bpsi(x) \gamma^{\lambda)}\psi(x)\nonumber\\
&+&\del_{(\mu}\bpsi(y) \gamma_{\nu)}\psi(y) \del^{(\mu}\bpsi(y) \gamma^{\nu)}\psi(y) \ \ \bpsi(x)\gamma_{(\rho}\del_{\lambda)}\psi(x) \bpsi(x)\gamma^{(\rho}\del^{\lambda)}\psi(x)\nonumber\\
&-&2\del_{(\mu}\bpsi(y) \gamma_{\nu)}\psi(y) \del^{(\mu}\bpsi(y) \gamma^{\nu)}\psi(y) \ \  \bpsi(x)\gamma_{(\rho}\del_{\lambda)}\psi(x) \del^{(\rho}\bpsi(x) \gamma^{\lambda)}\psi(x)\nonumber\\
&+&\del_{(\mu}\bpsi(y) \gamma_{\nu)}\psi(y) \del^{(\mu}\bpsi(y) \gamma^{\nu)}\psi(y) \ \ \del_{(\rho}\bpsi(x) \gamma_{\lambda)}\psi(x) \del^{(\rho}\bpsi(x) \gamma^{\lambda)}\psi(x) \ \Big] a^{\dagger}(p_1) b^{\dagger}(p_2)|0 \ran\ .
\label{TmunuPower4}
\eea
The $s$-channel amplitude (figure \ref{bubblediagram_twoderv_s}) can be computed by evaluating the contributions to the $s$-channel from each of the nine terms in  (\ref{TmunuPower4}). By the general $s$-channel amplitude (\ref{schannelbubblegeneral}) the contribution from the first term in (\ref{TmunuPower4}), $\bpsi(y)\gamma_{(\mu}\del_{\nu)}\psi(y)\bpsi(y)\gamma^{(\mu}\del^{\nu)}\psi(y) \times \bpsi(x)\gamma_{(\rho}\del_{\lambda)}\psi(x) \bpsi(x)\gamma^{(\rho}\del^{\lambda)}\psi(x)$, to the amplitude is given by,
\bea
\mathcal{A}_{\text{a}}^{(s_1)}&=&\frac{1}{2!}\big(i\frac{\lambda}{2}\big)^2 \frac{1}{16}\lan 0 | b(p_4) a(p_3) \tT\Big[\bpsi(y)\gamma_{(\mu}\del_{\nu)}\psi(y)\bpsi(y)\gamma^{(\mu}\del^{\nu)}\psi(y) \ \nonumber\\
&& \hspace{4.2cm} \times \bpsi(x)\gamma_{(\rho}\del_{\lambda)}\psi(x) \bpsi(x)\gamma^{(\rho}\del^{\lambda)}\psi(x)\Big] a^{\dagger}(p_1) b^{\dagger}(p_2)|0 \ran^{(s)}\nonumber\\
&=&-\frac{\lambda^2}{128}\gamma_{ef(\mu}\gamma^{(\mu}_{gh}\gamma_{ab(\rho}\gamma^{(\rho}_{cd} \ \times\nonumber\\
&\bigg[&-\bv_c(p_2)u_d(p_1)\int\frac{d^2q}{(2\pi)^2}
\Big(\bu_e(p_3)v_f(p_4)G_{ha}(\xi+q)G_{bg}(q)i p_{4\nu)}(-i)(\xi+q)^{\nu)}(-i) q_{\lambda)}(-i)p_1^{\lambda)}\nonumber\\
&-&\bu_e(p_3)v_h(p_4)G_{fa}(\xi+q)G_{bg}(q) (-i)(\xi+q)_{\nu)}i p_4^{\nu)}(-i) q_{\lambda)}(-i)p_1^{\lambda)} \nonumber\\
&-&\bu_{g}(p_3)v_f(p_4)G_{ha}(\xi+q)G_{be}(q)i p_{4\nu)}(-i)(\xi+q)^{\nu)}(-i) q_{\lambda)}(-i)p_1^{\lambda)}\nonumber\\
&+&\bu_g(p_3)v_h(p_4)G_{fa}(\xi+q)G_{be}(q)(-i)(\xi+q)_{\nu)}i p_4^{\nu)}(-i) q_{\lambda)}(-i)p_1^{\lambda)}\Big)
\nonumber\eea\bea
&-&\bv_a(p_2)u_b(p_1)\int\frac{d^2q}{(2\pi)^2}\Big(\bu_e(p_3)v_f(p_4)G_{hc}(\xi+q)G_{dg}(q) i p_{4\nu)}(-i)(\xi+q)^{\nu)}(-i) p_{1\lambda)}(-i)q^{\lambda)}\nonumber\\
&-&\bu_e(p_3)v_h(p_4)G_{fc}(\xi+q)G_{dg}(q)(-i)(\xi+q)_{\nu)}i p_4^{\nu)}(-i) p_{1\lambda)}(-i)q^{\lambda)}\nonumber\\
&-&\bu_{g}(p_3)v_f(p_4)G_{hc}(\xi+q)G_{de}(q)i p_{4\nu)}(-i)(\xi+q)^{\nu)}(-i) p_{1\lambda)}(-i)q^{\lambda)}\nonumber\\
&+&\bu_g(p_3)v_h(p_4)G_{fc}(\xi+q)G_{de}(q)(-i)(\xi+q)_{\nu)}i p_4^{\nu)}(-i) p_{1\lambda)}(-i)q^{\lambda)}\Big)
\nonumber\\
&+&\bv_a(p_2)u_d(p_1)\int\frac{d^2q}{(2\pi)^2}\Big(\bu_e(p_3)v_f(p_4)G_{hc}(\xi+q)G_{bg}(q)i p_{4\nu)}(-i)(\xi+q)^{\nu)}(-i) q_{\lambda)}(-i)p_1^{\lambda)}\nonumber\\
&-&\bu_e(p_3)v_h(p_4)G_{fc}(\xi+q)G_{bg}(q)(-i)(\xi+q)_{\nu)}i p_4^{\nu)}(-i) q_{\lambda)}(-i)p_1^{\lambda)}\nonumber\\
&-&\bu_{g}(p_3)v_f(p_4)G_{hc}(\xi+q)G_{be}(q)i p_{4\nu)}(-i)(\xi+q)^{\nu)}(-i) q_{\lambda)}(-i)p_1^{\lambda)}\nonumber\\
&+&\bu_g(p_3)v_h(p_4)G_{fc}(\xi+q)G_{be}(q)(-i)(\xi+q)_{\nu)}i p_4^{\nu)}(-i) q_{\lambda)}(-i)p_1^{\lambda)}\Big)
\nonumber\\
&+&\bv_c(p_2)u_b(p_1)\int\frac{d^2q}{(2\pi)^2}\Big(\bu_e(p_3)v_f(p_4)G_{ha}(\xi+q)G_{dg}(q)i p_{4\nu)}(-i)(\xi+q)^{\nu)}(-i) p_{1\lambda)}(-i)q^{\lambda)}\nonumber\\
&-&\bu_e(p_3)v_h(p_4)G_{fa}(\xi+q)G_{dg}(q)(-i)(\xi+q)_{\nu)}i p_4^{\nu)}(-i) p_{1\lambda)}(-i)q^{\lambda)}\nonumber\\
&-&\bu_{g}(p_3)v_f(p_4)G_{ha}(\xi+q)G_{de}(q)i p_{4\nu)}(-i)(\xi+q)^{\nu)}(-i) p_{1\lambda)}(-i)q^{\lambda)}\nonumber\\
&+&\bu_g(p_3)v_h(p_4)G_{fa}(\xi+q)G_{de}(q)(-i)(\xi+q)_{\nu)}i p_4^{\nu)}(-i) p_{1\lambda)}(-i)q^{\lambda)}\Big)\bigg]\ ,
\label{twoderv_s1}
\eea
where $\xi^{\mu}=(p_1+p_2)^{\mu}$.
The above expression, (\ref{twoderv_s1}), involves the one-loop integrals $(I_{\mu})^{(s)}_{abcd}(\xi)$ and $(I_{\mu\nu})^{(s)}_{abcd}(\xi)$ whose values are given by (\ref{I1_abcd}) and (\ref{I2_abcd}), respectively. Plugging the values of the integrals and Dirac spinors, (\ref{uv}) and  (\ref{ubarvbar}), into (\ref{twoderv_s1}) one finds the $s$-channel contribution from the first term in (\ref{TmunuPower4}) to be,
\bea
\mathcal{A}_{\text{a}}^{(s_1)}&=&-\frac{\lambda ^2 m^6}{768 \pi }\bigg[\frac{3}{2} \pi  \csch \theta \Big(87+24 \cosh \theta +17 \cosh 2 \theta +12 \cosh 3 \theta +4 \cosh 4 \theta \Big)
\nonumber\\
&+&i \Big(95-6 \frac{\Lambda ^2}{m^2}-24\log\frac{m}{\Lambda}+6 \cosh \theta  \big(1+25 \log \frac{m}{\Lambda }\big)-6 \cosh 2 \theta  \big(3+2 \frac{\Lambda ^2}{m^2}-12 \log \frac{m}{\Lambda }\big)\nonumber\\
&-&2\cosh3\theta\big(5-12\log\frac{m}{\Lambda}\big)+\frac{3}{2} \theta  \csch \theta \big(87+24 \cosh \theta +17 \cosh 2 \theta +12 \cosh 3 \theta \nonumber\\
&+&4 \cosh 4 \theta \big)\Big)\bigg]\ .
\label{twoderv_s1_final}
\eea
The contributions to the $s$-channel amplitude from the eight additional terms in (\ref{TmunuPower4}) can be obtained in a similar manner. Their values are given by,
\bea
\mathcal{A}_{\text{a}}^{(s_2)}&=&\frac{1}{2!}\big(i\frac{\lambda}{2}\big)^2 \frac{1}{16}\lan 0 | b(p_4) a(p_3) \tT\Big[-2\bpsi(y)\gamma_{(\mu}\del_{\nu)}\psi(y)\bpsi(y)\gamma^{(\mu}\del^{\nu)}\psi(y) \ \nonumber\\
&& \hspace{4.2cm} \times  \bpsi(x)\gamma_{(\rho}\del_{\lambda)}\psi(x) \del^{(\rho}\bpsi(x) \gamma^{\lambda)}\psi(x) \Big]a^{\dagger}(p_1) b^{\dagger}(p_2)|0 \ran^{(s)}\nonumber\\
&=&-\frac{\lambda ^2 m^6}{1536 \pi }\bigg[6\pi \csch \theta  \Big(29+4 \cosh \theta +3 \cosh 2 \theta +8\cosh 3 \theta +4 \cosh 4 \theta \Big)\nonumber\\
&+&i \Big(98-6 \frac{\Lambda ^2}{m^2}+\cosh \theta  \big(-47-42 \frac{\Lambda ^2}{m^2}+228  \log \frac{m}{\Lambda }\big)+ 2 \cosh 2\theta  \big(-46-21\frac{\Lambda^2}{m^2}\nonumber
\eea
\bea
&+&96\log \frac{m}{\Lambda }\big)+\cosh3\theta \big(-37+96\log\frac{m}{\Lambda}\big)+6\theta \csch \theta\big(29+4 \cosh \theta +3 \cosh 2 \theta \nonumber\\
&+&8\cosh 3 \theta +4 \cosh 4 \theta\big)
\Big)\bigg]\ ,
\label{twoderv_s2}
\eea
\bea
\mathcal{A}_{\text{a}}^{(s_3)}&=&\frac{1}{2!}\big(i\frac{\lambda}{2}\big)^2 \frac{1}{16}\lan 0 | b(p_4) a(p_3) \tT\Big[\bpsi(y)\gamma_{(\mu}\del_{\nu)}\psi(y)\bpsi(y)\gamma^{(\mu}\del^{\nu)}\psi(y) \ \nonumber\\
&& \hspace{4.2cm} \times  \del_{(\rho}\bpsi(x) \gamma_{\lambda)}\psi(x) \del^{(\rho}\bpsi(x) \gamma^{\lambda)}\psi(x) \Big] a^{\dagger}(p_1) b^{\dagger}(p_2)|0 \ran^{(s)}\nonumber\\
&=&-\frac{\lambda ^2 m^6}{1536 \pi }\bigg[3\pi \csch \theta  \Big(87+24 \cosh \theta +17 \cosh 2 \theta +12\cosh 3 \theta +4 \cosh 4 \theta \Big)\nonumber\\
&+&i \Big(70-240\log\frac{m}{\Lambda}+2\cosh \theta  \big(-74+15 \frac{\Lambda ^2}{m^2}+48  \log \frac{m}{\Lambda }\big)+ \cosh 2\theta  \big(-82+144\log \frac{m}{\Lambda }\big)\nonumber\\
&+&4\cosh3\theta \big(-5+12\log\frac{m}{\Lambda}\big)+3\theta \csch \theta\big(87+24 \cosh \theta +17 \cosh 2 \theta +12\cosh 3 \theta \nonumber\\
&+&4 \cosh 4 \theta\big)
\Big)\bigg]\ ,
\label{twoderv_s3}
\eea
\bea
\mathcal{A}_{\text{a}}^{(s_4)}&=&\frac{1}{2!}\big(i\frac{\lambda}{2}\big)^2 \frac{1}{16}\lan 0 | b(p_4) a(p_3) \tT\Big[-2\bpsi(y)\gamma_{(\mu}\del_{\nu)}\psi(y) \del^{(\mu}\bpsi(y) \gamma^{\nu)}\psi(y) \ \nonumber\\
&& \hspace{0.5cm} \times  \bpsi(x)\gamma_{(\rho}\del_{\lambda)}\psi(x) \bpsi(x)\gamma^{(\rho}\del^{\lambda)}\psi(x) \Big] a^{\dagger}(p_1) b^{\dagger}(p_2)|0 \ran^{(s)}
\ \ = \ \ \mathcal{A}_{\text{a}}^{(s_2)}\ ,
\label{twoderv_s4}
\eea
\bea
\mathcal{A}_{\text{a}}^{(s_5)}&=&\frac{1}{2!}\big(i\frac{\lambda}{2}\big)^2 \frac{1}{16}\lan 0 | b(p_4) a(p_3) \tT\Big[4\bpsi(y)\gamma_{(\mu}\del_{\nu)}\psi(y) \del^{(\mu}\bpsi(y) \gamma^{\nu)}\psi(y)  \ \nonumber\\
&& \hspace{4.2cm} \times  \bpsi(x)\gamma_{(\rho}\del_{\lambda)}\psi(x) \del^{(\rho}\bpsi(x) \gamma^{\lambda)}\psi(x) \Big] a^{\dagger}(p_1) b^{\dagger}(p_2)|0 \ran^{(s)}\nonumber\\
&=&-\frac{\lambda ^2 m^6}{76800\pi }\bigg[600 \pi  \csch \theta \Big(15-7 \cosh 2\theta +4 \cosh 3 \theta +4 \cosh 4 \theta \Big)
\nonumber\\
&+&i \Big(4252-120 \frac{\Lambda ^2}{m^2}-9600\log\frac{m}{\Lambda}+ \cosh \theta  \big(1111-5640\frac{\Lambda^2}{m^2}-1350\frac{\Lambda^4}{m^4}+10800 \log \frac{m}{\Lambda }\big)\nonumber\\
&-&8 \cosh 2 \theta  \big(986+615 \frac{\Lambda ^2}{m^2}-1200 \log \frac{m}{\Lambda }\big)
+\cosh3\theta\big(-3247+9600\log\frac{m}{\Lambda}\big)\nonumber\\
&+&600\theta  \csch \theta \big(15-7 \cosh 2\theta +4 \cosh 3 \theta +4 \cosh 4 \theta  \big)\Big)\bigg]\ ,
\label{twoderv_s5}
\eea
\bea
\mathcal{A}_{\text{a}}^{(s_6)}&=&\frac{1}{2!}\big(i\frac{\lambda}{2}\big)^2 \frac{1}{16}\lan 0 | b(p_4) a(p_3) \tT\Big[-2\bpsi(y)\gamma_{(\mu}\del_{\nu)}\psi(y) \del^{(\mu}\bpsi(y) \gamma^{\nu)}\psi(y) \ \nonumber\\
&& \hspace{0.5cm} \times  \del_{(\rho}\bpsi(x) \gamma_{\lambda)}\psi(x) \del^{(\rho}\bpsi(x) \gamma^{\lambda)}\psi(x) \Big] a^{\dagger}(p_1) b^{\dagger}(p_2)|0 \ran^{(s)}
 \ \ = \ \ \mathcal{A}_{\text{a}}^{(s_2)}\ ,
\label{twoderv_s6}
\eea
\bea
\mathcal{A}_{\text{a}}^{(s_7)}&=&\frac{1}{2!}\big(i\frac{\lambda}{2}\big)^2 \frac{1}{16}\lan 0 | b(p_4) a(p_3) \tT\Big[\del_{(\mu}\bpsi(y) \gamma_{\nu)}\psi(y) \del^{(\mu}\bpsi(y) \gamma^{\nu)}\psi(y) \ \nonumber\\
&& \hspace{0.5cm} \times  \bpsi(x)\gamma_{(\rho}\del_{\lambda)}\psi(x) \bpsi(x)\gamma^{(\rho}\del^{\lambda)}\psi(x) \Big] a^{\dagger}(p_1) b^{\dagger}(p_2)|0 \ran^{(s)}
\ \ = \ \ \mathcal{A}_{\text{a}}^{(s_3)}\ ,
\label{twoderv_s7}
\eea
\bea
\mathcal{A}_{\text{a}}^{(s_8)}&=&\frac{1}{2!}\big(i\frac{\lambda}{2}\big)^2 \frac{1}{16}\lan 0 | b(p_4) a(p_3) \tT\Big[-2\del_{(\mu}\bpsi(y) \gamma_{\nu)}\psi(y) \del^{(\mu}\bpsi(y) \gamma^{\nu)}\psi(y) \ \nonumber\\
&& \hspace{0.5cm} \times  \bpsi(x)\gamma_{(\rho}\del_{\lambda)}\psi(x) \del^{(\rho}\bpsi(x) \gamma^{\lambda)}\psi(x) \Big] a^{\dagger}(p_1) b^{\dagger}(p_2)|0 \ran^{(s)}
\ \ = \ \ \mathcal{A}_{\text{a}}^{(s_2)}\ ,
\label{twoderv_s8}
\eea
\bea
\mathcal{A}_{\text{a}}^{(s_9)}&=&\frac{1}{2!}\big(i\frac{\lambda}{2}\big)^2 \frac{1}{16}\lan 0 | b(p_4) a(p_3) \tT\Big[\del_{(\mu}\bpsi(y) \gamma_{\nu)}\psi(y) \del^{(\mu}\bpsi(y) \gamma^{\nu)}\psi(y) \ \nonumber\\
&& \hspace{0.5cm} \times  \del_{(\rho}\bpsi(x) \gamma_{\lambda)}\psi(x) \del^{(\rho}\bpsi(x) \gamma^{\lambda)}\psi(x) \Big] a^{\dagger}(p_1) b^{\dagger}(p_2)|0 \ran^{(s)}
\ \ = \ \ \mathcal{A}_{\text{a}}^{(s_1)}\ .
\label{twoderv_s9}
\eea
Combining the $s$-channel contributions, (\ref{twoderv_s1_final}) - (\ref{twoderv_s9}), gives the total $s$-channel amplitude from figure \ref{bubblediagram_twoderv_s},
\bea
\mathcal{A}_{\text{a}}^{(s)}&=&2\Big(\mathcal{A}_{\text{a}}^{(s_1)}+\mathcal{A}_{\text{a}}^{(s_2)}+\cdots+\mathcal{A}_{\text{a}}^{(s_9)}\Big)\nonumber\\
&=&-\frac{\lambda ^2 m^6}{38400\pi }\bigg[9600 \pi  \csch \theta \Big(10+2\cosh\theta +\cosh 2\theta +2 \cosh 3 \theta + \cosh 4 \theta \Big)
\nonumber\\
&+&i \Big(49852-2520 \frac{\Lambda ^2}{m^2}-38400\log\frac{m}{\Lambda}+ \cosh \theta  \big(-21889-11040\frac{\Lambda^2}{m^2}-1350\frac{\Lambda^4}{m^4}\nonumber\\
&+&96000 \log \frac{m}{\Lambda }\big)
+24 \cosh 2 \theta  \big(-1587-655 \frac{\Lambda ^2}{m^2}+3200 \log \frac{m}{\Lambda }\big)
+\cosh3\theta\big(-14647\nonumber\\
&+&38400\log\frac{m}{\Lambda}\big)
+9600\theta  \csch \theta \big(10+2\cosh\theta +\cosh 2\theta +2 \cosh 3 \theta + \cosh 4 \theta \big)\Big)\bigg]\ .\nonumber\\
\label{twoderv_s_amplitude}
\eea
An extra multiplicative factor $2$ arises from the identical contribution of the diagram with the two vertices $x$ and $y$ exchanged. The same factor will be included while computing the $t$ and $u$-channel amplitudes.
\\\\
Next, the amplitude due to the same quartic interactions will be computed in the $t$-channel case (figure \ref{bubblediagram_twoderv_t}).
By the general  $t$-channel amplitude (\ref{tchannelbubblegeneral}) the contribution from the first term in (\ref{TmunuPower4}), $\bpsi(y)\gamma_{(\mu}\del_{\nu)}\psi(y)\bpsi(y)\gamma^{(\mu}\del^{\nu)}\psi(y) \times \bpsi(x)\gamma_{(\rho}\del_{\lambda)}\psi(x) \bpsi(x)\gamma^{(\rho}\del^{\lambda)}\psi(x)$, to the amplitude is given by,
\bea
\mathcal{A}_{\text{a}}^{(t_1)}&=&\frac{1}{2!} \big(i\frac{\lambda}{2}\big)^2 \frac{1}{16}\lan 0 | b(p_4) a(p_3) \tT\Big[\bpsi(y)\gamma_{(\mu}\del_{\nu)}\psi(y)\bpsi(y)\gamma^{(\mu}\del^{\nu)}\psi(y) \ \nonumber\\
&& \hspace{4.2cm} \times  \bpsi(x)\gamma_{(\rho}\del_{\lambda)}\psi(x) \bpsi(x)\gamma^{(\rho}\del^{\lambda)}\psi(x)\Big] a^{\dagger}(p_1) b^{\dagger}(p_2)|0 \ran^{(t)}\nonumber\\
&=&-\frac{\lambda^2}{128}\gamma_{ef(\mu}\gamma^{(\mu}_{gh}\gamma_{ab(\rho}\gamma^{(\rho}_{cd} \ \times\nonumber\\
&\bigg[&
\bu_c(p_3)u_d(p_1)\int\frac{d^2q}{(2\pi)^2}
\Big(\bv_g(p_2)v_h(p_4)G_{fa}(q)G_{be}(q)(-i) q_{\nu)}i p_4^{\nu)}(-i) q_{\lambda)}(-i)p_1^{\lambda)}\nonumber\\
&-&\bv_g(p_2)v_f(p_4)G_{ha}(q)G_{be}(q)(-i) q^{\nu)}i p_{4\nu)}(-i) q_{\lambda)}(-i)p_1^{\lambda)}\nonumber\\
&-&\bv_{e}(p_2)v_h(p_4)G_{fa}(q)G_{bg}(q)(-i) q_{\nu)}i p_4^{\nu)}(-i) q_{\lambda)}(-i)p_1^{\lambda)}\nonumber\\
&+&\bv_e(p_2)v_f(p_4)G_{ha}(q)G_{bg}(q)(-i) q^{\nu)}i p_{4\nu)}(-i) q_{\lambda)}(-i)p_1^{\lambda)}\Big) \nonumber\eea\bea
&+&\bu_a(p_3)u_b(p_1)\int\frac{d^2q}{(2\pi)^2}\Big(\bv_g(p_2)v_h(p_4)G_{fc}(q)G_{de}(q)(-i) q_{\nu)}i p_4^{\nu)}(-i) p_{1\lambda)}(-i)q^{\lambda)}\nonumber\\
&-&\bv_g(p_2)v_f(p_4)G_{hc}(q)G_{de}(q)i p_{4\nu)}(-i) q^{\nu)}(-i) p_{1\lambda)}(-i)q^{\lambda)}\nonumber\\
&-&\bv_{e}(p_2)v_h(p_4)G_{fc}(q)G_{dg}(q)(-i) q_{\nu)}i p_4^{\nu)}(-i) p_{1\lambda)}(-i)q^{\lambda)}\nonumber\\
&+&\bv_e(p_2)v_f(p_4)G_{hc}(q)G_{dg}(q)i p_{4\nu)}(-i) q^{\nu)}(-i) p_{1\lambda)}(-i)q^{\lambda)}\Big)
\nonumber\\
&-&\bu_a(p_3)u_d(p_1)\int\frac{d^2q}{(2\pi)^2}\Big(\bv_g(p_2)v_h(p_4)G_{fc}(q)G_{be}(q)(-i) q_{\nu)}i p_4^{\nu)}(-i)q_{\lambda)}(-i) p_1^{\lambda)}\nonumber\\
&-&\bv_g(p_2)v_f(p_4)G_{hc}(q)G_{be}(q) i p_{4\nu)}(-i) q^{\nu)}(-i)q_{\lambda)}(-i) p_1^{\lambda)}\nonumber\\
&-&\bv_{e}(p_2)v_h(p_4)G_{fc}(q)G_{bg}(q)(-i) q_{\nu)}i p_4^{\nu)}(-i)q_{\lambda)}(-i) p_1^{\lambda)}\nonumber\\
&+&\bv_e(p_2)v_f(p_4)G_{hc}(q)G_{bg}(q)i p_{4\nu)}(-i) q^{\nu)}(-i)q_{\lambda)}(-i) p_1^{\lambda)}\Big)
\nonumber\\
&-&\bu_c(p_3)u_b(p_1)\int\frac{d^2q}{(2\pi)^2}\Big(\bv_g(p_2)v_h(p_4)G_{fa}(q)G_{de}(q)(-i) q_{\nu)}i p_4^{\nu)}(-i)p_{1\lambda)}(-i) q^{\lambda)}\nonumber\\
&-&\bv_g(p_2)v_f(p_4)G_{ha}(q)G_{de}(q)i p_{4\nu)}(-i) q^{\nu)}(-i)p_{1\lambda)}(-i) q^{\lambda)}\nonumber\\
&-&\bv_{e}(p_2)v_h(p_4)G_{fa}(q)G_{dg}(q)(-i) q_{\nu)}i p_4^{\nu)}(-i)p_{1\lambda)}(-i) q^{\lambda)}\nonumber\\
&+&\bv_e(p_2)v_f(p_4)G_{ha}(q)G_{dg}(q)i p_{4\nu)}(-i) q^{\nu)}(-i)p_{1\lambda)}(-i) q^{\lambda)}\Big)
\bigg]\ .
\label{twoderv_t1}
\eea
The above expression, (\ref{twoderv_t1}), involves the one-loop integral $(I_{\mu\nu})^{(t)}_{abcd}$ given by  (\ref{I_t_abcd}). Plugging the value of this integral and Dirac spinors into (\ref{twoderv_t1}) one finds the t-channel contribution from the first term in (\ref{TmunuPower4}) to be,
\bea
\mathcal{A}_{\text{a}}^{(t_1)}&=&-i\frac{\lambda ^2 m^6}{256 \pi }\bigg[1+ \frac{\Lambda ^2}{m^2}+6\log\frac{m}{\Lambda}- \cosh \theta  \big(4+5\frac{\Lambda^2}{m^2}+28 \log \frac{m}{\Lambda }\big)+4\cosh 2 \theta  \big(1+2 \log \frac{m}{\Lambda }\big)\bigg]\ .\nonumber\\
\label{twoderv_t1_final}
\eea
The contributions to the $t$-channel amplitude from the eight additional terms in (\ref{TmunuPower4}) can be obtained in a similar manner. Their values are given by, 
\bea
\mathcal{A}_{\text{a}}^{(t_2)}&=&\frac{1}{2!} \big(i\frac{\lambda}{2}\big)^2 \frac{1}{16}\lan 0 | b(p_4) a(p_3) \tT\Big[-2\bpsi(y)\gamma_{(\mu}\del_{\nu)}\psi(y)\bpsi(y)\gamma^{(\mu}\del^{\nu)}\psi(y) \ \nonumber\\
&& \hspace{4.3cm} \times  \bpsi(x)\gamma_{(\rho}\del_{\lambda)}\psi(x) \del^{(\rho}\bpsi(x) \gamma^{\lambda)}\psi(x) \Big] a^{\dagger}(p_1) b^{\dagger}(p_2)|0 \ran^{(t)}\nonumber\\
&=&-i\frac{\lambda ^2 m^6}{256 \pi }\bigg[-2-2 \frac{\Lambda ^2}{m^2}-12\log\frac{m}{\Lambda}- 3\cosh \theta  \big(1+\frac{\Lambda^2}{m^2}+6 \log \frac{m}{\Lambda }\big)+8\cosh 2 \theta  \big(1+2 \log \frac{m}{\Lambda }\big)\bigg]\ ,\nonumber\\
\label{twoderv_t2}
\eea
\bea
\mathcal{A}_{\text{a}}^{(t_3)}&=&\frac{1}{2!} \big(i\frac{\lambda}{2}\big)^2 \frac{1}{16}\lan 0 | b(p_4) a(p_3) \tT\Big[\bpsi(y)\gamma_{(\mu}\del_{\nu)}\psi(y)\bpsi(y)\gamma^{(\mu}\del^{\nu)}\psi(y) \ \nonumber\\
&& \hspace{4.4cm} \times  \del_{(\rho}\bpsi(x) \gamma_{\lambda)}\psi(x) \del^{(\rho}\bpsi(x) \gamma^{\lambda)}\psi(x) \Big] a^{\dagger}(p_1) b^{\dagger}(p_2)|0 \ran^{(t)}\nonumber\\
&=&-i\frac{\lambda ^2 m^6}{256 \pi }\bigg[1+ \frac{\Lambda ^2}{m^2}+6\log\frac{m}{\Lambda}- \cosh \theta  \big(5+4\frac{\Lambda^2}{m^2}+26 \log \frac{m}{\Lambda }\big)+4\cosh 2 \theta  \big(1+2 \log \frac{m}{\Lambda }\big)\bigg]\ ,\nonumber\\
\label{twoderv_t3}
\eea
\bea
\mathcal{A}_{\text{a}}^{(t_4)}&=&\frac{1}{2!}\big(i\frac{\lambda}{2}\big)^2 \frac{1}{16}\lan 0 | b(p_4) a(p_3) \tT\Big[4\bpsi(y)\gamma_{(\mu}\del_{\nu)}\psi(y) \del^{(\mu}\bpsi(y) \gamma^{\nu)}\psi(y) \ \nonumber\\
&& \hspace{0.5cm} \times \bpsi(x)\gamma_{(\rho}\del_{\lambda)}\psi(x) \bpsi(x)\gamma^{(\rho}\del^{\lambda)}\psi(x) \Big] a^{\dagger}(p_1) b^{\dagger}(p_2)|0 \ran^{(t)}
\ \ = \ \ \mathcal{A}_{\text{a}}^{(t_2)}\ ,
\label{twoderv_t4}
\eea
\bea
\mathcal{A}_{\text{a}}^{(t_5)}&=&\frac{1}{2!} \big(i\frac{\lambda}{2}\big)^2 \frac{1}{16}\lan 0 | b(p_4) a(p_3) \tT\Big[-2\bpsi(y)\gamma_{(\mu}\del_{\nu)}\psi(y) \del^{(\mu}\bpsi(y) \gamma^{\nu)}\psi(y)  \ \nonumber\\
&& \hspace{4.2cm} \times  \bpsi(x)\gamma_{(\rho}\del_{\lambda)}\psi(x) \del^{(\rho}\bpsi(x) \gamma^{\lambda)}\psi(x) \Big] a^{\dagger}(p_1) b^{\dagger}(p_2)|0 \ran^{(t)}\nonumber\\
&=&i\frac{\lambda ^2 m^6}{512\pi }\bigg[-8+10 \frac{\Lambda ^2}{m^2}+9\frac{\Lambda^4}{m^4}-48\log\frac{m}{\Lambda}+2 \cosh \theta  \big(1+3\frac{\Lambda^2}{m^2}+12 \log \frac{m}{\Lambda }\big)\nonumber\\
&-&32\cosh 2\theta  \big(1+2 \log \frac{m}{\Lambda }\big)\bigg]\ ,
\label{twoderv_t5}
\eea
\bea
\mathcal{A}_{\text{a}}^{(t_6)}&=&\frac{1}{2!} \big(i\frac{\lambda}{2}\big)^2 \frac{1}{16}\lan 0 | b(p_4) a(p_3) \tT\Big[-2\bpsi(y)\gamma_{(\mu}\del_{\nu)}\psi(y) \del^{(\mu}\bpsi(y) \gamma^{\nu)}\psi(y) \ \nonumber\\
&& \hspace{0.5cm} \times  \del_{(\rho}\bpsi(x) \gamma_{\lambda)}\psi(x) \del^{(\rho}\bpsi(x) \gamma^{\lambda)}\psi(x) \Big] a^{\dagger}(p_1) b^{\dagger}(p_2)|0 \ran^{(t)}
\ \ = \ \ \mathcal{A}_{\text{a}}^{(t_2)}\ ,
\label{twoderv_t6}
\eea
\bea
\mathcal{A}_{\text{a}}^{(t_7)}&=&\frac{1}{2!}\big(i\frac{\lambda}{2}\big)^2 \frac{1}{16}\lan 0 | b(p_4) a(p_3) \tT\Big[\del_{(\mu}\bpsi(y) \gamma_{\nu)}\psi(y) \del^{(\mu}\bpsi(y) \gamma^{\nu)}\psi(y) \ \nonumber\\
&& \hspace{0.5cm} \times  \bpsi(x)\gamma_{(\rho}\del_{\lambda)}\psi(x) \bpsi(x)\gamma^{(\rho}\del^{\lambda)}\psi(x) \Big] a^{\dagger}(p_1) b^{\dagger}(p_2)|0 \ran^{(s)}
\ \ = \ \ \mathcal{A}_{\text{a}}^{(t_3)}\ ,
\label{twoderv_t7}
\eea
\bea
\mathcal{A}_{\text{a}}^{(t_8)}&=&\frac{1}{2!}\big(i\frac{\lambda}{2}\big)^2 \frac{1}{16}\lan 0 | b(p_4) a(p_3) \tT\Big[-2\del_{(\mu}\bpsi(y) \gamma_{\nu)}\psi(y) \del^{(\mu}\bpsi(y) \gamma^{\nu)}\psi(y) \ \nonumber\\
&& \hspace{0.5cm} \times  \bpsi(x)\gamma_{(\rho}\del_{\lambda)}\psi(x) \del^{(\rho}\bpsi(x) \gamma^{\lambda)}\psi(x) \Big] a^{\dagger}(p_1) b^{\dagger}(p_2)|0 \ran^{(s)}
\ \ = \ \ \mathcal{A}_{\text{a}}^{(t_2)}\ ,
\label{twoderv_t8}
\eea
\bea
\mathcal{A}_{\text{a}}^{(t_9)}&=&\frac{1}{2!} \big(i\frac{\lambda}{2}\big)^2 \frac{1}{16}\lan 0 | b(p_4) a(p_3) \tT\Big[\del_{(\mu}\bpsi(y) \gamma_{\nu)}\psi(y) \del^{(\mu}\bpsi(y) \gamma^{\nu)}\psi(y) \ \nonumber\\
&& \hspace{0.5cm} \times  \del_{(\rho}\bpsi(x) \gamma_{\lambda)}\psi(x) \del^{(\rho}\bpsi(x) \gamma^{\lambda)}\psi(x) \Big] a^{\dagger}(p_1) b^{\dagger}(p_2)|0 \ran^{(t)}
\ \ = \ \ \mathcal{A}_{\text{a}}^{(t_1)}\ .
\label{twoderv_t9}
\eea
Combining the $t$-channel contributions, (\ref{twoderv_t1_final}) - (\ref{twoderv_t9}), gives the total $t$-channel amplitude from figure \ref{bubblediagram_twoderv_t},
\bea
\mathcal{A}_{\text{a}}^{(t)}&=&2\Big(\mathcal{A}_{\text{a}}^{(t_1)}+\mathcal{A}_{\text{a}}^{(t_2)}+\cdots+\mathcal{A}_{\text{a}}^{(t_9)}\Big)\nonumber\\
&=&i\frac{\lambda ^2 m^6}{256 \pi }\bigg[18 \frac{\Lambda ^2}{m^2}+9\frac{\Lambda ^4}{m^4}+2\cosh \theta  \big(31+33\frac{\Lambda^2}{m^2}+192 \log \frac{m}{\Lambda }\big)-128\cosh 2 \theta  \big(1+2 \log \frac{m}{\Lambda }\big)\bigg]\ .\nonumber\\
\label{twoderv_t_amplitude}
\eea
Lastly, the amplitude due to the same quartic interaction will be computed in the $u$-channel case (figure \ref{bubblediagram_twoderv_u}). By the general $u$-channel amplitude (\ref{uchannelbubblegeneral}), the contribution from the first term in (\ref{TmunuPower4}), $\bpsi(y)\gamma_{(\mu}\del_{\nu)}\psi(y)\bpsi(y)\gamma^{(\mu}\del^{\nu)}\psi(y) \times \bpsi(x)\gamma_{(\rho}\del_{\lambda)}\psi(x) \bpsi(x)\gamma^{(\rho}\del^{\lambda)}\psi(x)$, to the amplitude is given by,
\bea
\mathcal{A}_{\text{a}}^{(u_1)}&=&\frac{1}{2!} \big(i\frac{\lambda}{2}\big)^2 \frac{1}{16}\lan 0 | b(p_4) a(p_3) \tT\Big[\bpsi(y)\gamma_{(\mu}\del_{\nu)}\psi(y)\bpsi(y)\gamma^{(\mu}\del^{\nu)}\psi(y) \ \nonumber\\
&& \hspace{4.1cm} \times  \bpsi(x)\gamma_{(\rho}\del_{\lambda)}\psi(x) \bpsi(x)\gamma^{(\rho}\del^{\lambda)}\psi(x)\Big] a^{\dagger}(p_1) b^{\dagger}(p_2)|0 \ran^{(u)}\nonumber\\
&=&-\frac{\lambda^2}{128}\gamma_{ef(\mu}\gamma^{(\mu}_{gh}\gamma_{ab(\rho}\gamma^{(\rho}_{cd}\ \times\nonumber\\
&\Big[&u_d(p_1)v_b(p_4)\bu_e(p_3)\bv_g(p_2)i p_{4\lambda)}(-i)p_1^{\lambda)}
-u_d(p_1)v_b(p_4)\bu_g(p_3)\bv_e(p_2)i p_{4\lambda)}(-i)p_1^{\lambda)}\nonumber\\
&-&u_b(p_1)v_d(p_4)\bu_e(p_3)\bv_g(p_2)(-i)p_{1\lambda)}i p_4^{\lambda)}
+u_b(p_1)v_d(p_4)\bu_g(p_3)\bv_e(p_2)(-i)p_{1\lambda)}i p_4^{\lambda)}\Big]\nonumber\\
&\times&\int\frac{d^2q}{(2\pi)^2}\Big[G_{hc}(\zeta-q)G_{fa}(q)(-i)q_{\nu)}(-i)(\zeta-q)^{\nu)}
-G_{ha}(\zeta-q)G_{fc}(q)(-i)q_{\nu)}(-i)(\zeta-q)^{\nu)}\Big]\ ,\nonumber\\
\label{twoderv_u1}
\eea
where $\zeta^{\mu}=(p_1-p_4)^{\mu}=(p_1-p_2)^{\mu}$. The above expression, (\ref{twoderv_u1}), involves the one-loop integrals $(I_{\mu})_{abcd}^{(u)}(\zeta)$ and  $(I_{\mu\nu})_{abcd}^{(u)}(\zeta)$ whose values are given by (\ref{I1_abcd_u}) and (\ref{I2_abcd_u}),  respectively. Plugging the values of the integrals and Dirac spinors into (\ref{twoderv_u1}) one finds the $u$-channel contribution from the first term in (\ref{TmunuPower4}) to be,
\bea
\mathcal{A}_{\text{a}}^{(u_1)}&=&-i\frac{\lambda ^2 m^6}{128 \pi}\cosh^2\frac{\theta}{2}\big(1-4\cosh\theta\big)\bigg[3 \frac{\Lambda ^2}{m^2}+10\log\frac{m}{\Lambda}+3\theta\coth \frac{\theta}{2}+4\theta \sinh \theta\nonumber\\
&-&\cosh \theta  \big(3-8\log \frac{m}{\Lambda }\big)\bigg]\ .
\label{twoderv_u1_final}
\eea
The contributions to the u-channel amplitude from the eight additional terms in (\ref{TmunuPower4}) can be obtained in a similar manner. Their values are given by, 
\bea
\mathcal{A}_{\text{a}}^{(u_2)}&=&\frac{1}{2!}\big(i\frac{\lambda}{2}\big)^2 \frac{1}{16}\lan 0 | b(p_4) a(p_3) \tT\Big[-2\bpsi(y)\gamma_{(\mu}\del_{\nu)}\psi(y)\bpsi(y)\gamma^{(\mu}\del^{\nu)}\psi(y) \ \nonumber\\
&& \hspace{4.1cm} \times  \bpsi(x)\gamma_{(\rho}\del_{\lambda)}\psi(x) \del^{(\rho}\bpsi(x) \gamma^{\lambda)}\psi(x) \Big] a^{\dagger}(p_1) b^{\dagger}(p_2)|0 \ran^{(u)}\nonumber\\
&=&i\frac{\lambda ^2 m^6}{192 \pi}\cosh^2\frac{\theta}{2}\bigg[-18-24 \frac{\Lambda ^2}{m^2}+9\theta\coth \frac{\theta}{2}+24\theta \sinh 2\theta+\cosh \theta  \big(25+42\frac{\Lambda^2}{m^2}+24\log \frac{m}{\Lambda }\big)\nonumber\\
&+&\cosh 2\theta  \big(-25+48\log \frac{m}{\Lambda }\big)\bigg]\ ,
\label{twoderv_u2}
\eea
\bea
\mathcal{A}_{\text{a}}^{(u_3)}&=&\frac{1}{2!}\big(i\frac{\lambda}{2}\big)^2 \frac{1}{16}\lan 0 | b(p_4) a(p_3) \tT\Big[\bpsi(y)\gamma_{(\mu}\del_{\nu)}\psi(y)\bpsi(y)\gamma^{(\mu}\del^{\nu)}\psi(y) \ \nonumber\\
&& \hspace{4.1cm} \times  \del_{(\rho}\bpsi(x) \gamma_{\lambda)}\psi(x) \del^{(\rho}\bpsi(x) \gamma^{\lambda)}\psi(x) \Big] a^{\dagger}(p_1) b^{\dagger}(p_2)|0 \ran^{(u)}\nonumber\\
&=&i\frac{\lambda ^2 m^6}{19200 \pi}\cosh^2\frac{\theta}{2}\bigg[-1311+30 \frac{\Lambda ^2}{m^2}-675\frac{\Lambda^4}{m^4}+8700\log\frac{m}{\Lambda}+150\theta\coth\frac{\theta}{2}\big(1-4\cosh\theta\big)^2\nonumber\\
&+&2\cosh \theta  \big(29+660\frac{\Lambda^2}{m^2}+1200 \log \frac{m}{\Lambda }\big)
+\cosh 2 \theta  \big(-1447+2400\log \frac{m}{\Lambda }\big)\bigg]\ ,
\label{twoderv_u3}
\eea
\bea
\mathcal{A}_{\text{a}}^{(u_4)}&=&\frac{1}{2!}\big(i\frac{\lambda}{2}\big)^2 \frac{1}{16}\lan 0 | b(p_4) a(p_3) \tT\Big[-2\bpsi(y)\gamma_{(\mu}\del_{\nu)}\psi(y) \del^{(\mu}\bpsi(y) \gamma^{\nu)}\psi(y) \ \nonumber\\
&& \hspace{4.1cm} \times  \bpsi(x)\gamma_{(\rho}\del_{\lambda)}\psi(x) \bpsi(x)\gamma^{(\rho}\del^{\lambda)}\psi(x) \Big] a^{\dagger}(p_1) b^{\dagger}(p_2)|0 \ran^{(u)}\nonumber\\
&=&i\frac{\lambda ^2 m^6}{64 \pi }\cosh \frac{\theta }{2} (-1+4 \cosh \theta )\bigg[\cosh \frac{3 \theta }{2}\big(-1+4\log \frac{m}{\Lambda}\big) +\cosh \frac{ \theta }{2}\big(1+\theta\coth\frac{\theta}{2}+4\theta\sinh\theta\big)\bigg]\ ,\nonumber\\
\label{twoderv_u4}
\eea
\bea
\mathcal{A}_{\text{a}}^{(u_5)}&=&\frac{1}{2!}\big(i\frac{\lambda}{2}\big)^2 \frac{1}{16}\lan 0 | b(p_4) a(p_3) \tT\Big[4\bpsi(y)\gamma_{(\mu}\del_{\nu)}\psi(y) \del^{(\mu}\bpsi(y) \gamma^{\nu)}\psi(y)  \ \nonumber\\
&& \hspace{4.1cm} \times  \bpsi(x)\gamma_{(\rho}\del_{\lambda)}\psi(x) \del^{(\rho}\bpsi(x) \gamma^{\lambda)}\psi(x) \Big] a^{\dagger}(p_1) b^{\dagger}(p_2)|0 \ran^{(u)}\nonumber\\
&=&i\frac{\lambda ^2 m^6}{768\pi }\bigg[-43-12 \frac{\Lambda ^2}{m^2}+72\log\frac{m}{\Lambda}+24\theta\cosh^2\frac{\theta}{2}\coth\frac{\theta}{2}\big(3-4\cosh\theta\big)^2\nonumber\\
&+&4 \cosh \theta  \big(7+3\frac{\Lambda^2}{m^2}-6 \log \frac{m}{\Lambda }\big)+\cosh 2\theta  \big(31
+24 \frac{\Lambda^2}{m^2}\big)-8\cosh 3\theta  \big(5-12 \log\frac{m}{\Lambda}\big)\bigg]\ ,\nonumber\\
\label{twoderv_u5}
\eea
\bea
\mathcal{A}_{\text{a}}^{(u_6)}&=&\frac{1}{2!}\big(i\frac{\lambda}{2}\big)^2 \frac{1}{16}\lan 0 | b(p_4) a(p_3) \tT\Big[-2\bpsi(y)\gamma_{(\mu}\del_{\nu)}\psi(y) \del^{(\mu}\bpsi(y) \gamma^{\nu)}\psi(y) \ \nonumber\\
&& \hspace{0.5cm} \times  \del_{(\rho}\bpsi(x) \gamma_{\lambda)}\psi(x) \del^{(\rho}\bpsi(x) \gamma^{\lambda)}\psi(x) \Big] a^{\dagger}(p_1) b^{\dagger}(p_2)|0 \ran^{(u)}
\ \ = \ \ \mathcal{A}_{\text{a}}^{(u_2)}\ ,
\label{twoderv_u6}
\eea
\bea
\mathcal{A}_{\text{a}}^{(u_7)}&=&\frac{1}{2!} \big(i\frac{\lambda}{2}\big)^2 \frac{1}{16}\lan 0 | b(p_4) a(p_3) \tT\Big[\del_{(\mu}\bpsi(y) \gamma_{\nu)}\psi(y) \del^{(\mu}\bpsi(y) \gamma^{\nu)}\psi(y) \ \nonumber\\
&& \hspace{4.1cm} \times  \bpsi(x)\gamma_{(\rho}\del_{\lambda)}\psi(x) \bpsi(x)\gamma^{(\rho}\del^{\lambda)}\psi(x) \Big] a^{\dagger}(p_1) b^{\dagger}(p_2)|0 \ran^{(u)}\nonumber\\
&=&i\frac{\lambda ^2 m^6}{128\pi }\big(\cosh\frac{\theta}{2}+2\cosh\frac{3\theta}{2}\big)^2\big(\theta\coth\frac{\theta}{2}+2\log\frac{m}{\Lambda}\big)\ ,
\label{twoderv_u7}
\eea
\bea
\mathcal{A}_{\text{a}}^{(u_8)}&=&\frac{1}{2!}\big(i\frac{\lambda}{2}\big)^2 \frac{1}{16}\lan 0 | b(p_4) a(p_3) \tT\Big[-2\del_{(\mu}\bpsi(y) \gamma_{\nu)}\psi(y) \del^{(\mu}\bpsi(y) \gamma^{\nu)}\psi(y) \ \nonumber\\
&& \hspace{0.5cm} \times \bpsi(x)\gamma_{(\rho}\del_{\lambda)}\psi(x) \del^{(\rho}\bpsi(x) \gamma^{\lambda)}\psi(x) \Big] a^{\dagger}(p_1) b^{\dagger}(p_2)|0 \ran^{(u)}
\ \ = \ \ \mathcal{A}_{\text{a}}^{(u_4)}\ ,
\label{twoderv_u8}
\eea
\bea
\mathcal{A}_{\text{a}}^{(u_9)}&=&\frac{1}{2!}\big(i\frac{\lambda}{2}\big)^2 \frac{1}{16}\lan 0 | b(p_4) a(p_3) \tT\Big[\del_{(\mu}\bpsi(y) \gamma_{\nu)}\psi(y) \del^{(\mu}\bpsi(y) \gamma^{\nu)}\psi(y) \ \nonumber\\
&& \hspace{0.5cm} \times  \del_{(\rho}\bpsi(x) \gamma_{\lambda)}\psi(x) \del^{(\rho}\bpsi(x) \gamma^{\lambda)}\psi(x) \Big] a^{\dagger}(p_1) b^{\dagger}(p_2)|0 \ran^{(u)}
\ \ = \ \ \mathcal{A}_{\text{a}}^{(u_1)}\ .
\label{twoderv_u9}
\eea
Combining the $u$-channel contributions, (\ref{twoderv_u1_final}) - (\ref{twoderv_u9}), gives the total $u$-channel amplitude from figure \ref{bubblediagram_twoderv_u},
\bea
\mathcal{A}_{\text{a}}^{(u)}&=&2\Big(\mathcal{A}_{\text{a}}^{(u_1)}+\mathcal{A}_{\text{a}}^{(u_2)}+\cdots+\mathcal{A}_{\text{a}}^{(u_9)}\Big)\nonumber\\
&=&-i\frac{\lambda ^2 m^6}{9600 \pi }\cosh^2\frac{\theta}{2}\bigg[18011+7470 \frac{\Lambda ^2}{m^2}+675\frac{\Lambda ^4}{m^4}+6\cosh\theta\big(-3843-2620\frac{\Lambda^2}{m^2}+3200\log\frac{m}{\Lambda}\big)\nonumber\\
&+&\cosh 2\theta  \big(14647-38400\log\frac{m}{\Lambda}\big)-9600  \big(\theta\coth\frac{\theta}{2}+4 \log \frac{m}{\Lambda }+2\theta\sinh2\theta\big)\bigg]\ .
\label{twoderv_u_amplitude}
\eea
Finally, adding together the contributions from the $s$, $t$ and $u$-channels to the amplitude, (\ref{twoderv_s_amplitude}), (\ref{twoderv_t_amplitude}) and (\ref{twoderv_u_amplitude}), one finds the total amplitude from the bubble diagrams where both vertices contain the quartic interaction $\tilde{X}_{\mu\nu}\tilde{X}^{\mu\nu}$,
\bea
\mathcal{A}_{\text{a}}&=&\mathcal{A}_{\text{a}}^{(s)}+\mathcal{A}_{\text{a}}^{(t)}+\mathcal{A}_{\text{a}}^{(u)}\nonumber\\
&=&-\frac{\lambda^2m^6}{19200\pi\sinh\theta}\bigg[4800\pi\Big(10+2\cosh\theta+\cosh2\theta+2\cosh3\theta+\cosh4\theta\Big)+i\Big(\big(34571\nonumber\\
&+&4860\frac{\Lambda^2}{m^2}\big)\sinh\theta
-\sinh2\theta\big(6659+9360\frac{\Lambda^2}{m^2}+9600\log\frac{m}{\Lambda}\big)+4800\big(8\theta-13\sinh\theta\log\frac{m}{\Lambda}\big)\nonumber\\
&-&\sinh3\theta\big(3163+7860\frac{\Lambda^2}{m^2}-14400\log\frac{m}{\Lambda}\big)
\Big)\bigg]\ .
\label{amplitude_twoderv}
\eea

\subsubsection*{(b) One vertex contains $\tilde{X}_{\rho\lambda}\tilde{X}^{\rho\lambda}$ while the other contains $(\tilde{X}_{\mu}^{\ \mu})^2$ :}
\begin{figure}[h!]
	\centering
	\subfigure[$s$-channel]{
		\includegraphics[width=0.31\columnwidth,height=0.21\columnwidth]{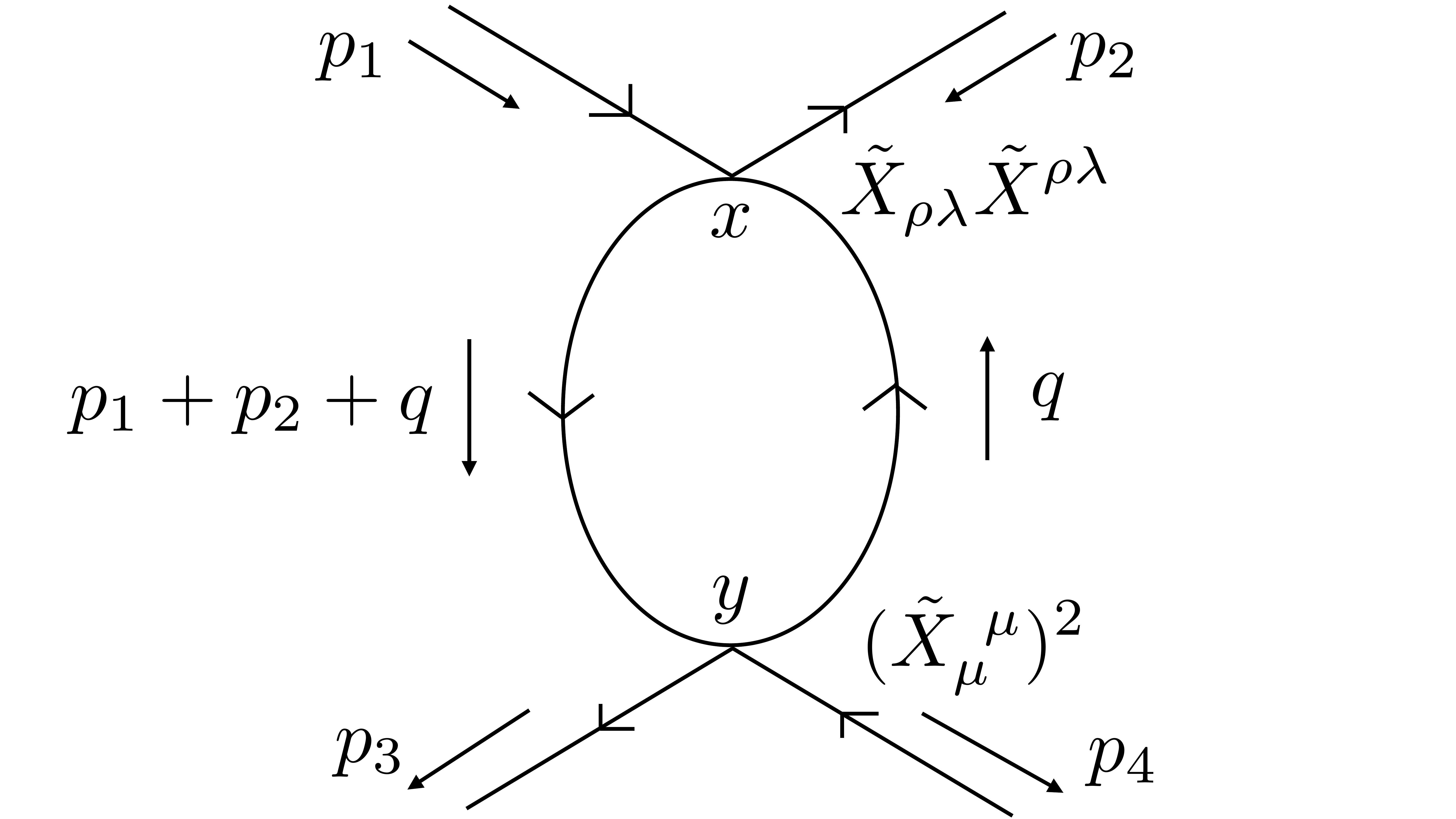}
		\label{bubblediagram_onederv_s} } 
	\subfigure[$t$-channel]{
		\includegraphics[width=0.31\columnwidth,height=0.21\columnwidth]{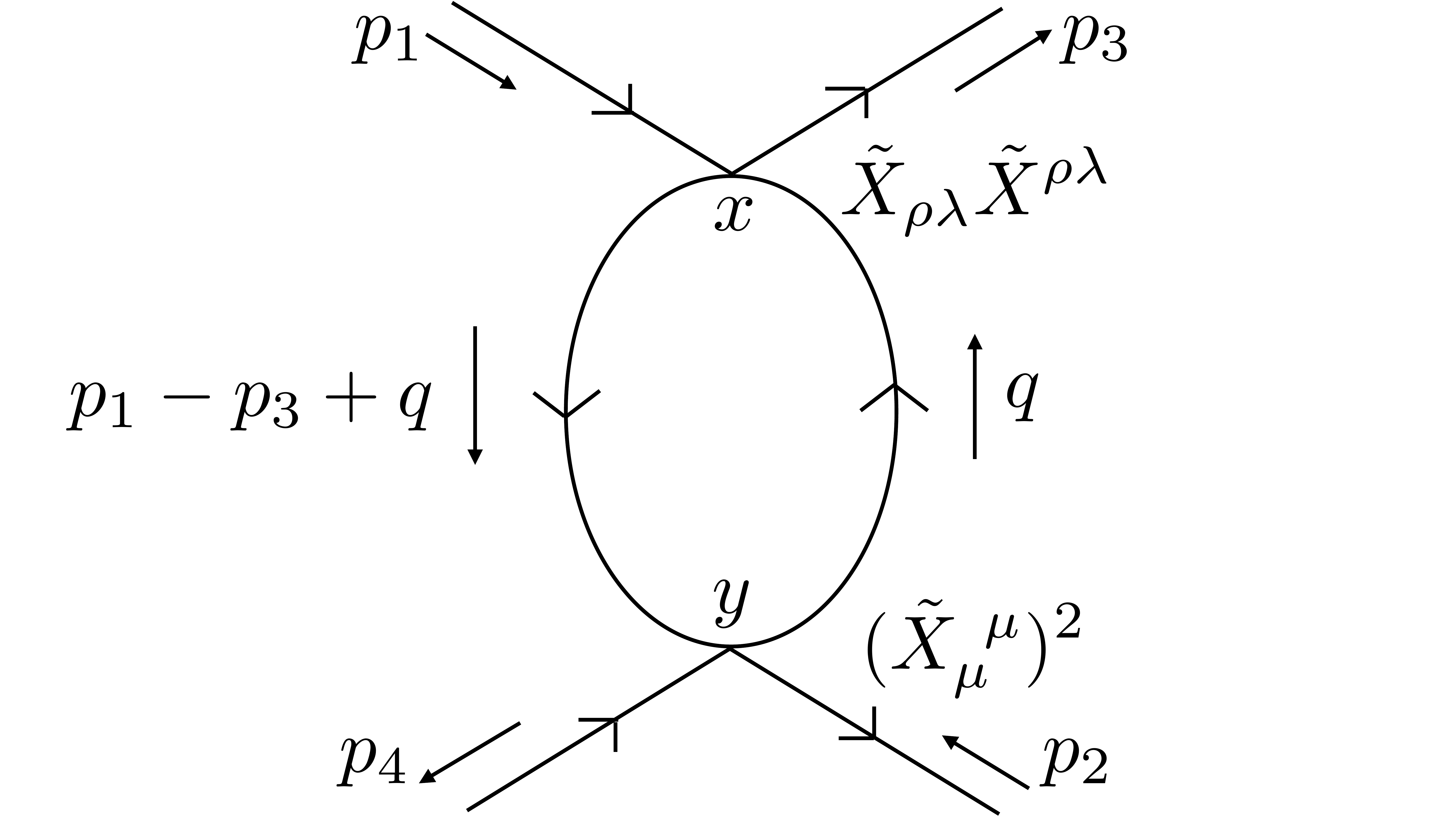}
		\label{bubblediagram_onederv_t} }
	\subfigure[$u$-channel]{
		\includegraphics[width=0.31\columnwidth,height=0.21\columnwidth]{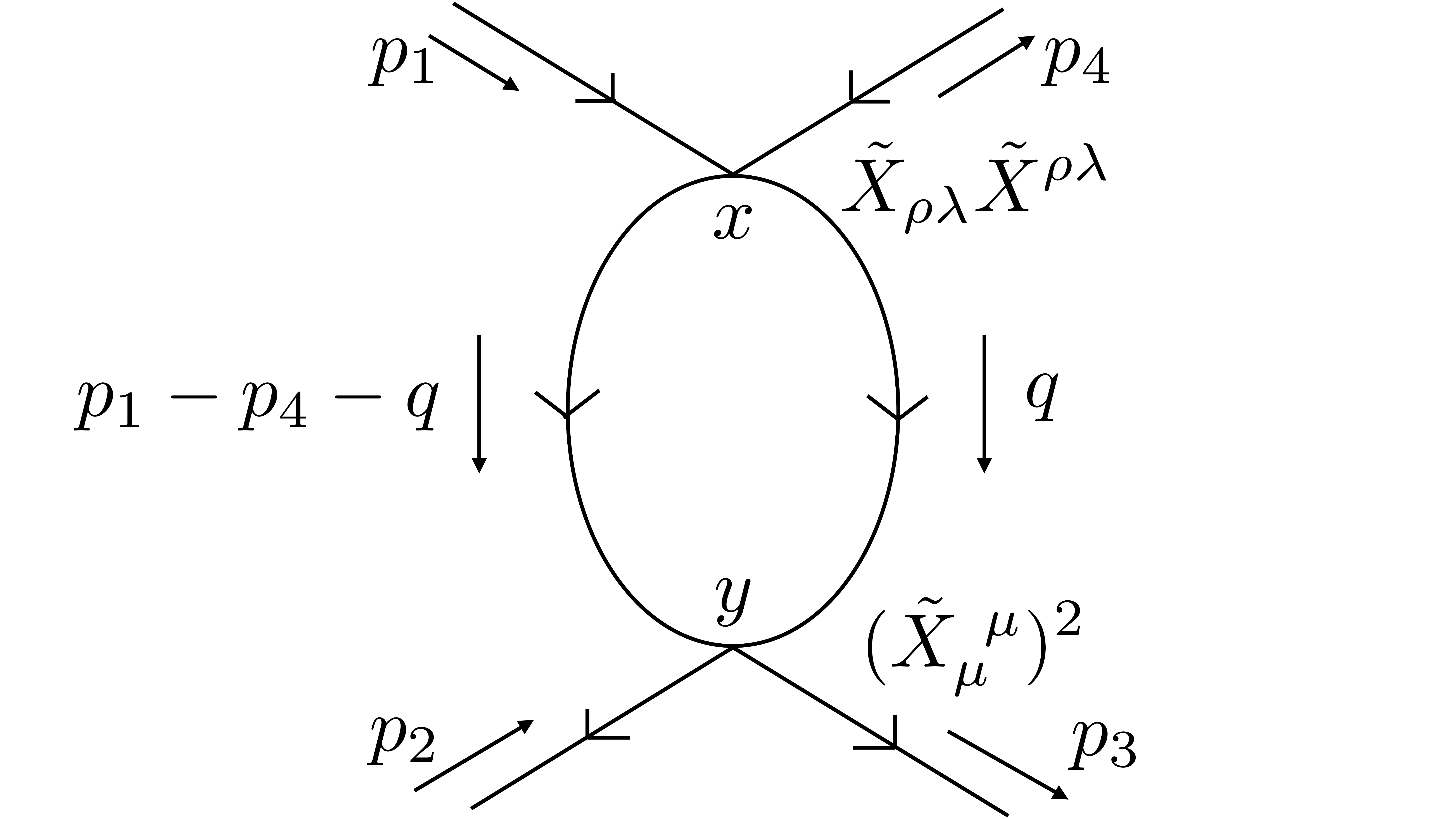}
		\label{bubblediagram_onederv_u} }
	\caption{\small The $s$, $t$ and $u$-channel amplitudes when one interaction vertex is $\tilde{X}_{\rho\lambda}\tilde{X}^{\rho\lambda}$ and the other vertex is $(\tilde{X}_{\mu}^{\ \mu})^2$.}
	\label{bubblediagram_onederv}
\end{figure}

Consider the bubble diagrams shown in figure \ref{bubblediagram_onederv} where one vertex contains the quartic coupling $\tilde{X}_{\rho\lambda}\tilde{X}^{\rho\lambda}$ while the second vertex contains the quartic coupling $(\tilde{X}_{\mu}^{\ \mu})^2$.

The amplitude due to these kinds of bubble diagrams is given by,
\bea
\mathcal{A}_{\text{b}}&=&-2\frac{1}{2!} \big(i\frac{\lambda}{2}\big)^2 \lan 0 | b(p_4) a(p_3) \tT\Big[\big(\tilde{X}_{\mu}^{\ \mu}(y)\big)^2 \tilde{X}_{\rho\lambda}(x)\tilde{X}^{\rho\lambda}(x)\Big] a^{\dagger}(p_1) b^{\dagger}(p_2)|0 \ran\nonumber\\
&=&-2\frac{1}{2!} \big(i\frac{\lambda}{2}\big)^2 \frac{1}{16}\lan 0 | b(p_4) a(p_3) \tT\Big[
\bpsi(y)\gamma^{\mu}\del_{\mu}\psi(y) \ \bpsi(y)\gamma^{\nu}\del_{\nu}\psi(y) \ \ \bpsi(x)\gamma_{(\rho}\del_{\lambda)}\psi(x) \bpsi(x)\gamma^{(\rho}\del^{\lambda)}\psi(x)\nonumber\\
&-&2\bpsi(y)\gamma^{\mu}\del_{\mu}\psi(y) \ \bpsi(y)\gamma^{\nu}\del_{\nu}\psi(y) \ \ \bpsi(x)\gamma_{(\rho}\del_{\lambda)}\psi(x) \del^{(\rho}\bpsi(x) \gamma^{\lambda)}\psi(x)\nonumber\\
&+&\bpsi(y)\gamma^{\mu}\del_{\mu}\psi(y) \ \bpsi(y)\gamma^{\nu}\del_{\nu}\psi(y) \ \
\del_{(\rho}\bpsi(x) \gamma_{\lambda)}\psi(x) \del^{(\rho}\bpsi(x) \gamma^{\lambda)}\psi(x)\nonumber\\
&-&2\bpsi(y)\gamma^{\mu}\del_{\mu}\psi(y) \ \del_{\nu}\bpsi(y) \gamma^{\nu}\psi(y) \ \ \bpsi(x)\gamma_{(\rho}\del_{\lambda)}\psi(x) \bpsi(x)\gamma^{(\rho}\del^{\lambda)}\psi(x)\nonumber\\
&+&4\bpsi(y)\gamma^{\mu}\del_{\mu}\psi(y) \ \del_{\nu}\bpsi(y) \gamma^{\nu}\psi(y) \ \ \bpsi(x)\gamma_{(\rho}\del_{\lambda)}\psi(x) \del^{(\rho}\bpsi(x) \gamma^{\lambda)}\psi(x)\nonumber\\
&-&2\bpsi(y)\gamma^{\mu}\del_{\mu}\psi(y) \ \del_{\nu}\bpsi(y) \gamma^{\nu}\psi(y) \ \ \del_{(\rho}\bpsi(x) \gamma_{\lambda)}\psi(x) \del^{(\rho}\bpsi(x) \gamma^{\lambda)}\psi(x)\nonumber\\
&+&\del_{\mu}\bpsi(y) \gamma^{\mu}\psi(y) \ \del_{\nu}\bpsi(y) \gamma^{\nu}\psi(y) \ \ \bpsi(x)\gamma_{(\rho}\del_{\lambda)}\psi(x) \bpsi(x)\gamma^{(\rho}\del^{\lambda)}\psi(x)\nonumber\\
&-&2\del_{\mu}\bpsi(y) \gamma^{\mu}\psi(y) \ \del_{\nu}\bpsi(y) \gamma^{\nu}\psi(y) \ \ \bpsi(x)\gamma_{(\rho}\del_{\lambda)}\psi(x) \del^{(\rho}\bpsi(x) \gamma^{\lambda)}\psi(x)\nonumber\\
&+&\del_{\mu}\bpsi(y) \gamma^{\mu}\psi(y) \ \del_{\nu}\bpsi(y) \gamma^{\nu}\psi(y) \ \ \del_{(\rho}\bpsi(x) \gamma_{\lambda)}\psi(x) \del^{(\rho}\bpsi(x) \gamma^{\lambda)}\psi(x) \ 
\Big] a^{\dagger}(p_1) b^{\dagger}(p_2)|0 \ran\ .
\label{TmumusqTmunusq}
\eea
The multiplicative factor of $2$ arises from the two identical types of cross-terms: 
\vspace{2mm}
\\
$\big(\tilde{X}_{\mu}^{\ \mu}(y)\big)^2 \tilde{X}_{\rho\lambda}(x)\tilde{X}^{\rho\lambda}(x)$ \hspace{1mm} and \hspace{1mm} $\tilde{X}_{\mu\nu}(y)\tilde{X}^{\mu\nu}(y)\big(\tilde{X}_{\rho}^{\ \rho}(x)\big)^2 $.
\vspace{3mm}
\\
The $s$-channel contribution to the amplitude from figure \ref{bubblediagram_onederv_s} can be computed by evaluating the contributions from each of the above terms in (\ref{TmumusqTmunusq}) and adding them together. By the general s-channel amplitude (\ref{schannelbubblegeneral}), the contributions from all nine terms in (\ref{TmumusqTmunusq}) are,
\bea
\mathcal{A}_{\text{b}}^{(s_1)}
&=&-2\frac{1}{2!} \big(i\frac{\lambda}{2}\big)^2 \frac{1}{16} \lan 0 | b(p_4) a(p_3) \tT\Big[ \bpsi(y)\gamma^{\mu}\del_{\mu}\psi(y)  \bpsi(y)\gamma^{\nu}\del_{\nu}\psi(y) \ \nonumber\\
&& \hspace{4.5cm} \times  \bpsi(x)\gamma_{(\rho}\del_{\lambda)}\psi(x) \bpsi(x)\gamma^{(\rho}\del^{\lambda)}\psi(x)\Big] a^{\dagger}(p_1) b^{\dagger}(p_2)|0 \ran^{(s)}\nonumber\\
&=&\frac{\lambda^2 m^6}{192\pi\sinh\theta}\bigg[\frac{3\pi}{2}\big(29+10\cosh\theta+7\cosh2\theta+2\cosh 3\theta\big)
+i\Big(\sinh\theta\big(36-6\frac{\Lambda^2}{m^2}-24\log\frac{m}{\Lambda}\big)\nonumber\\
&+&\sinh\theta\cosh 2\theta \big(1+12\log\frac{m}{\Lambda}\big)+5\sinh2\theta(1+3\log\frac{m}{\Lambda})+\frac{3}{2}\theta\big(29+10\cosh\theta+7\cosh2\theta\nonumber\\
&+&2\cosh 3\theta\big)\Big)\bigg]\ ,
\label{onederv_s1_final}
\eea
\bea
\mathcal{A}_{\text{b}}^{(s_2)}&=&-2\frac{1}{2!} \big(i\frac{\lambda}{2}\big)^2 \frac{1}{16}\lan 0 | b(p_4) a(p_3) \tT\Big[-2\bpsi(y)\gamma^{\mu}\del_{\mu}\psi(y) \ \bpsi(y)\gamma^{\nu}\del_{\nu}\psi(y) \ \nonumber\\
&& \hspace{4.5cm} \times  \bpsi(x)\gamma_{(\rho}\del_{\lambda)}\psi(x) \del^{(\rho}\bpsi(x) \gamma^{\lambda)}\psi(x)\Big] a^{\dagger}(p_1) b^{\dagger}(p_2)|0 \ran^{(s)}\nonumber\\
&=&\frac{\lambda^2 m^6}{192\pi\sinh\theta}\bigg[3\pi\big(7+2\cosh\theta+5\cosh2\theta+2\cosh 3\theta\big)
+i\Big(13\sinh\theta-\sinh2\theta\big(7+\frac{3}{2}\frac{\Lambda^2}{m^2}\nonumber\\
&-&15\log\frac{m}{\Lambda}\big)
+24\sinh2\theta\cosh\theta \log\frac{m}{\Lambda}+3\theta\big(7+2\cosh\theta+5\cosh2\theta+2\cosh 3\theta\big)\Big)\bigg]\ , \nonumber\\
\label{onederv_s2_final}
\eea
\bea
\mathcal{A}_{\text{b}}^{(s_3)}&=&-2\frac{1}{2!} \big(i\frac{\lambda}{2}\big)^2 \frac{1}{16}\lan 0 | b(p_4) a(p_3) \tT\Big[ \bpsi(y)\gamma^{\mu}\del_{\mu}\psi(y) \ \bpsi(y)\gamma^{\nu}\del_{\nu}\psi(y) \ \nonumber\\
&& \hspace{4.5cm} \times
\del_{(\rho}\bpsi(x) \gamma_{\lambda)}\psi(x) \del^{(\rho}\bpsi(x) \gamma^{\lambda)}\psi(x)\Big] a^{\dagger}(p_1) b^{\dagger}(p_2)|0 \ran^{(s)}\nonumber\\
&=&\frac{\lambda^2 m^6}{128\pi\sinh\theta}\bigg[\pi\big(29+10\cosh\theta+7\cosh2\theta+2\cosh 3\theta\big)
+i\Big(8\sinh\theta\big(1-4\log\frac{m}{\Lambda}\nonumber\\
&+&\cosh2\theta\log\frac{m}{\Lambda}\big)+\sinh2\theta\big(-6+3\frac{\Lambda^2}{m^2}+4\log\frac{m}{\Lambda}\big)
+\theta\big(29+10\cosh\theta+7\cosh2\theta\nonumber\\
&+&2\cosh 3\theta\big)\Big)\bigg]\ ,
\label{onederv_s3_final}
\eea
\bea
\mathcal{A}_{\text{b}}^{(s_4)}&=&-2\frac{1}{2!} \big(i\frac{\lambda}{2}\big)^2 \frac{1}{16}\lan 0 | b(p_4) a(p_3) \tT\Big[ -2\bpsi(y)\gamma^{\mu}\del_{\mu}\psi(y) \ \del_{\nu}\bpsi(y) \gamma^{\nu}\psi(y)\ \nonumber\\
&& \hspace{4.5cm} \times \bpsi(x)\gamma_{(\rho}\del_{\lambda)}\psi(x) \bpsi(x)\gamma^{(\rho}\del^{\lambda)}\psi(x)\Big] a^{\dagger}(p_1) b^{\dagger}(p_2)|0 \ran^{(s)}\nonumber\\
&=&\frac{\lambda^2 m^6}{384\pi\sinh\theta}\bigg[6\pi\big(29+10\cosh\theta+7\cosh2\theta+2\cosh 3\theta\big)
+i\Big(\sinh\theta\big(115-24\frac{\Lambda^2}{m^2}\nonumber\eea\bea
&-&144\log\frac{m}{\Lambda}+\cosh3\theta\big)
+\sinh\theta\cosh2\theta\big(13-6\frac{\Lambda^2}{m^2}+48\log\frac{m}{\Lambda}\big)+\sinh2\theta\big(\frac{13}{2}+42\log\frac{m}{\Lambda}\big)\nonumber\\
&+&6\theta\big(29+10\cosh\theta+7\cosh2\theta+2\cosh 3\theta\big)\Big)\bigg]\ ,
\label{onederv_s4_final}
\eea
\bea
\mathcal{A}_{\text{b}}^{(s_5)}&=&-2\frac{1}{2!} \big(i\frac{\lambda}{2}\big)^2 \frac{1}{16}\lan 0 | b(p_4) a(p_3) \tT\Big[4\bpsi(y)\gamma^{\mu}\del_{\mu}\psi(y) \ \del_{\nu}\bpsi(y) \gamma^{\nu}\psi(y)\  \nonumber\\
&& \hspace{4.5cm} \times \bpsi(x)\gamma_{(\rho}\del_{\lambda)}\psi(x) \del^{(\rho}\bpsi(x) \gamma^{\lambda)}\psi(x)\Big] a^{\dagger}(p_1) b^{\dagger}(p_2)|0 \ran^{(s)}\nonumber\\
&=&\frac{\lambda^2 m^6}{19200\pi\sinh\theta}\bigg[600\pi\big(7+2\cosh\theta+5\cosh2\theta+2\cosh 3\theta\big)
+i\Big(12\sinh\theta\big(177+30\frac{\Lambda^2}{m^2}\nonumber\\
&+&400\log\frac{m}{\Lambda}+\frac{151}{12}\cosh3\theta\big)
+48\sinh\theta\cosh2\theta\big(8-5\frac{\Lambda^2}{m^2}+100\log\frac{m}{\Lambda}\big)-\sinh2\theta\big(\frac{3343}{2}\nonumber\\
&+&540\frac{\Lambda^2}{m^2}+225\frac{\Lambda^4}{m^4}
-3000\log\frac{m}{\Lambda}\big)
+600 \ \theta\big(7+2\cosh\theta+5\cosh2\theta+2\cosh 3\theta\big)\Big)\bigg]\ ,\nonumber\\
\label{onederv_s5_final}
\eea
\bea
\mathcal{A}_{\text{b}}^{(s_6)}&=&-2\frac{1}{2!} \big(i\frac{\lambda}{2}\big)^2 \frac{1}{16}\lan 0 | b(p_4) a(p_3) \tT\Big[-2\bpsi(y)\gamma^{\mu}\del_{\mu}\psi(y) \ \del_{\nu}\bpsi(y) \gamma^{\nu}\psi(y) \ \nonumber\\
&& \hspace{0.1cm} \times  \del_{(\rho}\bpsi(x) \gamma_{\lambda)}\psi(x) \del^{(\rho}\bpsi(x) \gamma^{\lambda)}\psi(x) \Big] a^{\dagger}(p_1) b^{\dagger}(p_2)|0 \ran^{(s)}
\ \ = \ \ \mathcal{A}_{\text{b}}^{(s_4)}\ ,
\label{onederv_s6_final}
\eea
\bea
\mathcal{A}_{\text{b}}^{(s_7)}&=&-2\frac{1}{2!} \big(i\frac{\lambda}{2}\big)^2 \frac{1}{16}\lan 0 | b(p_4) a(p_3) \tT\Big[\del_{\mu}\bpsi(y) \gamma^{\mu}\psi(y) \ \del_{\nu}\bpsi(y) \gamma^{\nu}\psi(y) \ \nonumber\\
&& \hspace{0.1cm} \times \bpsi(x)\gamma_{(\rho}\del_{\lambda)}\psi(x) \bpsi(x)\gamma^{(\rho}\del^{\lambda)}\psi(x) \Big] a^{\dagger}(p_1) b^{\dagger}(p_2)|0 \ran^{(s)}
\ \ = \ \ \mathcal{A}_{\text{b}}^{(s_3)}\ ,
\label{onederv_s7_final}
\eea
\bea
\mathcal{A}_{\text{b}}^{(s_8)}&=&-2\frac{1}{2!} \big(i\frac{\lambda}{2}\big)^2 \frac{1}{16} \lan 0 | b(p_4) a(p_3) \tT\Big[-2\del_{\mu}\bpsi(y) \gamma^{\mu}\psi(y) \ \del_{\nu}\bpsi(y) \gamma^{\nu}\psi(y) \ \nonumber\\
&& \hspace{0.1cm} \times  \bpsi(x)\gamma_{(\rho}\del_{\lambda)}\psi(x) \del^{(\rho}\bpsi(x) \gamma^{\lambda)}\psi(x)\Big] a^{\dagger}(p_1) b^{\dagger}(p_2)|0 \ran^{(s)}
\ \ = \ \ \mathcal{A}_{\text{b}}^{(s_2)}\ ,
\label{onederv_s8_final}
\eea
\bea
\mathcal{A}_{\text{b}}^{(s_9)}&=&-2\frac{1}{2!} \big(i\frac{\lambda}{2}\big)^2 \frac{1}{16}\lan 0 | b(p_4) a(p_3) \tT\Big[\del_{\mu}\bpsi(y) \gamma^{\mu}\psi(y) \ \del_{\nu}\bpsi(y) \gamma^{\nu}\psi(y) \ \nonumber\\
&& \hspace{0.1cm} \times  \del_{(\rho}\bpsi(x) \gamma_{\lambda)}\psi(x) \del^{(\rho}\bpsi(x) \gamma^{\lambda)}\psi(x)\Big] a^{\dagger}(p_1) b^{\dagger}(p_2)|0 \ran^{(s)}
\ \ = \ \ \mathcal{A}_{\text{b}}^{(s_1)}\ ,
\label{onederv_s9_final}
\eea
where the computational method that was employed was exactly the same as when $\mathcal{A}_{\text{a}}^{(s_1)}$ was computed in (\ref{twoderv_s1}).
\vspace{2mm}
\\
Combining the $s$-channel contributions to the amplitude, (\ref{onederv_s1_final}) - (\ref{onederv_s9_final}), gives the total $s$-channel contribution to the amplitude from the diagrams where one vertex contains the quartic coupling $\tilde{X}_{\rho\lambda}\tilde{X}^{\rho\lambda}$ while the other vertex contains the quartic coupling $(\tilde{X}_{\mu}^{\ \mu})^2$,
\bea
\mathcal{A}_{\text{b}}^{(s)}&=& 2\Big(\mathcal{A}_{\text{b}}^{(s_1)}+\mathcal{A}_{\text{b}}^{(s_2)}+\cdots+\mathcal{A}_{\text{b}}^{(s_9)}\Big)\nonumber\\
&=&\frac{\lambda^2m^6}{19200\pi\sinh\theta}\bigg[9600\pi\big(9+3\cosh\theta+3\cosh2\theta+\cosh 3\theta\big)+i\Big(9600 \ \theta\big(9+3\cosh\theta\nonumber\\
&+&3\cosh2\theta+\cosh3\theta\big)+12\sinh\theta\big(4147-470\frac{\Lambda^2}{m^2}-4800\log\frac{m}{\Lambda}\big)
-2\sinh2\theta\big(3347\nonumber\\
&-&60\frac{\Lambda^2}{m^2}+225\frac{\Lambda^4}{m^4}
-14400\log\frac{m}{\Lambda}\big)
+12\sinh3\theta\big(157-70\frac{\Lambda^2}{m^2}+1600\log\frac{m}{\Lambda}\big)+251\sinh4\theta\Big)\bigg]\ .\nonumber\\
\label{onederv_s_amplitude}
\eea
Just as in the computation of the first type of bubble diagram, (\ref{twoderv_s_amplitude}), a factor of $2$ arises in the first line of the above expression, (\ref{onederv_s_amplitude}), because the same contribution would be found if the vertices $x$ and $y$ were exchanged. A similar factor would appear while evaluating the $t$ and $u$-channel amplitudes as well.
\vspace{2mm}
\\
The contribution to the amplitude from the $t$-channel diagram, figure \ref{bubblediagram_onederv_t}, can be computed using the general $t$-channel amplitude, (\ref{tchannelbubblegeneral}). The nine terms in (\ref{TmumusqTmunusq}) contribute to the amplitude as,
\bea
\mathcal{A}_{\text{b}}^{(t_1)}
&=&-2\frac{1}{2!} \big(i\frac{\lambda}{2}\big)^2 \frac{1}{16} \lan 0 | b(p_4) a(p_3) \tT\Big[ \bpsi(y)\gamma^{\mu}\del_{\mu}\psi(y)  \bpsi(y)\gamma^{\nu}\del_{\nu}\psi(y) \ \nonumber\\
&& \hspace{4.5cm} \times  \bpsi(x)\gamma_{(\rho}\del_{\lambda)}\psi(x) \bpsi(x)\gamma^{(\rho}\del^{\lambda)}\psi(x)\Big] a^{\dagger}(p_1) b^{\dagger}(p_2)|0 \ran^{(t)}\nonumber\\
&=&i\frac{\lambda^2m^6}{64\pi}\Big[1+\frac{\Lambda^2}{m^2}+6\log\frac{m}{\Lambda}-\cosh\theta\big(2+\frac{\Lambda^2}{m^2}+8\log\frac{m}{\Lambda}\big)\Big]\ ,
\label{onederv_t1_final}
\eea
\bea
\mathcal{A}_{\text{b}}^{(t_2)}&=&-2\frac{1}{2!} \big(i\frac{\lambda}{2}\big)^2 \frac{1}{16}\lan 0 | b(p_4) a(p_3) \tT\Big[-2\bpsi(y)\gamma^{\mu}\del_{\mu}\psi(y) \ \bpsi(y)\gamma^{\nu}\del_{\nu}\psi(y) \ \nonumber\\
&& \hspace{4.5cm} \times  \bpsi(x)\gamma_{(\rho}\del_{\lambda)}\psi(x) \del^{(\rho}\bpsi(x) \gamma^{\lambda)}\psi(x)\Big] a^{\dagger}(p_1) b^{\dagger}(p_2)|0 \ran^{(t)}\nonumber\\
&=&-i\frac{\lambda^2 m^6}{64\pi}(2+\cosh\theta)\Big[1+\frac{\Lambda^2}{m^2}+6\log\frac{m}{\Lambda}\Big]\ ,
\label{onederv_t2_final}
\eea
\bea
\mathcal{A}_{\text{b}}^{(t_3)}&=&-2\frac{1}{2!} \big(i\frac{\lambda}{2}\big)^2 \frac{1}{16}\lan 0 | b(p_4) a(p_3) \tT\Big[ \bpsi(y)\gamma^{\mu}\del_{\mu}\psi(y) \ \bpsi(y)\gamma^{\nu}\del_{\nu}\psi(y) \ \nonumber\\
&& \hspace{4.5cm} \times
\del_{(\rho}\bpsi(x) \gamma_{\lambda)}\psi(x) \del^{(\rho}\bpsi(x) \gamma^{\lambda)}\psi(x)\Big] a^{\dagger}(p_1) b^{\dagger}(p_2)|0 \ran^{(t)}\nonumber\\
&=&i\frac{\lambda^2m^6}{64\pi}\Big[1+\frac{\Lambda^2}{m^2}+6\log\frac{m}{\Lambda}-\cosh\theta\big(1+2\frac{\Lambda^2}{m^2}+10\log\frac{m}{\Lambda}\big)\Big]\ ,
\label{onederv_t3_final}
\eea
\bea
\mathcal{A}_{\text{b}}^{(t_4)}&=&-2\frac{1}{2!} \big(i\frac{\lambda}{2}\big)^2 \frac{1}{16}\lan 0 | b(p_4) a(p_3) \tT\Big[ -2\bpsi(y)\gamma^{\mu}\del_{\mu}\psi(y) \ \del_{\nu}\bpsi(y) \gamma^{\nu}\psi(y)\ \nonumber\\
&& \hspace{4.5cm} \times \bpsi(x)\gamma_{(\rho}\del_{\lambda)}\psi(x) \bpsi(x)\gamma^{(\rho}\del^{\lambda)}\psi(x)\Big] a^{\dagger}(p_1) b^{\dagger}(p_2)|0 \ran^{(t)}\nonumber\\
&=&-i\frac{\lambda^2 m^6}{64\pi}(-2+3\cosh\theta)\Big[1+\frac{\Lambda^2}{m^2}+6\log\frac{m}{\Lambda}\Big]\ ,
\label{onederv_t4_final}
\eea
\bea
\mathcal{A}_{\text{b}}^{(t_5)}&=&-2\frac{1}{2!} \big(i\frac{\lambda}{2}\big)^2 \frac{1}{16}\lan 0 | b(p_4) a(p_3) \tT\Big[4\bpsi(y)\gamma^{\mu}\del_{\mu}\psi(y) \ \del_{\nu}\bpsi(y) \gamma^{\nu}\psi(y)\  \nonumber\\
&& \hspace{4.5cm} \times \bpsi(x)\gamma_{(\rho}\del_{\lambda)}\psi(x) \del^{(\rho}\bpsi(x) \gamma^{\lambda)}\psi(x)\Big] a^{\dagger}(p_1) b^{\dagger}(p_2)|0 \ran^{(t)}\nonumber\\
&=&-i\frac{\lambda^2m^6}{128\pi}\Big[8+14\frac{\Lambda^2}{m^2}+3\frac{\Lambda^4}{m^4}+48\log\frac{m}{\Lambda}+2\cosh\theta\big(1+3\frac{\Lambda^2}{m^2}+12\log\frac{m}{\Lambda}\big)\Big]\ ,
\label{onederv_t5_final}
\eea
\bea
\mathcal{A}_{\text{b}}^{(t_6)}&=&-2\frac{1}{2!} \big(i\frac{\lambda}{2}\big)^2 \frac{1}{16}\lan 0 | b(p_4) a(p_3) \tT\Big[-2\bpsi(y)\gamma^{\mu}\del_{\mu}\psi(y) \ \del_{\nu}\bpsi(y) \gamma^{\nu}\psi(y) \ \nonumber\\
&& \hspace{0.1cm} \times  \del_{(\rho}\bpsi(x) \gamma_{\lambda)}\psi(x) \del^{(\rho}\bpsi(x) \gamma^{\lambda)}\psi(x) \Big] a^{\dagger}(p_1) b^{\dagger}(p_2)|0 \ran^{(t)}
\ \ = \ \ \mathcal{A}_{\text{b}}^{(t_4)}\ ,
\label{onederv_t6_final}
\eea
\bea
\mathcal{A}_{\text{b}}^{(t_7)}&=&-2\frac{1}{2!} \big(i\frac{\lambda}{2}\big)^2 \frac{1}{16}\lan 0 | b(p_4) a(p_3) \tT\Big[\del_{\mu}\bpsi(y) \gamma^{\mu}\psi(y) \ \del_{\nu}\bpsi(y) \gamma^{\nu}\psi(y) \ \nonumber\\
&& \hspace{0.1cm} \times \bpsi(x)\gamma_{(\rho}\del_{\lambda)}\psi(x) \bpsi(x)\gamma^{(\rho}\del^{\lambda)}\psi(x) \Big] a^{\dagger}(p_1) b^{\dagger}(p_2)|0 \ran^{(s)}
\ \ = \ \ \mathcal{A}_{\text{b}}^{(t_3)}\ ,
\label{onederv_t7_final}
\eea
\bea
\mathcal{A}_{\text{b}}^{(t_8)}&=&-2\frac{1}{2!} \big(i\frac{\lambda}{2}\big)^2 \frac{1}{16} \lan 0 | b(p_4) a(p_3) \tT\Big[-2\del_{\mu}\bpsi(y) \gamma^{\mu}\psi(y) \ \del_{\nu}\bpsi(y) \gamma^{\nu}\psi(y) \ \nonumber\\
&& \hspace{0.1cm} \times  \bpsi(x)\gamma_{(\rho}\del_{\lambda)}\psi(x) \del^{(\rho}\bpsi(x) \gamma^{\lambda)}\psi(x)\Big] a^{\dagger}(p_1) b^{\dagger}(p_2)|0 \ran^{(s)}
\ \ = \ \ \mathcal{A}_{\text{b}}^{(t_2)}\ ,
\label{onederv_t8_final}
\eea
\bea
\mathcal{A}_{\text{b}}^{(t_9)}&=&-2\frac{1}{2!} \big(i\frac{\lambda}{2}\big)^2 \frac{1}{16}\lan 0 | b(p_4) a(p_3) \tT\Big[\del_{\mu}\bpsi(y) \gamma^{\mu}\psi(y) \ \del_{\nu}\bpsi(y) \gamma^{\nu}\psi(y) \ \nonumber\\
&& \hspace{0.1cm} \times  \del_{(\rho}\bpsi(x) \gamma_{\lambda)}\psi(x) \del^{(\rho}\bpsi(x) \gamma^{\lambda)}\psi(x)\Big] a^{\dagger}(p_1) b^{\dagger}(p_2)|0 \ran^{(s)}
\ \ = \ \ \mathcal{A}_{\text{b}}^{(t_1)}\ ,
\label{onederv_t9_final}
\eea
where the computational method that was employed was exactly the same as was used to evaluate  $\mathcal{A}_{\text{a}}^{(t_1)}$ in (\ref{twoderv_t1}).
\vspace{2mm}
\\
Combining the $t$-channel contributions to the amplitude, (\ref{onederv_t1_final}) - (\ref{onederv_t9_final}), gives the total $t$-channel contribution to the amplitude from the diagram where one vertex contains the quartic coupling $\tilde{X}_{\rho\lambda}\tilde{X}^{\rho\lambda}$ while the other vertex contains the quartic coupling $(\tilde{X}_{\mu}^{\ \mu})^2$,
\bea
\mathcal{A}_{\text{b}}^{(t)}&=& 2\Big(\mathcal{A}_{\text{b}}^{(t_1)}+\mathcal{A}_{\text{b}}^{(t_2)}+\cdots+\mathcal{A}_{\text{b}}^{(t_9)}\Big)\nonumber\\
&=&-i\frac{\lambda^2 m^6}{64\pi}\bigg[3\frac{\Lambda^4}{m^4}+6\frac{\Lambda^2}{m^2}+2\cosh\theta\Big(15+17\frac{\Lambda^2}{m^2}+96\log\frac{m}{\Lambda}\Big)\bigg]\ .
\label{onederv_t_amplitude}
\eea
The contribution to the amplitude of the $u$-channel diagram, figure \ref{bubblediagram_onederv_u}, can be computed using the general $u$-channel amplitude, (\ref{uchannelbubblegeneral}). The contributions from the nine terms in (\ref{TmumusqTmunusq}), to the amplitude, are given by,
\bea
\mathcal{A}_{\text{b}}^{(u_1)}
&=&-2\frac{1}{2!} \big(i\frac{\lambda}{2}\big)^2 \frac{1}{16} \lan 0 | b(p_4) a(p_3) \tT\Big[ \bpsi(y)\gamma^{\mu}\del_{\mu}\psi(y)  \bpsi(y)\gamma^{\nu}\del_{\nu}\psi(y) \ \nonumber\\
&& \hspace{4.5cm} \times  \bpsi(x)\gamma_{(\rho}\del_{\lambda)}\psi(x) \bpsi(x)\gamma^{(\rho}\del^{\lambda)}\psi(x)\Big] a^{\dagger}(p_1) b^{\dagger}(p_2)|0 \ran^{(u)}\nonumber\\
&=&i\frac{\lambda^2m^6}{96\pi}\cosh^2\frac{\theta}{2} \ \big(1-4\cosh\theta\big)\Big[-4+3\frac{\Lambda^2}{m^2}+18\log\frac{m}{\Lambda}+\cosh\theta+3\theta\coth\frac{\theta}{2} \ \Big]\ ,\nonumber\\
\label{onederv_u1_final}
\eea
\bea
\mathcal{A}_{\text{b}}^{(u_2)}&=&-2\frac{1}{2!} \big(i\frac{\lambda}{2}\big)^2 \frac{1}{16}\lan 0 | b(p_4) a(p_3) \tT\Big[-2\bpsi(y)\gamma^{\mu}\del_{\mu}\psi(y) \ \bpsi(y)\gamma^{\nu}\del_{\nu}\psi(y) \ \nonumber\\
&& \hspace{4.5cm} \times  \bpsi(x)\gamma_{(\rho}\del_{\lambda)}\psi(x) \del^{(\rho}\bpsi(x) \gamma^{\lambda)}\psi(x)\Big] a^{\dagger}(p_1) b^{\dagger}(p_2)|0 \ran^{(u)}\nonumber\\
&=&-i\frac{\lambda^2m^6}{96\pi}\coth\frac{\theta}{2}\bigg[\frac{3}{2}\sinh\theta\big(1+2\frac{\Lambda^2}{m^2}\big)-2\cosh\frac{3\theta}{2}\sinh\frac{\theta}{2}\big(8-3\frac{\Lambda^2}{m^2}-24\log\frac{m}{\Lambda}\big)\nonumber\\
&+&\cosh\frac{5\theta}{2}\sinh\frac{\theta}{2}-3\theta\big(1-\cosh\theta-2\cosh2\theta\big)\bigg]\ ,
\label{onederv_u2_final}
\eea
\bea
\mathcal{A}_{\text{b}}^{(u_3)}&=&-2\frac{1}{2!} \big(i\frac{\lambda}{2}\big)^2 \frac{1}{16}\lan 0 | b(p_4) a(p_3) \tT\Big[ \bpsi(y)\gamma^{\mu}\del_{\mu}\psi(y) \ \bpsi(y)\gamma^{\nu}\del_{\nu}\psi(y) \ \nonumber\\
&& \hspace{4.5cm} \times
\del_{(\rho}\bpsi(x) \gamma_{\lambda)}\psi(x) \del^{(\rho}\bpsi(x) \gamma^{\lambda)}\psi(x)\Big] a^{\dagger}(p_1) b^{\dagger}(p_2)|0 \ran^{(u)}\nonumber\\
&=&-i\frac{\lambda^2m^6}{1600\pi}\bigg[25\theta\coth\frac{\theta}{2}\big(1+3\cosh\theta+2\cosh2\theta\big)+\cosh^2\frac{\theta}{2}\Big(-79+270\frac{\Lambda^2}{m^2}-75\frac{\Lambda^4}{m^4}\nonumber\\
&&+1100\log\frac{m}{\Lambda}
+2\cosh\theta\big(-119-60\frac{\Lambda^2}{m^2}+200\log\frac{m}{\Lambda}\big)+17\cosh2\theta\Big)\bigg]\ ,
\label{onederv_u3_final}
\eea
\bea
\mathcal{A}_{\text{b}}^{(u_4)}&=&-2\frac{1}{2!} \big(i\frac{\lambda}{2}\big)^2 \frac{1}{16}\lan 0 | b(p_4) a(p_3) \tT\Big[ -2\bpsi(y)\gamma^{\mu}\del_{\mu}\psi(y) \ \del_{\nu}\bpsi(y) \gamma^{\nu}\psi(y)\ \nonumber\\
&& \hspace{4.5cm} \times \bpsi(x)\gamma_{(\rho}\del_{\lambda)}\psi(x) \bpsi(x)\gamma^{(\rho}\del^{\lambda)}\psi(x)\Big] a^{\dagger}(p_1) b^{\dagger}(p_2)|0 \ran^{(u)}\nonumber\\
&=&i\frac{\lambda^2m^6}{16\pi}\cosh^2\frac{\theta}{2} \ \big(1-4\cosh\theta\big)\Big[4\log\frac{m}{\Lambda}+\theta\coth\frac{\theta}{2} \ \Big]\ ,
\label{onederv_u4_final}
\eea
\bea
\mathcal{A}_{\text{b}}^{(u_5)}&=&-2\frac{1}{2!} \big(i\frac{\lambda}{2}\big)^2 \frac{1}{16}\lan 0 | b(p_4) a(p_3) \tT\Big[4\bpsi(y)\gamma^{\mu}\del_{\mu}\psi(y) \ \del_{\nu}\bpsi(y) \gamma^{\nu}\psi(y)\  \nonumber\\
&& \hspace{4.5cm} \times \bpsi(x)\gamma_{(\rho}\del_{\lambda)}\psi(x) \del^{(\rho}\bpsi(x) \gamma^{\lambda)}\psi(x)\Big] a^{\dagger}(p_1) b^{\dagger}(p_2)|0 \ran^{(u)}\nonumber\\
&=&-i\frac{\lambda^2m^6}{96\pi}\coth\frac{\theta}{2} \Big[\sinh\theta \ \big(\frac{1}{2}+6\frac{\Lambda^2}{m^2}\big)-6\theta\big(1-\cosh\theta-2\cosh2\theta\big)\nonumber\\
&&+\sinh\frac{\theta}{2}\cosh\frac{3\theta}{2}\big(-13+72\log\frac{m}{\Lambda}\big) \ \Big]\ ,
\label{onederv_u5_final}
\eea
\bea
\mathcal{A}_{\text{b}}^{(u_6)}&=&-2\frac{1}{2!} \big(i\frac{\lambda}{2}\big)^2 \frac{1}{16}\lan 0 | b(p_4) a(p_3) \tT\Big[-2\bpsi(y)\gamma^{\mu}\del_{\mu}\psi(y) \ \del_{\nu}\bpsi(y) \gamma^{\nu}\psi(y) \ \nonumber\\
&& \hspace{4.5cm} \times  \del_{(\rho}\bpsi(x) \gamma_{\lambda)}\psi(x) \del^{(\rho}\bpsi(x) \gamma^{\lambda)}\psi(x) \Big] a^{\dagger}(p_1) b^{\dagger}(p_2)|0 \ran^{(u)}\nonumber\\
&=&-i\frac{\lambda^2m^6}{16\pi}\cosh^2\frac{\theta}{2} \ \bigg[-1+6\frac{\Lambda^2}{m^2}+16\log\frac{m}{\Lambda}+\theta\big(4\sinh\theta+3\coth\frac{\theta}{2}\big)\nonumber\\
&&+\cosh\theta\big(-5+8\log\frac{m}{\Lambda}\big)\bigg]\ ,
\label{onederv_u6_final}
\eea
\bea
\mathcal{A}_{\text{b}}^{(u_7)}&=&-2\frac{1}{2!} \big(i\frac{\lambda}{2}\big)^2 \frac{1}{16}\lan 0 | b(p_4) a(p_3) \tT\Big[\del_{\mu}\bpsi(y) \gamma^{\mu}\psi(y) \ \del_{\nu}\bpsi(y) \gamma^{\nu}\psi(y) \ \nonumber\\
&& \hspace{4.5cm} \times \bpsi(x)\gamma_{(\rho}\del_{\lambda)}\psi(x) \bpsi(x)\gamma^{(\rho}\del^{\lambda)}\psi(x) \Big] a^{\dagger}(p_1) b^{\dagger}(p_2)|0 \ran^{(u)}\nonumber\\
&=&i\frac{\lambda^2m^6}{32\pi}\cosh^2\frac{\theta}{2} \ \big(1-4\cosh\theta\big)\Big[2\log\frac{m}{\Lambda}+\theta\coth\frac{\theta}{2} \ \Big]\ ,
\label{onederv_u7_final}
\eea
\bea
\mathcal{A}_{\text{b}}^{(u_8)}&=&-2\frac{1}{2!} \big(i\frac{\lambda}{2}\big)^2 \frac{1}{16} \lan 0 | b(p_4) a(p_3) \tT\Big[-2\del_{\mu}\bpsi(y) \gamma^{\mu}\psi(y) \ \del_{\nu}\bpsi(y) \gamma^{\nu}\psi(y) \ \nonumber\\
&& \hspace{4.5cm} \times  \bpsi(x)\gamma_{(\rho}\del_{\lambda)}\psi(x) \del^{(\rho}\bpsi(x) \gamma^{\lambda)}\psi(x)\Big] a^{\dagger}(p_1) b^{\dagger}(p_2)|0 \ran^{(u)}\nonumber\\
&=&-i\frac{\lambda^2m^6}{32\pi}\bigg[1+4\cosh\theta \ \log\frac{m}{\Lambda}-\cosh2\theta\big(1-4\log\frac{m}{\Lambda}\big)\nonumber\\
&&+\theta\big(5\sinh\theta+2\sinh2\theta+2\coth\frac{\theta}{2}\big)\bigg]\ ,
\label{onederv_u8_final}
\eea
\bea
\mathcal{A}_{\text{b}}^{(u_9)}&=&-2\frac{1}{2!} \big(i\frac{\lambda}{2}\big)^2 \frac{1}{16}\lan 0 | b(p_4) a(p_3) \tT\Big[\del_{\mu}\bpsi(y) \gamma^{\mu}\psi(y) \ \del_{\nu}\bpsi(y) \gamma^{\nu}\psi(y) \ \nonumber\\
&& \hspace{4.5cm} \times  \del_{(\rho}\bpsi(x) \gamma_{\lambda)}\psi(x) \del^{(\rho}\bpsi(x) \gamma^{\lambda)}\psi(x)\Big] a^{\dagger}(p_1) b^{\dagger}(p_2)|0 \ran^{(u)}\nonumber\\
&=&-i\frac{\lambda^2m^6}{32\pi}\cosh^2\frac{\theta}{2} \ \bigg[3\frac{\Lambda^2}{m^2}+10\log\frac{m}{\Lambda}+\cosh\theta\big(-3+8\log\frac{m}{\Lambda}\big)+\theta\big(4\sinh\theta+3\coth\frac{\theta}{2}\big)\bigg]\ ,\nonumber\\
\label{onederv_u9_final}
\eea
where the computational method that was employed was exactly the same as when $\mathcal{A}_{\text{a}}^{(u_1)}$ was computed in (\ref{twoderv_u1}).
\vspace{2mm}
\\
Combining the $u$-channel contributions to the amplitude, (\ref{onederv_u1_final}) - (\ref{onederv_u9_final}), gives the total $u$-channel contribution to the amplitude from the diagram where one vertex contains the quartic coupling $\tilde{X}_{\rho\lambda}\tilde{X}^{\rho\lambda}$ while the other vertex contains the quartic coupling $(\tilde{X}_{\mu}^{\ \mu})^2$,
\bea
\mathcal{A}_{\text{b}}^{(u)}&=& 2\Big(\mathcal{A}_{\text{b}}^{(u_1)}+\mathcal{A}_{\text{b}}^{(u_2)}+\cdots+\mathcal{A}_{\text{b}}^{(u_9)}\Big)\nonumber\\
&=&-i\frac{\lambda^2m^6}{2400\pi}\cosh\frac{\theta}{2}
\bigg[4800 \ \theta\cosh\frac{3\theta}{2}\coth\frac{\theta}{2}+\cosh\frac{\theta}{2}\big(2063+3510\frac{\Lambda^2}{m^2}-225\frac{\Lambda^4}{m^4}+251\cosh2\theta\big)\nonumber\\
&+&\cosh\frac{\theta}{2}\cosh\theta\big(-7114+840\frac{\Lambda^2}{m^2}+28800\log\frac{m}{\Lambda}\big)\bigg]\ .
\label{onederv_u_amplitude}
\eea
Finally, adding together the contributions from the $s$, $t$ and $u$-channels to the amplitude, (\ref{onederv_s_amplitude}), (\ref{onederv_t_amplitude}) and (\ref{onederv_u_amplitude}), one finds the total amplitude from the bubble diagrams where one vertex contains the quartic coupling $\tilde{X}_{\rho\lambda}\tilde{X}^{\rho\lambda}$ while the other vertex contains quartic coupling $(\tilde{X}_{\mu}^{\ \mu})^2$,
\bea
\mathcal{A}_{\text{b}}&=&\mathcal{A}_{\text{b}}^{(s)}+\mathcal{A}_{\text{b}}^{(t)}+\mathcal{A}_{\text{b}}^{(u)}\nonumber\\
&=&\frac{\lambda^2m^6}{800\pi\sinh\theta}\Big[400\pi\big(9+3\cosh\theta+3\cosh2\theta+\cosh3\theta\big)-i\Big(\sinh\theta\big(-2047+930\frac{\Lambda^2}{m^2}\nonumber\\
&+&3600\log\frac{m}{\Lambda}\big)
+\sinh2\theta\big(\frac{91}{2}+570\frac{\Lambda^2}{m^2}+2400\log\frac{m}{\Lambda}\big)+2\sinh3\theta\big(-177+35\frac{\Lambda^2}{m^2}\nonumber\\
&+&200\log\frac{m}{\Lambda}\big)-3200 \ \theta\Big)\Big]\ .
\label{amplitude_onederv}
\eea

\subsubsection*{(c) Both vertices contain interactions $(\tilde{X}_{\mu}^{\ \mu})^2$ :}
\begin{figure}[h!]
	\centering
	\subfigure[$s$-channel]{
		\includegraphics[width=0.31\columnwidth,height=0.21\columnwidth]{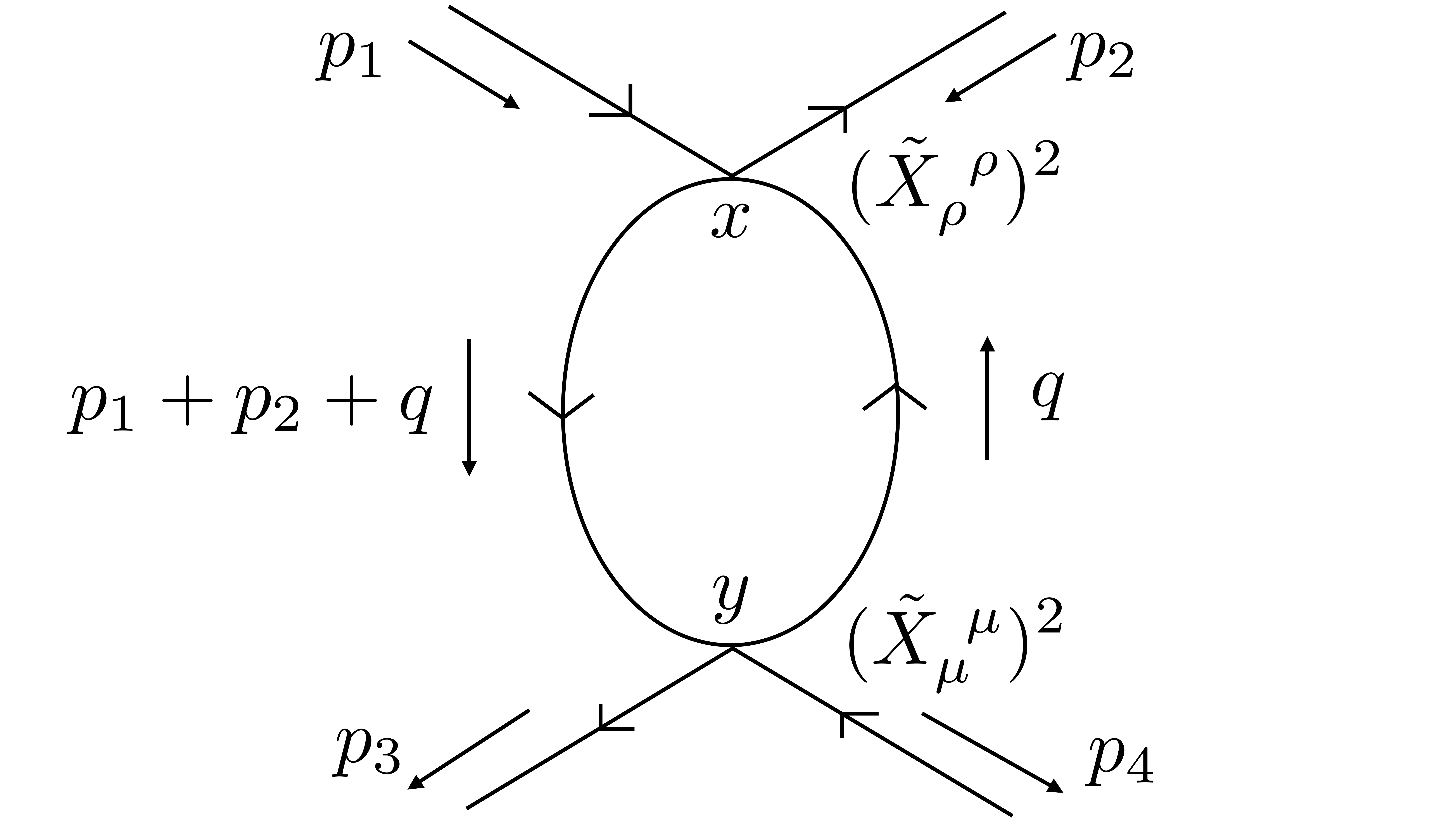}
		\label{bubblediagram_noderv_s} } 
	\subfigure[$t$-channel]{
		\includegraphics[width=0.31\columnwidth,height=0.21\columnwidth]{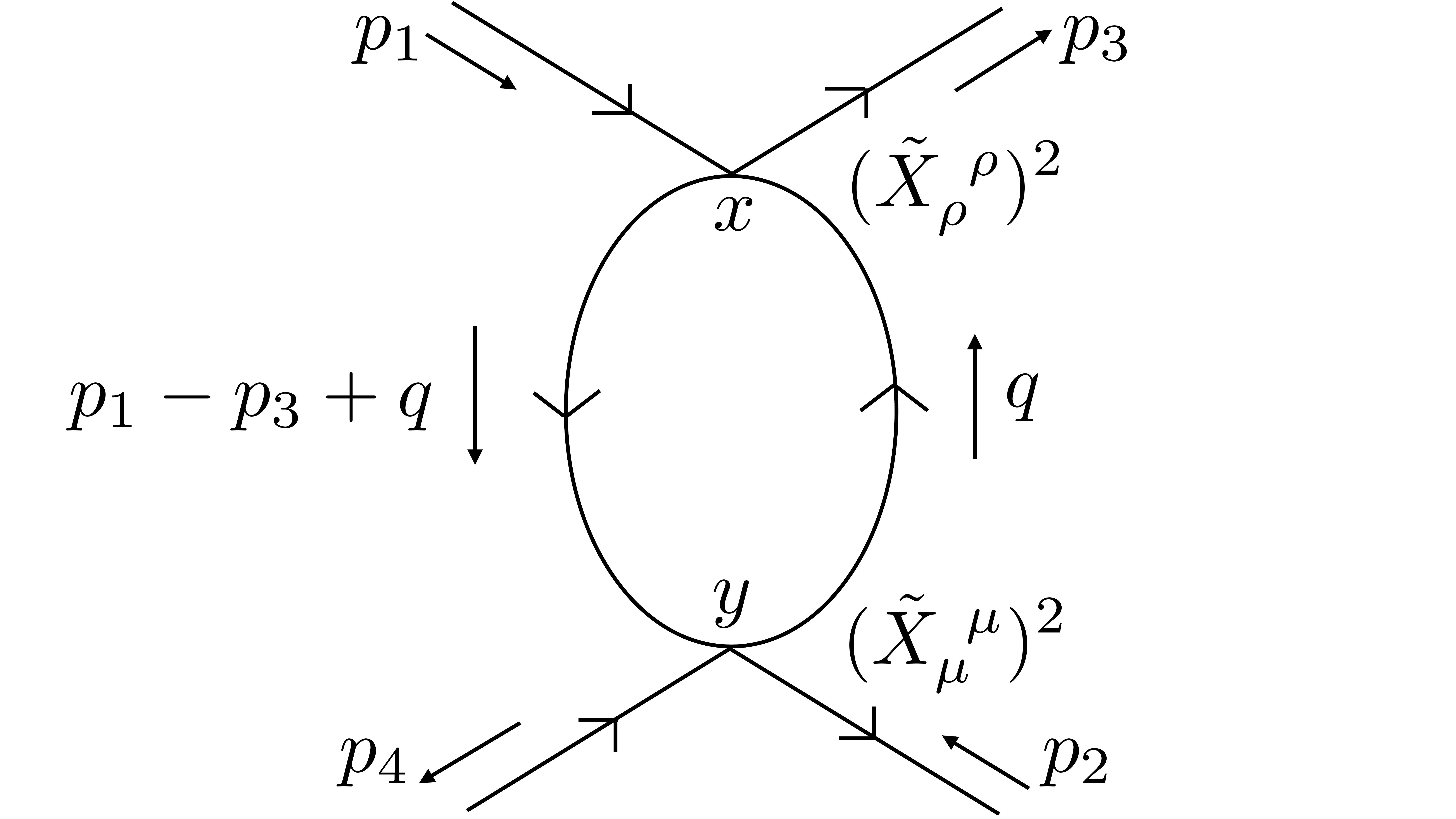}
		\label{bubblediagram_noderv_t} }
	\subfigure[$u$-channel]{
		\includegraphics[width=0.31\columnwidth,height=0.21\columnwidth]{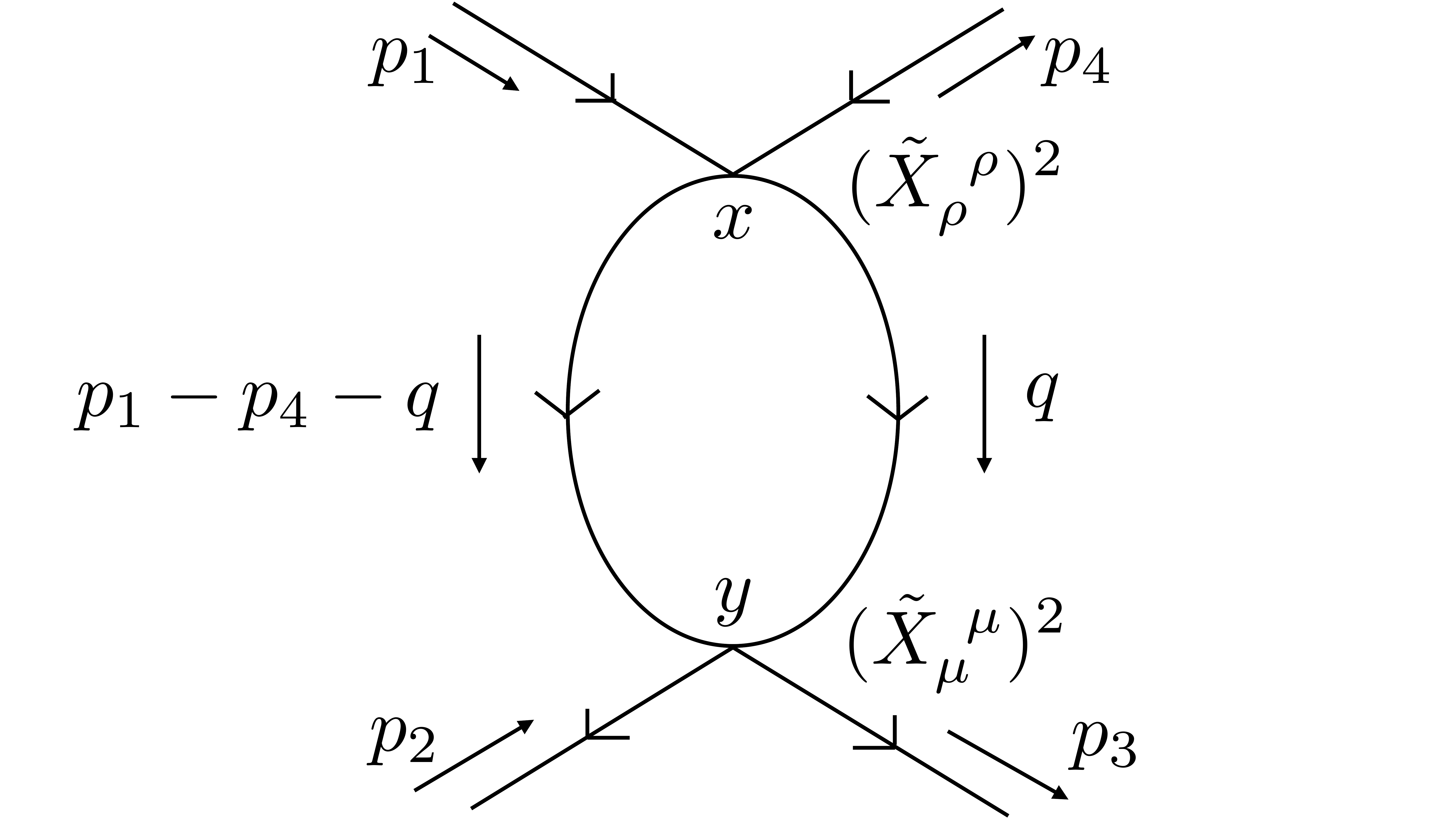}
		\label{bubblediagram_noderv_u} }
	\caption{\small The $s$, $t$ and $u$-channel amplitudes when both of the interaction vertices are $(\tilde{X}_{\mu}^{\ \mu})^2$.}
	\label{bubblediagram_noderv}
\end{figure}

Finally consider bubble diagrams shown in figure \ref{bubblediagram_noderv} where both vertices contain  the quartic coupling $(\tilde{X}_{\mu}^{\ \mu})^2$ .

The amplitude due to these kinds of bubble diagrams is given by,
\bea
\mathcal{A}_{\text{c}}&=&\frac{1}{2!} \big(-i\frac{\lambda}{2}\big)^2\lan 0 | b(p_4) a(p_3) \tT\Big[\big(\tilde{X}_{\mu}^{\ \mu}(y)\big)^2 \big(\tilde{X}_{\rho}^{\ \rho}(x)\big)^2\Big] a^{\dagger}(p_1) b^{\dagger}(p_2)|0 \ran\nonumber\\
&=&\frac{1}{2!} \big(-i\frac{\lambda}{2}\big)^2 \frac{1}{16}\lan 0 | b(p_4) a(p_3) \tT\Big[
\bpsi(y)\gamma^{\mu}\del_{\mu}\psi(y) \ \bpsi(y)\gamma^{\nu}\del_{\nu}\psi(y) \ \ \bpsi(x)\gamma^{\rho}\del_{\rho}\psi(x) \ \bpsi(x)\gamma^{\lambda}\del_{\lambda}\psi(x)\nonumber\\
&-&2\bpsi(y)\gamma^{\mu}\del_{\mu}\psi(y) \ \bpsi(y)\gamma^{\nu}\del_{\nu}\psi(y) \ \ \bpsi(x)\gamma^{\rho}\del_{\rho}\psi(x) \ \del_{\lambda}\bpsi(x) \gamma^{\lambda}\psi(x)\nonumber\\
&+&\bpsi(y)\gamma^{\mu}\del_{\mu}\psi(y) \ \bpsi(y)\gamma^{\nu}\del_{\nu}\psi(y) \ \
\del_{\rho}\bpsi(x) \gamma^{\rho}\psi(x) \ \del_{\lambda}\bpsi(x) \gamma^{\lambda}\psi(x)\nonumber\\
&-&2\bpsi(y)\gamma^{\mu}\del_{\mu}\psi(y) \ \del_{\nu}\bpsi(y) \gamma^{\nu}\psi(y)\ \ \bpsi(x)\gamma^{\rho}\del_{\rho}\psi(x) \ \bpsi(x)\gamma^{\lambda}\del_{\lambda}\psi(x)\nonumber\\
&+&4\bpsi(y)\gamma^{\mu}\del_{\mu}\psi(y) \ \del_{\nu}\bpsi(y) \gamma^{\nu}\psi(y)\ \ \bpsi(x)\gamma^{\rho}\del_{\rho}\psi(x) \ \del_{\lambda}\bpsi(x)\gamma^{\lambda}\psi(x)\nonumber\\
&-&2\bpsi(y)\gamma^{\mu}\del_{\mu}\psi(y) \ \del_{\nu}\bpsi(y) \gamma^{\nu}\psi(y) \ \ \del_{\rho}\bpsi(x) \gamma^{\rho}\psi(x) \ \del_{\lambda}\bpsi(x) \gamma^{\lambda}\psi(x)\nonumber\\
&+&\del_{\mu}\bpsi(y) \gamma^{\mu}\psi(y) \ \del_{\nu}\bpsi(y) \gamma^{\nu}\psi(y) \ \ \bpsi(x)\gamma^{\rho}\del_{\rho}\psi(x) \ \bpsi(x)\gamma^{\lambda}\del_{\lambda}\psi(x)\nonumber\\
&-&2\del_{\mu}\bpsi(y) \gamma^{\mu}\psi(y) \ \del_{\nu}\bpsi(y) \gamma^{\nu}\psi(y) \ \ \bpsi(x)\gamma^{\rho}\del_{\rho}\psi(x) \ \del_{\lambda}\bpsi(x)\gamma^{\lambda}\psi(x)\nonumber\\
&+&\del_{\mu}\bpsi(y) \gamma^{\mu}\psi(y) \ \del_{\nu}\bpsi(y) \gamma^{\nu}\psi(y) \ \ \del_{\rho}\bpsi(x) \gamma^{\rho}\psi(x) \ \del_{\lambda}\bpsi(x) \gamma^{\lambda}\psi(x) \
\Big] a^{\dagger}(p_1) b^{\dagger}(p_2)|0 \ran\ .
\label{TmumuPower4}
\eea
The $s$-channel contribution to the amplitude from figure \ref{bubblediagram_noderv_s} can be computed by evaluating the contributions from each of the above terms in (\ref{TmumuPower4}) and adding the results together. By the general $s$-channel amplitude (\ref{schannelbubblegeneral}), the contributions from all nine terms in (\ref{TmumuPower4}) are,
\bea
\mathcal{A}_{\text{c}}^{(s_1)}
&=&\frac{1}{2!} \big(-i\frac{\lambda}{2}\big)^2\frac{1}{16}\lan 0 | b(p_4) a(p_3) \tT\Big[ \bpsi(y)\gamma^{\mu}\del_{\mu}\psi(y) \ \bpsi(y)\gamma^{\nu}\del_{\nu}\psi(y) \  \nonumber\\
&& \hspace{4.7cm} \times  \bpsi(x)\gamma^{\rho}\del_{\rho}\psi(x) \ \bpsi(x)\gamma^{\lambda}\del_{\lambda}\psi(x)\Big] a^{\dagger}(p_1) b^{\dagger}(p_2)|0 \ran^{(s)}\nonumber\\
&=&-\frac{\lambda^2 m^6}{384 \pi \sinh\theta}\bigg[3\pi\big(11+4\cosh\theta+\cosh2\theta\big)+i\Big(\sinh \theta \big(34-12\frac{\Lambda^2}{m^2}-48\log \frac{m}{\Lambda }\big) \nonumber\\
&+&2\sinh2\theta\big(4+3\log \frac{m}{\Lambda }\big)+3 \theta \big(11+4\cosh\theta+\cosh2\theta\big)\Big)\bigg]\ ,
\label{noderv_s1_final}
\eea
\bea
\mathcal{A}_{\text{c}}^{(s_2)}&=&\frac{1}{2!} \big(-i\frac{\lambda}{2}\big)^2\frac{1}{16}\lan 0 | b(p_4) a(p_3) \tT\Big[-2\bpsi(y)\gamma^{\mu}\del_{\mu}\psi(y) \ \bpsi(y)\gamma^{\nu}\del_{\nu}\psi(y) \ \nonumber\\
&& \hspace{4.7cm} \times  \bpsi(x)\gamma^{\rho}\del_{\rho}\psi(x) \ \del_{\lambda}\bpsi(x) \gamma^{\lambda}\psi(x)\Big] a^{\dagger}(p_1) b^{\dagger}(p_2)|0 \ran^{(s)}\nonumber
\eea\bea
&=&-\frac{\lambda^2 m^6}{384 \pi \sinh\theta}\bigg[6\pi\big(11+4\cosh\theta+\cosh2\theta\big)+i\Big(\sinh \theta \big(53-18\frac{\Lambda^2}{m^2}-96\log \frac{m}{\Lambda }\big) \nonumber\\
&+&\sinh2\theta\big(13+6\log \frac{m}{\Lambda }\big)+3\cosh2\theta\sinh\theta+6 \theta \big(11+4\cosh\theta+\cosh2\theta\big)\Big)\bigg]\ ,
\label{noderv_s2_final}
\eea
\bea
\mathcal{A}_{\text{c}}^{(s_3)}&=&\frac{1}{2!} \big(-i\frac{\lambda}{2}\big)^2\frac{1}{16}\lan 0 | b(p_4) a(p_3) \tT\Big[ \bpsi(y)\gamma^{\mu}\del_{\mu}\psi(y) \ \bpsi(y)\gamma^{\nu}\del_{\nu}\psi(y) \ \nonumber\\
&& \hspace{4.7cm} \times \del_{\rho}\bpsi(x) \gamma^{\rho}\psi(x) \ \del_{\lambda}\bpsi(x) \gamma^{\lambda}\psi(x)\Big] a^{\dagger}(p_1) b^{\dagger}(p_2)|0 \ran^{(s)}\nonumber\\
&=&-\frac{\lambda^2 m^6}{384 \pi \sinh\theta}\bigg[3\pi\big(11+4\cosh\theta+\cosh2\theta\big)+i\Big(\sinh \theta \big(9-48\log \frac{m}{\Lambda }\big) +3\frac{\Lambda^2}{m^2}\sinh2\theta\nonumber\\
&+&\sinh3\theta+3 \theta \big(11+4\cosh\theta+\cosh2\theta\big)\Big)\bigg]\ ,
\label{noderv_s3_final}
\eea
\bea
\mathcal{A}_{\text{c}}^{(s_4)}&=&\frac{1}{2!} \big(-i\frac{\lambda}{2}\big)^2\frac{1}{16}\lan 0 | b(p_4) a(p_3) \tT\Big[ -2\bpsi(y)\gamma^{\mu}\del_{\mu}\psi(y) \ \del_{\nu}\bpsi(y) \gamma^{\nu}\psi(y)\ \nonumber\\
&& \hspace{0.2cm} \times \bpsi(x)\gamma^{\rho}\del_{\rho}\psi(x) \ \bpsi(x)\gamma^{\lambda}\del_{\lambda}\psi(x)\Big] a^{\dagger}(p_1) b^{\dagger}(p_2)|0 \ran^{(s)}
\ \ = \ \ \mathcal{A}_{\text{c}}^{(s_2)} \ ,
\label{noderv_s4_final}
\eea
\bea
\mathcal{A}_{\text{c}}^{(s_5)}&=&\frac{1}{2!} \big(-i\frac{\lambda}{2}\big)^2\frac{1}{16}\lan 0 | b(p_4) a(p_3) \tT\Big[4\bpsi(y)\gamma^{\mu}\del_{\mu}\psi(y) \ \del_{\nu}\bpsi(y) \gamma^{\nu}\psi(y)\  \nonumber\\
&& \hspace{4.7cm} \times \bpsi(x)\gamma^{\rho}\del_{\rho}\psi(x) \ \del_{\lambda}\bpsi(x)\gamma^{\lambda}\psi(x)\Big] a^{\dagger}(p_1) b^{\dagger}(p_2)|0 \ran^{(s)}\nonumber\\
&=&-\frac{\lambda^2 m^6}{19200 \pi \sinh\theta}\bigg[600\pi\big(11+4\cosh\theta+\cosh2\theta\big)+i\Big(\sinh \theta \big(5948-2280\frac{\Lambda^2}{m^2}-9600\log \frac{m}{\Lambda }\big) \nonumber\\
&+&\sinh2\theta\big(\frac{2839}{2}-180\frac{\Lambda^2}{m^2}-75\frac{\Lambda^4}{m^4}+600\log \frac{m}{\Lambda }\big)+8\sinh\theta\cosh2\theta\big(26+15\frac{\Lambda^2}{m^2}\big)\nonumber\\
&+&17\sinh\theta\cosh3\theta
+600 \  \theta \big(11+4\cosh\theta+\cosh2\theta\big)\Big)\bigg]\ ,
\label{noderv_s5_final}
\eea
\bea
\mathcal{A}_{\text{c}}^{(s_6)}&=&\frac{1}{2!} \big(-i\frac{\lambda}{2}\big)^2\frac{1}{16}\lan 0 | b(p_4) a(p_3) \tT\Big[-2\bpsi(y)\gamma^{\mu}\del_{\mu}\psi(y) \ \del_{\nu}\bpsi(y) \gamma^{\nu}\psi(y) \  \nonumber\\
&& \hspace{0.2cm} \times \del_{\rho}\bpsi(x) \gamma^{\rho}\psi(x) \ \del_{\lambda}\bpsi(x) \gamma^{\lambda}\psi(x)\Big] a^{\dagger}(p_1) b^{\dagger}(p_2)|0 \ran^{(s)}
\ \ = \ \ \mathcal{A}_{\text{c}}^{(s_2)}\ ,
\label{noderv_s6_final}
\eea
\bea
\mathcal{A}_{\text{c}}^{(s_7)}&=&\frac{1}{2!} \big(-i\frac{\lambda}{2}\big)^2\frac{1}{16}\lan 0 | b(p_4) a(p_3) \tT\Big[\del_{\mu}\bpsi(y) \gamma^{\mu}\psi(y) \ \del_{\nu}\bpsi(y) \gamma^{\nu}\psi(y) \ \nonumber\\
&& \hspace{0.2cm} \times  \bpsi(x)\gamma^{\rho}\del_{\rho}\psi(x) \ \bpsi(x)\gamma^{\lambda}\del_{\lambda}\psi(x) \Big] a^{\dagger}(p_1) b^{\dagger}(p_2)|0 \ran^{(s)}
\ \ = \ \ \mathcal{A}_{\text{c}}^{(s_3)}\ ,
\label{noderv_s7_final}
\eea\bea
\mathcal{A}_{\text{c}}^{(s_8)}&=&\frac{1}{2!} \big(-i\frac{\lambda}{2}\big)^2\frac{1}{16}\lan 0 | b(p_4) a(p_3) \tT\Big[-2\del_{\mu}\bpsi(y) \gamma^{\mu}\psi(y) \ \del_{\nu}\bpsi(y) \gamma^{\nu}\psi(y) \ \nonumber\\
&& \hspace{0.2cm} \times  \bpsi(x)\gamma^{\rho}\del_{\rho}\psi(x) \ \del_{\lambda}\bpsi(x)\gamma^{\lambda}\psi(x)\Big] a^{\dagger}(p_1) b^{\dagger}(p_2)|0 \ran^{(s)}
\ \ = \ \ \mathcal{A}_{\text{c}}^{(s_2)}\ ,
\label{noderv_s8_final}
\eea\bea
\mathcal{A}_{\text{c}}^{(s_9)}&=&\frac{1}{2!} \big(-i\frac{\lambda}{2}\big)^2\frac{1}{16}\lan 0 | b(p_4) a(p_3) \tT\Big[\del_{\mu}\bpsi(y) \gamma^{\mu}\psi(y) \ \del_{\nu}\bpsi(y) \gamma^{\nu}\psi(y) \ \nonumber\\
&& \hspace{0.2cm} \times \del_{\rho}\bpsi(x) \gamma^{\rho}\psi(x) \ \del_{\lambda}\bpsi(x) \gamma^{\lambda}\psi(x)\Big] a^{\dagger}(p_1) b^{\dagger}(p_2)|0 \ran^{(s)}
\ \ = \ \ \mathcal{A}_{\text{c}}^{(s_1)}\ ,
\label{noderv_s9_final}
\eea
where the employed computational method was exactly the same as when $\mathcal{A}_{\text{a}}^{(s_1)}$ was computed in (\ref{twoderv_s1}).
\vspace{1mm}
\\
Combining the $s$-channel contributions to the amplitude, (\ref{noderv_s1_final}) - (\ref{noderv_s9_final}), gives the total $s$-channel contribution to the amplitude from the diagrams where both vertices contain interactions $(\tilde{X}_{\mu}^{\ \mu})^2$,
\bea
\mathcal{A}_{\text{c}}^{(s)}&=& 2\Big(\mathcal{A}_{\text{c}}^{(s_1)}+\mathcal{A}_{\text{c}}^{(s_2)}+\cdots+\mathcal{A}_{\text{c}}^{(s_9)}\Big)\nonumber\\
&=&-\frac{\lambda^2m^6}{19200\pi\sinh\theta}\bigg[4800\pi\big(11+4\cosh\theta+\cosh2\theta\big)+i\Big(4800 \ \theta\big(11+4\cosh\theta+\cosh2\theta\big)\nonumber\\
&-&8\sinh\theta\big(-5111+1785\frac{\Lambda^2}{m^2}+9600\log\frac{m}{\Lambda}\big)
+2\sinh2\theta\big(4811+120\frac{\Lambda^2}{m^2}-75\frac{\Lambda^4}{m^4}\nonumber\\
&+&2400\log\frac{m}{\Lambda}\big)
+24\sinh3\theta\big(42+5\frac{\Lambda^2}{m^2}\big)+17\sinh4\theta\Big)\bigg]\ .
\label{noderv_s_amplitude}
\eea
Just as in the computation of the first and second type of bubble diagrams, (\ref{twoderv_s_amplitude}) and (\ref{onederv_s_amplitude}), a factor of $2$ arises in the first line of the above expression, (\ref{noderv_s_amplitude}), because the same contribution would be found if the vertices $x$ and $y$ were exchanged. A similar factor would appear while evaluating the $t$ and $u$-channel amplitudes as well.
\vspace{2mm}
\\
The contribution to the amplitude from the $t$-channel diagram, figure \ref{bubblediagram_noderv_t}, can be computed using the general $t$-channel amplitude (\ref{tchannelbubblegeneral}). The nine terms in (\ref{TmumuPower4}) yield the following contributions to the amplitude,
\bea
\mathcal{A}_{\text{c}}^{(t_1)}
&=&\frac{1}{2!} \big(-i\frac{\lambda}{2}\big)^2\frac{1}{16}\lan 0 | b(p_4) a(p_3) \tT\Big[ \bpsi(y)\gamma^{\mu}\del_{\mu}\psi(y) \ \bpsi(y)\gamma^{\nu}\del_{\nu}\psi(y) \  \nonumber\\
&& \hspace{4.7cm} \times  \bpsi(x)\gamma^{\rho}\del_{\rho}\psi(x) \ \bpsi(x)\gamma^{\lambda}\del_{\lambda}\psi(x)\Big] a^{\dagger}(p_1) b^{\dagger}(p_2)|0 \ran^{(t)}\nonumber\\
&=&-i\frac{\lambda^2 m^6}{64\pi}\bigg[1+\frac{\Lambda^2}{m^2}+6\log\frac{m}{\Lambda}-\cosh\theta\big(\frac{\Lambda^2}{m^2}+4\log\frac{m}{\Lambda}\big)\bigg]\ ,
\label{noderv_t1_final}
\eea
\bea
\mathcal{A}_{\text{c}}^{(t_2)}&=&\frac{1}{2!} \big(-i\frac{\lambda}{2}\big)^2\frac{1}{16}\lan 0 | b(p_4) a(p_3) \tT\Big[-2\bpsi(y)\gamma^{\mu}\del_{\mu}\psi(y) \ \bpsi(y)\gamma^{\nu}\del_{\nu}\psi(y) \ \nonumber\\
&& \hspace{4.7cm} \times  \bpsi(x)\gamma^{\rho}\del_{\rho}\psi(x) \ \del_{\lambda}\bpsi(x) \gamma^{\lambda}\psi(x)\Big] a^{\dagger}(p_1) b^{\dagger}(p_2)|0 \ran^{(t)}\nonumber\\
&=&-i\frac{\lambda^2 m^6}{64\pi}(2-\cosh\theta)\bigg[1+\frac{\Lambda^2}{m^2}+6\log\frac{m}{\Lambda}\bigg]\ ,
\label{noderv_t2_final}
\eea
\bea
\mathcal{A}_{\text{c}}^{(t_3)}&=&\frac{1}{2!} \big(-i\frac{\lambda}{2}\big)^2\frac{1}{16}\lan 0 | b(p_4) a(p_3) \tT\Big[ \bpsi(y)\gamma^{\mu}\del_{\mu}\psi(y) \ \bpsi(y)\gamma^{\nu}\del_{\nu}\psi(y) \ \nonumber\\
&& \hspace{4.7cm} \times \del_{\rho}\bpsi(x) \gamma^{\rho}\psi(x) \ \del_{\lambda}\bpsi(x) \gamma^{\lambda}\psi(x)\Big] a^{\dagger}(p_1) b^{\dagger}(p_2)|0 \ran^{(t)}\nonumber\\
&=&-i\frac{\lambda^2 m^6}{64\pi}\bigg[1+\frac{\Lambda^2}{m^2}+6\log\frac{m}{\Lambda}-\cosh\theta\big(1+2\log\frac{m}{\Lambda}\big)\bigg]\ ,
\label{noderv_t3_final}
\eea
\bea
\mathcal{A}_{\text{c}}^{(t_4)}&=&\frac{1}{2!} \big(-i\frac{\lambda}{2}\big)^2\frac{1}{16}\lan 0 | b(p_4) a(p_3) \tT\Big[ -2\bpsi(y)\gamma^{\mu}\del_{\mu}\psi(y) \ \del_{\nu}\bpsi(y) \gamma^{\nu}\psi(y)\ \nonumber\\
&& \hspace{0.2cm} \times \bpsi(x)\gamma^{\rho}\del_{\rho}\psi(x) \ \bpsi(x)\gamma^{\lambda}\del_{\lambda}\psi(x)\Big] a^{\dagger}(p_1) b^{\dagger}(p_2)|0 \ran^{(t)}
\ \ = \ \ \mathcal{A}_{\text{c}}^{(t_2)}\ ,
\label{noderv_t4_final}
\eea
\bea
\mathcal{A}_{\text{c}}^{(t_5)}&=&\frac{1}{2!} \big(-i\frac{\lambda}{2}\big)^2\frac{1}{16}\lan 0 | b(p_4) a(p_3) \tT\Big[4\bpsi(y)\gamma^{\mu}\del_{\mu}\psi(y) \ \del_{\nu}\bpsi(y) \gamma^{\nu}\psi(y)\  \nonumber\\
&& \hspace{4.7cm} \times \bpsi(x)\gamma^{\rho}\del_{\rho}\psi(x) \ \del_{\lambda}\bpsi(x)\gamma^{\lambda}\psi(x)\Big] a^{\dagger}(p_1) b^{\dagger}(p_2)|0 \ran^{(t)}\nonumber\\
&=&-i\frac{\lambda^2 m^6}{128\pi}\bigg[8+6\frac{\Lambda^2}{m^2}-\frac{\Lambda^4}{m^4}+48\log\frac{m}{\Lambda}-2\cosh\theta\big(1+3\frac{\Lambda^2}{m^2}+12\log\frac{m}{\Lambda}\big)\bigg]\ ,
\label{noderv_t5_final}
\eea
\bea
\mathcal{A}_{\text{c}}^{(t_6)}&=&\frac{1}{2!} \big(-i\frac{\lambda}{2}\big)^2\frac{1}{16}\lan 0 | b(p_4) a(p_3) \tT\Big[-2\bpsi(y)\gamma^{\mu}\del_{\mu}\psi(y) \ \del_{\nu}\bpsi(y) \gamma^{\nu}\psi(y) \  \nonumber\\
&& \hspace{0.2cm} \times \del_{\rho}\bpsi(x) \gamma^{\rho}\psi(x) \ \del_{\lambda}\bpsi(x) \gamma^{\lambda}\psi(x)\Big] a^{\dagger}(p_1) b^{\dagger}(p_2)|0 \ran^{(t)}
\ \ = \ \ \mathcal{A}_{\text{c}}^{(t_2)}\ ,
\label{noderv_t6_final}
\eea\bea
\mathcal{A}_{\text{c}}^{(t_7)}&=&\frac{1}{2!} \big(-i\frac{\lambda}{2}\big)^2\frac{1}{16}\lan 0 | b(p_4) a(p_3) \tT\Big[\del_{\mu}\bpsi(y) \gamma^{\mu}\psi(y) \ \del_{\nu}\bpsi(y) \gamma^{\nu}\psi(y) \ \nonumber\\
&& \hspace{0.2cm} \times  \bpsi(x)\gamma^{\rho}\del_{\rho}\psi(x) \ \bpsi(x)\gamma^{\lambda}\del_{\lambda}\psi(x) \Big] a^{\dagger}(p_1) b^{\dagger}(p_2)|0 \ran^{(t)}
\ \ = \ \ \mathcal{A}_{\text{c}}^{(t_3)}\ ,
\label{noderv_t7_final}
\eea\bea
\mathcal{A}_{\text{c}}^{(t_8)}&=&\frac{1}{2!} \big(-i\frac{\lambda}{2}\big)^2\frac{1}{16}\lan 0 | b(p_4) a(p_3) \tT\Big[-2\del_{\mu}\bpsi(y) \gamma^{\mu}\psi(y) \ \del_{\nu}\bpsi(y) \gamma^{\nu}\psi(y) \ \nonumber\\
&& \hspace{0.2cm} \times  \bpsi(x)\gamma^{\rho}\del_{\rho}\psi(x) \ \del_{\lambda}\bpsi(x)\gamma^{\lambda}\psi(x)\Big] a^{\dagger}(p_1) b^{\dagger}(p_2)|0 \ran^{(t)}
\ \ = \ \ \mathcal{A}_{\text{c}}^{(t_2)}\ ,
\label{noderv_t8_final}
\eea\bea
\mathcal{A}_{\text{c}}^{(t_9)}&=&\frac{1}{2!} \big(-i\frac{\lambda}{2}\big)^2\frac{1}{16}\lan 0 | b(p_4) a(p_3) \tT\Big[\del_{\mu}\bpsi(y) \gamma^{\mu}\psi(y) \ \del_{\nu}\bpsi(y) \gamma^{\nu}\psi(y) \ \nonumber\\
&& \hspace{0.2cm} \times \del_{\rho}\bpsi(x) \gamma^{\rho}\psi(x) \ \del_{\lambda}\bpsi(x) \gamma^{\lambda}\psi(x)\Big] a^{\dagger}(p_1) b^{\dagger}(p_2)|0 \ran^{(t)}
\ \ = \ \ \mathcal{A}_{\text{c}}^{(t_1)}\ ,
\label{noderv_t9_final}
\eea
where the employed computational method was exactly the same as was used in  evaluating $\mathcal{A}_{\text{a}}^{(t_1)}$ in (\ref{twoderv_t1}).
\vspace{2mm}
\\
Combining the $t$-channel contributions to the amplitude,  (\ref{noderv_t1_final}) - (\ref{noderv_t9_final}), gives the total $t$-channel contribution to the amplitude from the diagram where both vertices contain interactions $(\tilde{X}_{\mu}^{\ \mu})^2$,
\bea
\mathcal{A}_{\text{c}}^{(t)}&=& 2\Big(\mathcal{A}_{\text{c}}^{(t_1)}+\mathcal{A}_{\text{c}}^{(t_2)}+\cdots+\mathcal{A}_{\text{c}}^{(t_9)}\Big)\nonumber\\
&=&-i\frac{\lambda^2 m^6}{64\pi}\bigg[32+30\frac{\Lambda^2}{m^2}-\frac{\Lambda^4}{m^4}+192\log\frac{m}{\Lambda}-2\cosh\theta\big(7+9\frac{\Lambda^2}{m^2}+48\log\frac{m}{\Lambda}\big)\bigg]\ .\nonumber\\
\label{noderv_t_amplitude}
\eea
The contribution to the amplitude of the $u$-channel diagram, figure \ref{bubblediagram_noderv_u}, can be computed using the general $u$-channel amplitude (\ref{uchannelbubblegeneral}). The contributions to the amplitude from the nine terms in (\ref{TmumuPower4}) are given by,
\bea
\mathcal{A}_{\text{c}}^{(u_1)}
&=&\frac{1}{2!} \big(-i\frac{\lambda}{2}\big)^2\frac{1}{16}\lan 0 | b(p_4) a(p_3) \tT\Big[ \bpsi(y)\gamma^{\mu}\del_{\mu}\psi(y) \ \bpsi(y)\gamma^{\nu}\del_{\nu}\psi(y) \  \nonumber\\
&& \hspace{4.7cm} \times  \bpsi(x)\gamma^{\rho}\del_{\rho}\psi(x) \ \bpsi(x)\gamma^{\lambda}\del_{\lambda}\psi(x)\Big] a^{\dagger}(p_1) b^{\dagger}(p_2)|0 \ran^{(u)}\nonumber\\
&=&-i\frac{\lambda^2 m^6}{96\pi}\cosh^2\frac{\theta}{2}\bigg[4-3\frac{\Lambda^2}{m^2}-18\log\frac{m}{\Lambda}-\cosh\theta-3\theta\coth\frac{\theta}{2}\bigg]\ ,
\label{noderv_u1_final}
\eea
\bea
\mathcal{A}_{\text{c}}^{(u_2)}&=&\frac{1}{2!} \big(-i\frac{\lambda}{2}\big)^2\frac{1}{16}\lan 0 | b(p_4) a(p_3) \tT\Big[-2\bpsi(y)\gamma^{\mu}\del_{\mu}\psi(y) \ \bpsi(y)\gamma^{\nu}\del_{\nu}\psi(y) \ \nonumber\\
&& \hspace{4.7cm} \times  \bpsi(x)\gamma^{\rho}\del_{\rho}\psi(x) \ \del_{\lambda}\bpsi(x) \gamma^{\lambda}\psi(x)\Big] a^{\dagger}(p_1) b^{\dagger}(p_2)|0 \ran^{(u)}\nonumber\\
&=&-i\frac{\lambda^2 m^6}{48\pi}\cosh^2\frac{\theta}{2}\bigg[5-6\frac{\Lambda^2}{m^2}-24\log\frac{m}{\Lambda}+\cosh\theta-3\theta\coth\frac{\theta}{2}\bigg]\ ,
\label{noderv_u2_final}
\eea
\bea
\mathcal{A}_{\text{c}}^{(u_3)}&=&\frac{1}{2!} \big(-i\frac{\lambda}{2}\big)^2\frac{1}{16}\lan 0 | b(p_4) a(p_3) \tT\Big[ \bpsi(y)\gamma^{\mu}\del_{\mu}\psi(y) \ \bpsi(y)\gamma^{\nu}\del_{\nu}\psi(y) \ \nonumber\\
&& \hspace{4.7cm} \times \del_{\rho}\bpsi(x) \gamma^{\rho}\psi(x) \ \del_{\lambda}\bpsi(x) \gamma^{\lambda}\psi(x)\Big] a^{\dagger}(p_1) b^{\dagger}(p_2)|0 \ran^{(u)}\nonumber\\
&=&-i\frac{\lambda^2 m^6}{4800\pi}\cosh^2\frac{\theta}{2}\bigg[279-270\frac{\Lambda^2}{m^2}+75\frac{\Lambda^4}{m^4}-1500\log\frac{m}{\Lambda}+2\cosh\theta\big(19+60\frac{\Lambda^2}{m^2}\big)\nonumber\\
&-&17\cosh2\theta-150 \ \theta\coth\frac{\theta}{2}\bigg]\ ,
\label{noderv_u3_final}
\eea
\bea
\mathcal{A}_{\text{c}}^{(u_4)}&=&\frac{1}{2!} \big(-i\frac{\lambda}{2}\big)^2\frac{1}{16}\lan 0 | b(p_4) a(p_3) \tT\Big[ -2\bpsi(y)\gamma^{\mu}\del_{\mu}\psi(y) \ \del_{\nu}\bpsi(y) \gamma^{\nu}\psi(y)\ \nonumber\\
&& \hspace{4.7cm} \times \bpsi(x)\gamma^{\rho}\del_{\rho}\psi(x) \ \bpsi(x)\gamma^{\lambda}\del_{\lambda}\psi(x)\Big] a^{\dagger}(p_1) b^{\dagger}(p_2)|0 \ran^{(u)}\nonumber\\
&=&i\frac{\lambda^2 m^6}{16\pi}\cosh^2\frac{\theta}{2}\bigg[4\log\frac{m}{\Lambda}+\theta\coth\frac{\theta}{2}\bigg]\ ,
\label{noderv_u4_final}
\eea
\bea
\mathcal{A}_{\text{c}}^{(u_5)}&=&\frac{1}{2!} \big(-i\frac{\lambda}{2}\big)^2\frac{1}{16}\lan 0 | b(p_4) a(p_3) \tT\Big[4\bpsi(y)\gamma^{\mu}\del_{\mu}\psi(y) \ \del_{\nu}\bpsi(y) \gamma^{\nu}\psi(y)\  \nonumber\\
&& \hspace{4.7cm} \times \bpsi(x)\gamma^{\rho}\del_{\rho}\psi(x) \ \del_{\lambda}\bpsi(x)\gamma^{\lambda}\psi(x)\Big] a^{\dagger}(p_1) b^{\dagger}(p_2)|0 \ran^{(u)}\nonumber\\
&=&-i\frac{\lambda^2 m^6}{48\pi}\cosh^2\frac{\theta}{2}\bigg[5-6\frac{\Lambda^2}{m^2}-36\log\frac{m}{\Lambda}+\cosh\theta-6\theta\coth\frac{\theta}{2}\bigg]\ ,
\label{noderv_u5_final}
\eea
\bea
\mathcal{A}_{\text{c}}^{(u_6)}&=&\frac{1}{2!} \big(-i\frac{\lambda}{2}\big)^2\frac{1}{16}\lan 0 | b(p_4) a(p_3) \tT\Big[-2\bpsi(y)\gamma^{\mu}\del_{\mu}\psi(y) \ \del_{\nu}\bpsi(y) \gamma^{\nu}\psi(y) \  \nonumber\\
&& \hspace{0.2cm} \times \del_{\rho}\bpsi(x) \gamma^{\rho}\psi(x) \ \del_{\lambda}\bpsi(x) \gamma^{\lambda}\psi(x)\Big] a^{\dagger}(p_1) b^{\dagger}(p_2)|0 \ran^{(u)}
\ \ = \ \ \mathcal{A}_{\text{c}}^{(u_2)}\ ,
\label{noderv_u6_final}
\eea\bea
\mathcal{A}_{\text{c}}^{(u_7)}&=&\frac{1}{2!} \big(-i\frac{\lambda}{2}\big)^2\frac{1}{16}\lan 0 | b(p_4) a(p_3) \tT\Big[\del_{\mu}\bpsi(y) \gamma^{\mu}\psi(y) \ \del_{\nu}\bpsi(y) \gamma^{\nu}\psi(y) \ \nonumber\\
&& \hspace{4.7cm} \times  \bpsi(x)\gamma^{\rho}\del_{\rho}\psi(x) \ \bpsi(x)\gamma^{\lambda}\del_{\lambda}\psi(x) \Big] a^{\dagger}(p_1) b^{\dagger}(p_2)|0 \ran^{(u)}\nonumber\\
&=&i\frac{\lambda^2 m^6}{32\pi}\cosh^2\frac{\theta}{2}\bigg[2\log\frac{m}{\Lambda}+\theta\coth\frac{\theta}{2}\bigg]\ ,
\label{noderv_u7_final}
\eea
\bea
\mathcal{A}_{\text{c}}^{(u_8)}&=&\frac{1}{2!} \big(-i\frac{\lambda}{2}\big)^2\frac{1}{16}\lan 0 | b(p_4) a(p_3) \tT\Big[-2\del_{\mu}\bpsi(y) \gamma^{\mu}\psi(y) \ \del_{\nu}\bpsi(y) \gamma^{\nu}\psi(y) \ \nonumber\\
&& \hspace{0.2cm} \times  \bpsi(x)\gamma^{\rho}\del_{\rho}\psi(x) \ \del_{\lambda}\bpsi(x)\gamma^{\lambda}\psi(x)\Big] a^{\dagger}(p_1) b^{\dagger}(p_2)|0 \ran^{(u)}
\ \ = \ \ \mathcal{A}_{\text{c}}^{(u_4)}\ ,
\label{noderv_u8_final}
\eea\bea
\mathcal{A}_{\text{c}}^{(u_9)}&=&\frac{1}{2!} \big(-i\frac{\lambda}{2}\big)^2\frac{1}{16}\lan 0 | b(p_4) a(p_3) \tT\Big[\del_{\mu}\bpsi(y) \gamma^{\mu}\psi(y) \ \del_{\nu}\bpsi(y) \gamma^{\nu}\psi(y) \ \nonumber\\
&& \hspace{0.2cm} \times \del_{\rho}\bpsi(x) \gamma^{\rho}\psi(x) \ \del_{\lambda}\bpsi(x) \gamma^{\lambda}\psi(x)\Big] a^{\dagger}(p_1) b^{\dagger}(p_2)|0 \ran^{(u)}
\ \ = \ \ \mathcal{A}_{\text{c}}^{(u_1)}\ .
\label{noderv_u9_final}
\eea
Combining the $u$-channel contributions to the amplitude, (\ref{noderv_u1_final}) - (\ref{noderv_u9_final}), gives the total $u$-channel contribution to the amplitude from the diagram where both vertices contain interactions $(\tilde{X}_{\mu}^{\ \mu})^2$,
\bea
\mathcal{A}_{\text{c}}^{(u)}&=& 2\Big(\mathcal{A}_{\text{c}}^{(u_1)}+\mathcal{A}_{\text{c}}^{(u_2)}+\cdots+\mathcal{A}_{\text{c}}^{(u_9)}\Big)\nonumber\\
&=&-i\frac{\lambda^2 m^6}{2400\pi}\cosh^2\frac{\theta}{2}\bigg[2179-2370\frac{\Lambda^2}{m^2}+75\frac{\Lambda^4}{m^4}-14400\log\frac{m}{\Lambda}+2\cosh\theta\big(119+60\frac{\Lambda^2}{m^2}\big)\nonumber\\
&-&17\cosh2\theta-2400 \ \theta\coth\frac{\theta}{2}\bigg]\ .
\label{noderv_u_amplitude}
\eea
Finally, adding together the contributions from the $s$, $t$ and $u$-channels to the amplitude, (\ref{noderv_s_amplitude}), (\ref{noderv_t_amplitude}) and (\ref{noderv_u_amplitude}), one finds the total amplitude from the bubble diagrams where both vertices contain interactions $(\tilde{X}_{\mu}^{\ \mu})^2$,
\bea
\mathcal{A}_{\text{c}}&=&\mathcal{A}_{\text{c}}^{(s)}+\mathcal{A}_{\text{c}}^{(t)}+\mathcal{A}_{\text{c}}^{(u)}\nonumber\\
&=&-\frac{\lambda^2m^6}{2400\pi\sinh\theta}\bigg[600\pi\big(11+4\cosh\theta+\cosh2\theta\big)
-i\Big(\sinh\theta\big(-7586+1800\frac{\Lambda^2}{m^2}+9600\log\frac{m}{\Lambda}\big)\nonumber\\
&-&3\sinh\theta\cosh2\theta\big(101+20\frac{\Lambda^2}{m^2}\big)+\sinh2\theta\big(-\frac{3089}{2}+870\frac{\Lambda^2}{m^2}+4800\log\frac{m}{\Lambda}\big)-4800 \ \theta\Big)\bigg]\ .\nonumber\\
\label{amplitude_noderv}
\eea

\subsubsection*{Total bubble diagram contribution to the amplitude}
The total contribution to the amplitude from the bubble diagrams is the sum of (\ref{amplitude_twoderv}), (\ref{amplitude_onederv}) and (\ref{amplitude_noderv}),
\bea
\mathcal{A}^{\text{bubble}}&=&\mathcal{A}_{\text{a}}+\mathcal{A}_{\text{b}}+\mathcal{A}_{\text{c}}\nonumber\\
&=&-2\lambda^2m^6\sinh^3\theta-i\frac{\lambda^2m^6}{4800\pi}\cosh^2\frac{\theta}{2}\bigg[27683+9240\frac{\Lambda^2}{m^2}-38400\log\frac{m}{\Lambda}\nonumber\\
&-&2\cosh\theta\big(10447+5940\frac{\Lambda^2}{m^2}-24000\log\frac{m}{\Lambda}\big)\bigg]\ .
\label{amplitude_bubble}
\eea
\subsubsection*{Total second order $S$-matrix}
Adding the tadpole contribution (\ref{tadpoleamplitudefinal}) and the bubble contribution (\ref{amplitude_bubble}) to the amplitude gives the total second order amplitude,
\bea
\mathcal{A}^{(2)}&=&\mathcal{A}^{\text{tad}}+\mathcal{A}^{\text{bubble}}\nonumber\\
&=&-2\lambda^2m^6\sinh^3\theta-i\frac{\lambda^2m^6}{4800\pi}\cosh^2\frac{\theta}{2}\bigg[27683+12840\frac{\Lambda^2}{m^2}-19200\log\frac{m}{\Lambda}\nonumber\\
&-&2\cosh\theta\big(10447+5940\frac{\Lambda^2}{m^2}-19200\log\frac{m}{\Lambda}\big)\bigg]\ .
\label{amplitude_secondorder}
\eea
Therefore, by (\ref{AtoS}) the second order $S$-matrix is,
\bea
S^{(2)}(\theta)&=&\frac{\mathcal{A}^{(2)}}{4 m^2 \sinh \theta}\nonumber\\
&=&-\frac{1}{2}\lambda^2m^4\sinh^2\theta-i\frac{\lambda^2m^4}{38400\pi}\coth\frac{\theta}{2}\bigg[27683+12840\frac{\Lambda^2}{m^2}-19200\log\frac{m}{\Lambda}\nonumber\\
&-&2\cosh\theta\big(10447+5940\frac{\Lambda^2}{m^2}-19200\log\frac{m}{\Lambda}\big)\bigg]\ .
\label{Smatrix_secondorder}
\eea
It is interesting to note that at second order, the real part of the amplitude comes only from the $s$-channel, while the $t$ and $u$-channels give purely imaginary contributions to the amplitude.
\subsection{Renormalized Lagrangian}
Now that the $S$-matrix has been computed the \ttb-deformed theory can be renormalized by demanding that the $S$-matrix has the form (\ref{CDDSZ1}). 
\vspace{2mm}
\\
Discarding the imaginary finite pieces of second order $S$-matrix (\ref{Smatrix_secondorder}) yields, \footnote{The finite pieces are not essential as one can change them by rescaling the cut-off, their choices are just equivalent to using different regularization schemes.}
\bea
S^{(2)}(\theta)&=&-\frac{1}{2}\lambda^2m^4\sinh^2\theta-i\frac{\lambda^2m^4}{38400\pi}\coth\frac{\theta}{2}\bigg[12840\frac{\Lambda^2}{m^2}-19200\log\frac{m}{\Lambda}
-2\cosh\theta\big(5940\frac{\Lambda^2}{m^2}\nonumber\\
&-&19200\log\frac{m}{\Lambda}\big)\bigg]\nonumber\\
&=&-\frac{1}{2}\lambda^2m^4\sinh^2\theta-i\frac{\lambda^2 m^4}{\pi}\coth\frac{\theta}{2}\bigg[\frac{107-99\cosh\theta}{320}\frac{\Lambda^2}{m^2}
-\frac{1-2\cosh\theta}{2}\log\frac{m}{\Lambda}\bigg]\ .\nonumber\\
\label{Smatrix_secondorder_imaginary}
\eea
Observe how the first term in (\ref{Smatrix_secondorder_imaginary}) is exactly the expected second order $S$-matrix in an integrable field theory, but (\ref{Smatrix_secondorder_imaginary}) also contains an extra divergent imaginary part.
The $S$-matrix was computed using the classical (bare) Lagrangian (\ref{deformedLag1}), so counterterms must be added to the classical Lagrangian in order to get rid of the imaginary part of $S^{(2)}(\theta)$. In this process the Lagrangian will be perturbatively renormalized up to second order in the \ttb-coupling $\lambda$.
\vspace{2mm}
\\
It is important to mention that the integrable structure of the \ttb-deformed free massive Dirac fermion theory is what enables the theory to be renormalized perturbatively. In general, the scattering amplitude of a quantum field theory may contain logarithms of  functions of Mandelstam variables, which involves $\theta$\footnote{(\ref{eqn2}) is an expression for the logarithm of a function of Mandelstam variables which appear while computing the one-loop momentum integrals in the scattering amplitude. When expressed in terms of rapidity, it yields terms involving $\theta$.}. However, local counterterms can never cancel a term  involving $\theta$ and can only cancel terms involving powers of $\cosh^2\Big(\frac{\theta}{2}\Big)$\footnote{A polynomial in Mandelstam variable $s$ corresponds to a polynomial in $\cosh^2\Big(\frac{\theta}{2}\Big)$, when expressed in terms of the rapidity.}. Due to the integrable structure of the theory, our final amplitude does not contain terms involving $\theta$ although individual contributions to the amplitude have terms involving $\theta$. This is expected because the $S$-matrix of an integrable theory can not have branch cuts and can only have poles.
\vspace{2mm}
\\
Discarding the imaginary finite pieces of the second order amplitude (\ref{amplitude_secondorder}) yields,
\bea
\mathcal{A}^{(2)}&=&-2\lambda^2m^6\sinh^3\theta-i\frac{8\lambda^2m^6}{\pi}\cosh^2\frac{\theta}{2}\bigg[\frac{107-99\cosh\theta}{320}\frac{\Lambda^2}{m^2}
-\frac{1-2\cosh\theta}{2}\log\frac{m}{\Lambda}\bigg]\nonumber\\
&=&-2\lambda^2m^6\sinh^3\theta-i\frac{8\lambda^2 m^6}{\pi}\bigg[\Big(\frac{103}{160}\frac{\Lambda^2}{m^2}-\frac{3}{2}\log\frac{m}{\Lambda}\Big)\cosh^2\frac{\theta}{2}+\Big(-\frac{99}{160}\frac{\Lambda^2}{m^2}\nonumber\\
&+&2\log\frac{m}{\Lambda}\Big)\cosh^4\frac{\theta}{2}\bigg]\ .
\label{amplitude_secondorder_imaginary}
\eea
Let the renormalized Lagrangian be,
\bea
\mathcal{L}_{\text{ren}}(\lambda)&=&-i \bpsi \gamma^{\mu}\del_{\mu} \psi +m \bpsi \psi +\frac{\lambda}{2}\Big(\tilde{X}_{\mu\nu}\tilde{X}^{\mu\nu}-(\tilde{X}_{\mu}^{\ \mu})^2\Big)
+\frac{\lambda^2}{2}m\bpsi\psi\Big(\tilde{X}_{\mu\nu}\tilde{X}^{\mu\nu}-(\tilde{X}_{\mu}^{\ \mu})^2\Big)\nonumber\\
&+&\frac{7}{4}\lambda^2 m^2(\bar\psi\psi)^2 \ \tilde{X}_{\mu}^{\ \mu}-\frac{7}{4}\lambda^2 m^3 (\bar\psi\psi)^3+\alpha \lambda^2 (\tilde{X}_{\mu}^{\ \mu})^2+\beta \lambda^2 \tilde{X}_{\mu\nu}\tilde{X}^{\mu\nu}+\mathcal{O}(\lambda^3)\ , \nonumber\\
\label{renormLag}
\eea 
where $\alpha$ and $\beta$ are divergent coefficients which must be tuned to exactly cancel the imaginary divergent contributions to the second order $S$-matrix (\ref{Smatrix_secondorder_imaginary}).
\vspace{2mm}
\\
The term $\alpha \lambda^2  (\tilde{X}_{\mu}^{\ \mu})^2$ contributes to the amplitude as,
\bea
\mathcal{A}_{\alpha}^{\text{count}}=(i\alpha \lambda^2 ) 4m^4(1+\cosh\theta)
=8i\alpha \lambda^2 m^4\cosh^2\frac{\theta}{2}\ .
\label{counteralpha}
\eea
While the term $\beta \lambda^2 \tilde{X}_{\mu\nu}\tilde{X}^{\mu\nu}$ contributes to the amplitude as,
\bea
\mathcal{A}_{\beta}^{\text{count}}=(i\beta \lambda^2) 4m^4(\cosh\theta+\cosh2\theta)
=-8i\beta \lambda^2 m^4\big(3\cosh^2\frac{\theta}{2}-4\cosh^4\frac{\theta}{2}\big)\ .
\label{counterbeta}
\eea
Adding (\ref{counteralpha}) and (\ref{counterbeta}), gives the total contribution from the counterterms to the amplitude,
\bea
\mathcal{A}^{\text{count}}=\mathcal{A}_{\alpha}^{\text{count}}+\mathcal{A}_{\beta}^{\text{count}}
=8i\lambda^2 m^4\big(\alpha-3\beta \big)\cosh^2\frac{\theta}{2}+32i\beta\lambda^2 m^4\cosh^4\frac{\theta}{2}\ .
\label{counterAmplitude}
\eea
The condition that the sum amplitude of the counterterms, (\ref{counterAmplitude}), must exactly cancel the imaginary part of (\ref{amplitude_secondorder_imaginary}) gives the conditions,
\bea
m^4\big(\alpha-3\beta\big)&=&\frac{m^6}{\pi}\Big(\frac{103}{160}\frac{\Lambda^2}{m^2}-\frac{3}{2}\log\frac{m}{\Lambda}\Big)\ ,
\label{counterEq1}
\eea
\bea
32\beta m^4&=&\frac{8m^6}{\pi}\Big(-\frac{99}{160}\frac{\Lambda^2}{m^2}+2\log\frac{m}{\Lambda}\Big)\ .
\label{counterEq2}
\eea
Solving (\ref{counterEq1}) and (\ref{counterEq2}) for $\alpha$ and $\beta$ gives,
\bea
\alpha&=&\frac{23}{128\pi}\Lambda^2\ ,\\
\label{alphavalue}
\beta&=&\frac{m^2}{4\pi}\Big(-\frac{99}{160}\frac{\Lambda^2}{m^2}+2\log\frac{m}{\Lambda}\Big).
\label{betavalue}
\eea
Substituting the above values of $\alpha$ and $\beta$ into the renormalized Lagrangian (\ref{renormLag}) yields,
\bea
\mathcal{L}_{\text{ren}}(\lambda)&=&-i \bpsi \gamma^{\mu}\del_{\mu} \psi +m \bpsi \psi +\frac{\lambda}{2}\Big(\tilde{X}_{\mu\nu}\tilde{X}^{\mu\nu}-(\tilde{X}_{\mu}^{\ \mu})^2\Big)
+\frac{\lambda^2}{2}m\bpsi\psi\Big(\tilde{X}_{\mu\nu}\tilde{X}^{\mu\nu}-(\tilde{X}_{\mu}^{\ \mu})^2\Big)\nonumber\\
&+&\frac{7}{4}\lambda^2 m^2(\bar\psi\psi)^2 \ \tilde{X}_{\mu}^{\ \mu}-\frac{7}{4}\lambda^2 m^3 (\bar\psi\psi)^3\nonumber\\
&+&\frac{23\Lambda^2}{128\pi} \lambda^2 (\tilde{X}_{\mu}^{\ \mu})^2+\frac{m^2}{4\pi}\Big(-\frac{99}{160}\frac{\Lambda^2}{m^2}+2\log\frac{m}{\Lambda}\Big) \lambda^2 \tilde{X}_{\mu\nu}\tilde{X}^{\mu\nu}+\mathcal{O}(\lambda^3)\ .
\label{renormLagFinal}
\eea 
However, the above renormalized Lagrangian is written in terms of the redefined fields. The renormalized Lagrangian can be written in terms of the original fields as,
\bea
\mathcal{L}_{\text{ren}}(\lambda)&=&-i \bpsi \gamma^{\mu}\del_{\mu} \psi +m \bpsi \psi
+\frac{\lambda}{2}\Big(\tilde{X}_{\mu\nu}\tilde{X}^{\mu\nu}-(\tilde{X}_{\mu}^{\ \mu})^2+2m\bpsi\psi  \tilde{X}_{\mu}^{\ \mu}-2m^2(\bpsi\psi)^2\Big)\nonumber\\
&-&\frac{\lambda^2}{2}m\bpsi\psi\Big(\tilde{X}_{\mu\nu}\tilde{X}^{\mu\nu}-(\tilde{X}_{\mu}^{\ \mu})^2\Big)\nonumber\\
&+&\frac{23\Lambda^2}{128\pi} \lambda^2 (\tilde{X}_{\mu}^{\ \mu})^2+\frac{m^2}{4\pi}\Big(-\frac{99}{160}\frac{\Lambda^2}{m^2}+2\log\frac{m}{\Lambda}\Big) \lambda^2 \tilde{X}_{\mu\nu}\tilde{X}^{\mu\nu}+\mathcal{O}(\lambda^3) \ .
\label{renormLagFinalOriginal}
\eea 
Finally, the renormalized Lagrangian of the \ttb-deformed free massive Dirac fermion in two dimensional Euclidean spacetime is given by,
\bea
\mathcal{L}_{\text{ren}}(\lambda)&=&i \bpsi \gamma^{\mu}\del_{\mu} \psi -m \bpsi \psi -\frac{\lambda}{2}\Big(2m\bpsi\psi \ \tilde{X}_{\mu}^{\ \mu}-2m^2(\bpsi\psi)^2\Big)-\frac{g}{2}\tilde{X}_{\mu\nu}\tilde{X}^{\mu\nu}+\frac{h}{2}(\tilde{X}_{\mu}^{\ \mu})^2\nonumber\\
&+&\frac{\lambda^2}{2}m\bpsi\psi\Big(\tilde{X}_{\mu\nu}\tilde{X}^{\mu\nu}-(\tilde{X}_{\mu}^{\ \mu})^2\Big)+\mathcal{O}(\lambda^3)\ ,
\label{renormLagFinal1}
\eea
where the renormalized couplings are given by,
\bea
g&=&\lambda-\frac{\lambda^2 m^2}{2\pi}\Big(\frac{99}{160}\frac{\Lambda^2}{m^2}-2\log\frac{m}{\Lambda}\Big)\ ,\nonumber\\
h&=&\lambda-\frac{23\lambda^2}{64\pi}\Lambda^2\ .
\label{renormcouplings}
\eea 
It is important to notice the major qualitative difference between the renormalized Lagrangian (\ref{renormLagFinal1}) and the classical Lagrangian (\ref{deformedLag1}). The classical Lagrangian (\ref{deformedLag1}) has only one scale, $\lambda$, which appears to be the coupling for all the quartic terms. However, the renormalized Lagrangian contains three different couplings, $\lambda$, $g$ and $h$. In the renormalized Lagrangian, the two quartic terms $\bpsi\psi \ \tilde{X}_{\mu}^{\ \mu}$ and $(\bpsi\psi)^2$ share the old classical coupling $\lambda$, where as the other two quartic terms $\tilde{X}_{\mu\nu}\tilde{X}^{\mu\nu}$ and $(\tilde{X}_{\mu}^{\ \mu})^2$ have very different couplings $g$ and $h$, respectively.

\section{Discussion}
\label{Discussion}
In this paper the \ttb-deformed free massive Dirac fermion in two dimensions was studied. First, the Lagrangian of the deformed theory was stated and massaged into an easier form for amplitude calculations using a field redefinition. The two-to-two $S$-matrix of the fermion anti-fermion scattering process was computed to second order in the \ttb-coupling $\lambda$. At first order, the $S$-matrix exactly matches the expected result for an integrable field theory (\ref{CDDSZ1}). However, at second order the $S$-matrix matches the expected result up to some divergent imaginary second order terms. Counterterms were added to the Lagrangian to cancel these divergent pieces and ensure that the final second order $S$-matrix agrees with the expected result (\ref{CDDSZ1}), in the process the renormalized Lagrangian was obtained. Amazingly, integrability allows the naively non-renormalizable theory to be renormalized perturbatively. The renormalized Lagrangian was qualitatively very different from the classical Lagrangian as there are three different coupling constants ($\lambda$, $g$ and $h$) in the renormalized Lagrangian, while in the classical case there is only one coupling constant ($\lambda$). Thus, the quantum integrability here leads to a more complicated renormalized Lagrangian than the classical one. This is not what always happens in an  integrable theory, for example, the integrable $\sinh$-Gordon model gives rise to a renormalized Lagrangian which has the same functional form as the classical one \cite{Rosenhaus}. Further, the existence of the renormalized Lagrangian means that all quantities of the theory may now be computed using the standard QFT techniques. For example, one can compute the correlation functions of local fields perturbatively using the renormalized Lagrangian. 
\\
\\
 In this paper, renormalization was performed by computing the two-to-two $S$-matrix for the fermion anti-fermion scattering: $f_1+\bar{f}_2\rightarrow f_3+\bar{f}_4$, and adding counterterms to cancel the divergences. However, one may consider the other possible two-to-two scattering process in this theory, namely, the fermion-fermion scattering: $f_1+f_2\rightarrow f_3+f_4$. If this process had been chosen the same renormalized Lagrangian would be expected. The calculation of the $S$-matrix would involve an almost identical computation to what was done here except that only one plane wave solution to the Dirac equation, $u(k)$, would be present in the expressions.
\\
\\
The form of the renormalized Lagrangian is already exciting at second order in the \ttb-coupling. It would be interesting to see how the renormalized Lagrangian looks at higher orders, because of the simple structure of the $S$-matrix the renormalized Lagrangian may have a simple and compact form. Similar to \cite{Rosenhaus}, the second order real contribution to the $S$-matrix came from the $s$-channel only. It would be useful to understand why this occurs and whether this is a general property of the \ttb-deformed integrable theories. 

\section*{Acknowledgement}
We would like to thank Michael Smolkin for suggesting the problem to us, for many insightful discussions throughout the work and useful comments on the draft. We are also thankful to Mikhail Goykhman and Lorenzo Di Pietro for valuable discussions. This research was supported by the Israeli Science Foundation Center of Excellence (grant No. 2289/18) and the Quantum Universe I-CORE program of the Israel Planning and Budgeting Committee (grant No. 1937/12).  

\section*{Appendices}
\appendix
\section{Field redefinitions}
\label{AppendixA}
The  Lagrangian of the $T\bar{T}$-deformed free massive Dirac fermion in two dimensional Euclidean spacetime  is given by (see (\ref{deformedLag1})),
\bea
\mathcal{L}(\lambda)&=& \big(i \bpsi \gamma^{\mu}\del_{\mu} \psi -m \bpsi \psi \big)-\frac{\lambda}{2}\Big(\tilde{X}_{\mu\nu}\tilde{X}^{\mu\nu}-(\tilde{X}_{\mu}^{\ \mu})^2+2m\bpsi\psi  \tilde{X}_{\mu}^{\ \mu}-2m^2(\bpsi\psi)^2\Big)\nonumber\\
&+&\frac{\lambda^2}{2}m\bpsi\psi\Big(\tilde{X}_{\mu\nu}\tilde{X}^{\mu\nu}-(\tilde{X}_{\mu}^{ \ \mu})^2\Big). 
\label{LagOriginal}
\eea
One can perform the following field redefinitions to the Lagrangian,
\bea
\psi_{a}&\rightarrow& \psi_a+\alpha \lambda \psi_{a}(\bpsi\psi)\nonumber\\
\bpsi_{a}&\rightarrow& \bpsi_a+\alpha \lambda (\bpsi\psi) \bpsi_{a}\ ,
\label{redef1}
\eea
where `$\alpha$' is assumed to be a real constant with dimension $1$.
\vspace{2mm}
\\
Under the field redefinitions the following terms transform as
\bea
\bpsi\psi &\rightarrow& B \bpsi\psi\nonumber\\
\bpsi\gamma^{\mu}\del_{\mu}\psi &\rightarrow& B \  \bpsi\gamma^{\mu}\del_{\mu}\psi+\alpha\lambda(1+\alpha\lambda \bpsi\psi) \ \bpsi\gamma^{\mu}\psi\del_{\mu}(\bpsi\psi)\nonumber\\
\del_{\mu}\bpsi\gamma^{\mu}\psi &\rightarrow& B \ \del_{\mu}\bpsi\gamma^{\mu}\psi +\alpha\lambda(1+\alpha\lambda \bpsi\psi) \ \bpsi\gamma^{\mu}\psi\del_{\mu}(\bpsi\psi)\ ,
\label{redef2}
\eea
where
\bea
B=1+2\alpha \lambda \bpsi\psi+\alpha^2\lambda^2(\bpsi\psi)^2.
\label{Bval}
\eea
By (\ref{Xtmunu}) and (\ref{redef2}), $\tilde{X}_{\mu\nu}$ and  $\tilde{X}_{\mu}^{ \ \mu}$ transform as,
\bea
\tilde{X}_{\mu\nu}\rightarrow B \ \tilde{X}_{\mu\nu} \ \ \ \text{and} \ \ \ \tilde{X}_{\mu}^{ \ \mu}\rightarrow B \ \tilde{X}_{\mu}^{ \ \mu}.
\label{redef3}
\eea
Substituting the above redefined fields into the original Lagrangian (\ref{LagOriginal}) one obtains the redefined Lagrangian,
\bea
\mathcal{L}'(\lambda)=\mathcal{L}_0'(\lambda)+\mathcal{L}_1'(\lambda)+\mathcal{L}_2'(\lambda)
\label{redefLag}
\eea
where,
\bea
\mathcal{L}_0'(\lambda)&=&\big(i \bpsi \gamma^{\mu}\del_{\mu} \psi -m \bpsi \psi \big)+2\alpha\lambda\bpsi\psi\big[\tilde{X}_{\mu}^{ \ \mu}-m\bpsi\psi\big]+\alpha^2\lambda^2 (\bpsi\psi)^2\big[\tilde{X}_{\mu}^{ \ \mu}-m\bpsi\psi\big]\nonumber\\
\mathcal{L}_1'(\lambda)&=&\frac{\lambda}{2}\big[(\tilde{X}_{\mu}^{ \ \mu})^2-\tilde{X}_{\mu\nu}\tilde{X}^{\mu\nu}\big]-\lambda m \bpsi\psi \big[ \tilde{X}_{\mu}^{ \ \mu}-m\bpsi\psi\big]+2\alpha\lambda^2 \bpsi\psi\big[(\tilde{X}_{\mu}^{ \ \mu})^2-\tilde{X}_{\mu\nu}\tilde{X}^{\mu\nu}\nonumber\\
&-&2m\bpsi\psi \tilde{X}_{\mu}^{ \ \mu}+2m^2(\bpsi\psi)^2\big]\nonumber\\
\mathcal{L}_2'(\lambda)&=&\frac{\lambda^2}{2}m\bpsi\psi\Big(\tilde{X}_{\mu\nu}\tilde{X}^{\mu\nu}-(\tilde{X}_{\mu}^{ \ \mu})^2\Big)\ .
\label{L0L1L2new}
\eea
In order to simplify the computation of the $S$-matrix contributions from the bubble diagrams it is useful to get rid of the linear terms $\bpsi\psi \tilde{X}_{\mu}^{ \ \mu}$ and $(\bpsi\psi)^2$ from the new redefined Lagrangian (\ref{redefLag}). This can be done by choosing $\alpha=\frac{m}{2}$. Substituting $\alpha=\frac{m}{2}$ into the new redefined Lagrangian (\ref{L0L1L2new}) gives,
\bea
\mathcal{L}'(\lambda)&=& \big(i \bpsi \gamma^{\mu}\del_{\mu} \psi -m \bpsi \psi \big)-\frac{\lambda}{2}\Big(\tilde{X}_{\mu\nu}\tilde{X}^{\mu\nu}-(\tilde{X}_{\mu}^{\ \mu})^2\Big)\nonumber\\
&-&\frac{\lambda^2}{2}m\bpsi\psi\Big(\tilde{X}_{\mu\nu}\tilde{X}^{\mu\nu}-(\tilde{X}_{\mu}^{ \ \mu})^2\Big)-\frac{7}{4}\lambda^2 m^2(\bpsi\psi)^2 \tilde{X}_{\mu}^{ \ \mu}+\frac{7}{4}\lambda^2 m^3(\bpsi\psi)^3
\label{redefLagFinal}
\eea
Therefore, the Lorentzian action in terms of the redefined fields can be written as
\bea
-I=i\int dt dx\, &\Big[&\big(-i \bpsi \gamma^{\mu}\del_{\mu} \psi +m \bpsi \psi \big)+\frac{\lambda}{2}\Big(\tilde{X}_{\mu\nu}\tilde{X}^{\mu\nu}-(\tilde{X}_{\mu}^{\ \mu})^2\Big)\nonumber\\
&+&\frac{\lambda^2}{2}m\bpsi\psi\Big(\tilde{X}_{\mu\nu}\tilde{X}^{\mu\nu}-(\tilde{X}_{\mu}^{\ \mu})^2\Big)+\frac{7}{4}\lambda^2 m^2(\bar\psi\psi)^2  \tilde{X}_{\mu}^{\ \mu}-\frac{7}{4}\lambda^2 m^3 (\bar\psi\psi)^3\Big]\nonumber\\
\label{redefAction}
\eea 
\section{Propagators and some useful formula}
\label{AppendixB}
In this appendix the values of all the propagators that will arise in the \ttb-deformed theory will be derived.
In two dimensions, the propagator of the free massive Dirac fermion is given by,
\bea
\left\lan \psi_a(x)\bar{\psi}_b(y)\right\ran&=&\int\frac{d^2q}{(2\pi)^2}\frac{i(\gamma\cdot q+m)_{ab}}{q^2-m^2+i\epsilon} e^{-iq(x-y)}= \int\frac{d^2q}{(2\pi)^2}G_{ab}(q) e^{-iq(x-y)}\ ,
\label{prop}
\eea
where, 
\bea
G_{ab}(q)=\frac{i(\gamma\cdot q+m)_{ab}}{q^2-m^2+i\epsilon}.
\label{prop_q}
\eea
The value of the propagator needed in the computation of the tadpole diagrams can be computed as,
\bea
\left\lan \psi_a(x)\bar{\psi}_b(x)\right\ran&=&\int\frac{d^2q}{(2\pi)^2}\frac{i(\gamma\cdot q+m)_{ab}}{q^2-m^2+i\epsilon}\nonumber\\
&=&i\int\frac{d^2 q}{(2\pi)^2}\bigg[\frac{\gamma_{ab}\cdot q}{q^2-m^2+i\epsilon}+\frac{m \delta_{ab}}{q^2-m^2+i\epsilon}\bigg].
\label{prop_div}
\eea
By Lorentz symmetry the first term in (\ref{prop_div}) vanishes, thus
\bea
\left\lan \psi_a(x)\bar{\psi}_b(x)\right\ran&=&i m \delta_{ab}\int\frac{d^2 q}{(2\pi)^2}  \frac{1}{q^2-m^2+i\epsilon}\nonumber\\
&=&i m \delta_{ab} (-i)\int_0^{\Lambda}\frac{d^2 q_E}{(2\pi)^2}  \frac{1}{q_E^2-m^2+i\epsilon}\nonumber\\
&=&\frac{m}{4\pi} \log\big(\frac{\Lambda^2+m^2}{m^2}\big) \delta_{ab} \approx -\frac{m}{4\pi} \log\big(\frac{m^2}{\Lambda^2}\big) \delta_{ab}=N_0 \ \delta_{ab}\ ,
\label{prop_div_final}
\eea
where $N_0=-\frac{m}{4\pi} \log\big(\frac{m^2}{\Lambda^2}\big)$, in the second equality a Wick rotation to Euclidean signature was performed\footnote{$q^0=i q_E^0$, $q^2=-q_E^2$ and $d^2q\rightarrow i d^2 q_E$.} and the integral was regulated by a hard cutoff $\Lambda \ (\Lambda \gg m)$.
\vspace{2mm}
\\
From (\ref{prop}) the values of propagators with derivatives can be derived. When there is a derivative acting on $\psi$ one finds,
\bea
\left\lan\del^{\mu}\psi_a(x)\bar{\psi}_b(y)\right\ran&=&\frac{\del}{\del x_{\mu}}\left\lan\psi_a(x)\bar{\psi}_b(y)\right\ran=-i\int\frac{d^2q}{(2\pi)^2}\frac{i(\gamma\cdot q+m)_{ab}}{q^2-m^2+i\epsilon} \ q^{\mu} e^{-iq(x-y)}.
\label{prop_dervx}
\eea
Hence,
\bea
\left\lan \del^{\mu} \psi_a(x)\bar{\psi}_b(x)\right\ran&=&\int\frac{d^2 q}{(2\pi)^2}\bigg[\frac{q^{\mu}(\gamma_{ab}\cdot q)}{q^2-m^2+i\epsilon}+\frac{m \delta_{ab}q^{\mu}}{q^2-m^2+i\epsilon}\bigg].
\label{prop_dervx_div}
\eea
By Lorentz symmetry the second term vanishes yielding,
\bea
\left\lan \del^{\mu} \psi_a(x)\bar{\psi}_b(x)\right\ran&=&\int\frac{d^2 q}{(2\pi)^2}\frac{q^{\mu}q^{\nu}(\gamma_{ab})_{\nu}}{q^2-m^2+i\epsilon}=\frac{1}{2}\eta^{\mu\nu}(\gamma_{ab})_{\nu}\int\frac{d^2 q}{(2\pi)^2}\frac{q^2}{q^2-m^2+i\epsilon}.
\label{prop_dervx_div_1}
\eea
As before, perform the integral (\ref{prop_dervx_div_1}) using a Wick rotation,
\bea
\left\lan \del^{\mu} \psi_a(x)\bar{\psi}_b(x)\right\ran&=&\frac{i\Lambda^2}{8\pi}
\Big[1+\frac{m^2}{\Lambda^2}\ln\Big(\frac{m^2}{\Lambda^2}\Big)\Big]\gamma^{\nu}_{ab}= N_1\gamma^{\nu}_{ab}\ ,
\label{prop_dervx_div_final}
\eea
where $N_1=\frac{i\Lambda^2}{8\pi}\Big[1+\frac{m^2}{\Lambda^2}\ln\Big(\frac{m^2}{\Lambda^2}\Big)\Big]$.
\\
\\
Similarly when a derivative acts of $\bpsi$, 
\bea
\left\lan\psi_a(x)\del^{\mu}\bar{\psi}_b(x)\right\ran=-\left\lan \del^{\mu} \psi_a(x)\bar{\psi}_b(x)\right\ran=-N_1\delta_{ab}.
\label{prop_dervy_div_final}
\eea
If one derivative acts on $\psi$ and a second on $\bpsi$ one finds,
\bea
\left\lan\del^{\mu}\psi_a(x)\del^{\nu}\bar{\psi}_b(y)\right\ran&=&\frac{\del}{\del x_{\mu}}\frac{\del}{\del y_{\nu}}\left\lan\psi_a(x)\bar{\psi}_b(y)\right\ran=i\int\frac{d^2q}{(2\pi)^2}\frac{(\gamma\cdot q+m)_{ab}}{q^2-m^2+i\epsilon} \ q^{\mu}q^{\nu} e^{-iq(x-y)}.\nonumber\\
\label{prop_dervxy}
\eea
Implementing Lorentz symmetry and performing a Wick rotation yields, 
\bea
\left\lan\del^{\mu}\psi_a(x)\del^{\nu}\bar{\psi}_b(x)\right\ran&=&-\frac{m\Lambda^2}{8\pi} \Big(1+\frac{m^2}{\Lambda^2}\ln\big(\frac{m^2}{\Lambda^2}\big)\Big)\eta^{\mu\nu}\delta_{ab}= i m N_1\eta^{\mu\nu}\delta_{ab}.
\label{prop_dervxy_div_final}
\eea

\section{ One-loop integrals}
\label{AppendixC}
In this section the values of the one-loop integrals that arise in the computation of the amplitude from the bubble diagrams will be derived.
\\
\\
Begin with the simplest integral,
\bea
\int  \frac{d^2 q}{(2\pi)^2} \frac{1}{(q^2 - \alpha^2)^2} = i \int  \frac{d^2 q_E}{(2\pi)^2} \frac{1}{(q_E^2 +\alpha^2)^{2}} =  \frac{ i }{ 4\pi  \alpha^2}~,
\label{finite_int}
\eea
where a Wick rotation was performed and the integral was converted to polar coordinates.
\\
\\
Next evaluate the following divergent integral which will be used in the computation of future integrals,
\bea
\int  \frac{d^2 q}{(2\pi)^2} \frac{q^2}{(q^2 - \alpha^2)^{2}} =-\frac{i}{2\pi}\int_0^{ \Lambda}  d q_E \frac{q_E^3}{(q_E^2 +\alpha^2)^{2}}=\frac{i}{4\pi} \bigg[  \frac{\Lambda^2}{\Lambda^2 + \alpha^2} - \log\Big( \frac{\Lambda^2+\alpha^2}{\alpha^2}\Big) \bigg]
\label{div_int_simple}
\eea
where a Wick rotation was performed, the integral was converted to polar coordinates and the integral was regulated by a hard cut-off $\Lambda \ (\Lambda^2>>\alpha^2)$. 
\vspace{2mm}
\\
The following divergent integrals can be evaluated using the Lorentz symmetry and (\ref{div_int_simple}),
\bea
\int  \frac{d^2 q}{(2\pi)^2} \frac{q_{\mu} q_{\nu}}{(q^2 - \alpha^2)^{2}}=\frac{\eta_{\mu\nu}}{2}\int  \frac{d^2 q}{(2\pi)^2} \frac{q^2}{(q^2 - \alpha^2)^{2}}  = \frac{i}{8\pi}\eta_{\mu\nu}\bigg[\frac{\Lambda^2}{\Lambda^2+\alpha^2}-\log\Big(\frac{\Lambda^2+\alpha^2}{\alpha^2}\Big)\bigg]
\label{I2}
\eea
\bea
\int  \frac{d^2 q}{(2\pi)^2} \frac{q_{\mu} q_{\nu} q_{\rho} q_{\lambda}}{(q^2 - \alpha^2)^{2}} &=&\frac{1}{8}\big(\eta_{\mu\nu}\eta_{\rho\lambda}+\eta_{\mu\rho}\eta_{\nu\lambda}+\eta_{\mu\lambda}\eta_{\nu\rho}\big)\int  \frac{d^2 q}{(2\pi)^2}\frac{(q^2)^2}{(q^2 - \alpha^2)^{2}}\nonumber\\
&=&\frac{i}{16\pi}C_{\mu\nu\rho\lambda}\bigg[\frac{\Lambda^2(\Lambda^2+2\alpha^2)}{2(\Lambda^2+\alpha^2)}-\alpha^2\log\Big(\frac{\Lambda^2+\alpha^2}{\alpha^2}\Big)\bigg]\ ,
\label{I4}
\eea
where, 
\bea
C_{\mu\nu\rho\lambda}=\eta_{\mu\nu}\eta_{\rho\lambda}+\eta_{\mu\rho}\eta_{\nu\lambda}+\eta_{\mu\lambda}\eta_{\nu\rho}.
\label{C_mnrl}
\eea
Lastly,
\bea
\ \ \int  \frac{d^2 q}{(2\pi)^2} \frac{q_{\mu} q_{\nu} q_{\rho} q_{\lambda}q_{\delta}q_{\omega}}{(q^2 - \alpha^2)^{2}} &=&\frac{1}{48}\bigg[\eta_{\mu\nu}\big(\eta_{\rho\lambda}\eta_{\delta\omega}+\eta_{\rho\delta}\eta_{\lambda\omega}+\eta_{\rho\omega}\eta_{\lambda\delta}\big)+\big(\nu \leftrightarrow \rho\big)+\big(\nu \leftrightarrow \lambda\big)\nonumber\\
&&+\big(\nu \leftrightarrow \delta\big)
+\big(\nu \leftrightarrow \omega\big)\bigg]\int  \frac{d^2 q}{(2\pi)^2}\frac{(q^2)^3}{(q^2 - \alpha^2)^{2}}\nonumber\\
&=&-\frac{i}{96\pi}F_{\mu\nu\rho\lambda\delta\omega}\bigg[\frac{\Lambda^4(\Lambda^2-3\alpha^2-\frac{6}{\Lambda^2}\alpha^4)}{4(\Lambda^2+\alpha^2)}+\frac{3}{2}\alpha^4\log\Big(\frac{\Lambda^2+\alpha^2}{\alpha^2}\Big)\bigg]\ ,\nonumber\\
\label{I6}
\eea
where,
\bea
F_{\mu\nu\rho\lambda\delta\omega}&=&\eta_{\mu\nu}\big(\eta_{\rho\lambda}\eta_{\delta\omega}+\eta_{\rho\delta}\eta_{\lambda\omega}+\eta_{\rho\omega}\eta_{\lambda\delta}\big)+\big(\nu \leftrightarrow \rho\big)+\big(\nu \leftrightarrow \lambda\big)+\big(\nu \leftrightarrow \delta\big)+\big(\nu \leftrightarrow \omega\big).\nonumber\\
\label{F_mnrldo}
\eea
In the case where there is an odd number of momenta in the numerator, the integral vanishes as it is an odd function of the integration variable,
\bea
	\int\frac{d^2 q}{(2\pi)^2}\frac{\prod_{i=1}^{n+1}q_{\mu_i}}{(q^2-\alpha^2)^2}=&0\ ,
	\label{oddq}
\eea
where $n=\{0,2,4\}$.
\\\\
The values of the above integrals, (\ref{finite_int}) - (\ref{oddq}), will be used to derive useful results for evaluating the one-loop integrals that appear in section \ref{Bubble_Contribution}. 
\subsubsection*{\underline{ $s$-channel one-loop integrals}}
First, evaluate the one-loop integrals that appear while evaluating the $s$-channel amplitudes.
Evaluate the following finite integral,
\bea
\text{(i)} \ \ \ \ L^{(s)}(\xi)=\int  \frac{d^2 q}{(2\pi)^2} \frac{1}{\big[(\xi+q)^2-m^2\big]\big(q^2-m^2\big)}
\label{L_int}
\eea
where $\ \ \ \xi^{\mu}=(p_1+p_2)^{\mu}$.
By introducing Feynman parametrization (\ref{L_int}) can be written as,
\bea
L^{(s)}(\xi)&=&\int  \frac{d^2 q}{(2\pi)^2} \int_0^1 d x \ \frac{1}{\Big[(1-x)\big((\xi+q)^2-m^2\big) +x\big(q^2-m^2\big)\Big]^2}\nonumber\\
&=&\int_0^1 d x \int  \frac{d^2 q}{(2\pi)^2} \ \frac{1}{\Big[\big(q+\xi(1-x)\big)^2 +\xi^2  x(1-x)-m^2\Big]^2}\nonumber\\
&=&\int_0^1 d x \int  \frac{d^2 k}{(2\pi)^2} \ \frac{1}{\Big[k^2 +\xi^2  x(1-x)-m^2\Big]^2}\nonumber\\
&=&\frac{i}{4\pi}\int_0^1 d x  \frac{1}{\Big(m^2-\xi^2  x(1-x)\Big)}\ ,
\label{L_int_1}
\eea
where in the third line the momentum was shifted $q\rightarrow k=q+\xi(1-x)$ and in the fourth line (\ref{finite_int}) was used to evaluate the $k$ integral.
Evaluating the $x$-integral yields,
\bea
L^{(s)}(\xi)&=& \frac{i}{2\pi\xi^2\sqrt{1-\frac{4m^2}{\xi^2}}} \log\bigg(\frac{\sqrt{1-\frac{4m^2}{\xi^2}}-1}{\sqrt{1-\frac{4m^2}{\xi^2}}+1}\bigg)
.\label{L_int_2}
\eea
It is useful to express the integral (\ref{L_int_2}) in terms of the rapidity difference $\theta$. Recall that the  Mandelstam variables $s=(p_1+p_2)^2=\xi^2$, $t=0$ and $u=4m^2-s$ can be written in terms of $\theta$ as,
\bea
s&=&2m^2(1+\cosh\theta)=4m^2\cosh^2\frac{\theta}{2}\nonumber\\
u&=&4m^2-s=-4m^2\sinh^2\frac{\theta}{2}.
\label{s_and_u}
\eea
Hence,
\bea
\sqrt{1-\frac{4m^2}{\xi^2}}&=&\sqrt{1-\frac{4m^2}{s}}=\sqrt{-\frac{u}{s}}=\tanh \frac{\theta}{2}
\label{eqn1}\\
\implies \log\bigg(\frac{\sqrt{1-\frac{4m^2}{\xi^2}}-1}{\sqrt{1-\frac{4m^2}{\xi^2}}+1}\bigg)&=&\log\bigg(\frac{\sqrt{-u/s}-1}{\sqrt{-u/s}+1}\bigg)=\log\bigg(\frac{\tanh \frac{\theta}{2}-1}{\tanh \frac{\theta}{2}+1}\bigg)=\log\big(-e^{-\theta}\big)=i\pi-\theta\nonumber\\
\label{eqn2}
\eea
where we used that $s$ is really $s+i\epsilon$, in an attempt to pick the correct sheet.
\\
\\
Plugging everything into $L^{(s)}(\theta)$ gives,
\bea
L^{(s)}(\theta)=-\frac{\pi+i\theta}{4\pi m^2 \sinh\theta}.
\label{L_theta}
\eea
\bea
\text{(ii)} \ \ \ \ L^{(s)}_{\mu}(\xi)=\int  \frac{d^2 q}{(2\pi)^2} \frac{q_{\mu}}{\big[(\xi+q)^2-m^2\big]\big(q^2-m^2\big)}
\label{L1_int}
\eea
Introduce the Feynman parameter $x$ as before. Performing the $q$-integral and then the $x$-integral gives,
\bea
L^{(s)}_{\mu}(\xi)&=& -\frac{i \xi_{\mu}}{4\pi\xi^2\sqrt{1-\frac{4m^2}{\xi^2}}} \log\bigg(\frac{\sqrt{1-\frac{4m^2}{\xi^2}}-1}{\sqrt{1-\frac{4m^2}{\xi^2}}+1}\bigg).
\label{L1_int_1}
\eea
Using (\ref{eqn1}) and (\ref{eqn2}) in (\ref{L1_int_1}) one obtains the final expression for $L^{(s)}_{\mu}(\theta)$,
\bea
L^{(s)}_{\mu}(\theta)=\xi_{\mu}\frac{\pi+i\theta}{8\pi m^2 \sinh\theta}.
\label{L1_theta}
\eea
In a similar manner, using the Feynman parametrization one can derive the following integrals,
\bea
\text{(iii)} \ \ \ \ L^{(s)}_{\mu\nu}(\xi)&=&\int  \frac{d^2 q}{(2\pi)^2} \frac{q_{\mu}q_{\nu}}{\big[(\xi+q)^2-m^2\big]\big(q^2-m^2\big)}\nonumber\\
&=&-\frac{i}{8\pi}\bigg[1+\log \frac{\Lambda^2}{m^2}+\sqrt{1-\frac{4m^2}{\xi^2}} \log\bigg(\frac{\sqrt{1-\frac{4m^2}{\xi^2}}-1}{\sqrt{1-\frac{4m^2}{\xi^2}}+1}\bigg)\bigg]\eta_{\mu\nu}\nonumber\\
&+&\frac{i}{4\pi \xi^2}\bigg[1+\frac{1-2\frac{m^2}{\xi^2}}{\sqrt{1-\frac{4m^2}{\xi^2}}} \log\bigg(\frac{\sqrt{1-\frac{4m^2}{\xi^2}}-1}{\sqrt{1-\frac{4m^2}{\xi^2}}+1}\bigg)\bigg]\xi_{\mu}\xi_{\nu}
\label{L2_int}
\eea
By (\ref{eqn1}), (\ref{eqn2}) and (\ref{L2_int}),
\bea
L^{(s)}_{\mu\nu}(\theta)=-\frac{i}{8\pi}\bigg[1+\log \frac{\Lambda^2}{m^2}+\big(i\pi-\theta\big)\tanh\frac{\theta}{2}\bigg]\eta_{\mu\nu}+\frac{i}{16\pi m^2\cosh^2\frac{\theta}{2}}\bigg[1+\big(i\pi-\theta\big)\coth\theta\bigg]\xi_{\mu}\xi_{\nu}\nonumber\\
\label{L2_theta}
\eea
\bea
\text{(iv)} \ \ \ \ L^{(s)}_{\mu\nu\rho}(\xi)&=&\int  \frac{d^2 q}{(2\pi)^2} \frac{q_{\mu}q_{\nu}q_{\rho}}{\big[(\xi+q)^2-m^2\big]\big(q^2-m^2\big)}\nonumber\\
&=&\frac{i}{16\pi}\bigg[1+\log \frac{\Lambda^2}{m^2}+\sqrt{1-\frac{4m^2}{\xi^2}} \log\bigg(\frac{\sqrt{1-\frac{4m^2}{\xi^2}}-1}{\sqrt{1-\frac{4m^2}{\xi^2}}+1}\bigg)\bigg]\big(\eta_{\mu\nu}\xi_{\rho}+\eta_{\nu\rho}\xi_{\mu}+\eta_{\rho\mu}\xi_{\nu}\big)\nonumber\\
&-&\frac{i}{4\pi \xi^2}\bigg[\frac{3}{2}+\frac{1-3\frac{m^2}{\xi^2}}{\sqrt{1-\frac{4m^2}{\xi^2}}} \log\bigg(\frac{\sqrt{1-\frac{4m^2}{\xi^2}}-1}{\sqrt{1-\frac{4m^2}{\xi^2}}+1}\bigg)\bigg]\xi_{\mu}\xi_{\nu}\xi_{\rho}
\label{L3_int}
\eea
By (\ref{eqn1}), (\ref{eqn2}) and (\ref{L3_int}),
\bea
L^{(s)}_{\mu\nu\rho}(\theta)&=&\frac{i}{16\pi}\bigg[1+\log \frac{\Lambda^2}{m^2}+\big(i\pi-\theta\big)\tanh\frac{\theta}{2}\bigg]\big(\eta_{\mu\nu}\xi_{\rho}+\eta_{\nu\rho}\xi_{\mu}+\eta_{\rho\mu}\xi_{\nu}\big)\nonumber\\
&-&\frac{i}{16\pi m^2\cosh^2\frac{\theta}{2}}\bigg[\frac{3}{2}+\big(i\pi-\theta\big)\big(\coth\theta-\frac{1}{2\sinh\theta}\big)\bigg]\xi_{\mu}\xi_{\nu}\xi_{\rho}
\label{L3_theta}
\eea
\bea
\text{(v)} \ \ \ \ &&L^{(s)}_{\mu\nu\rho\lambda}(\xi)=\int  \frac{d^2 q}{(2\pi)^2} \frac{q_{\mu}q_{\nu}q_{\rho}q_{\lambda}}{\big[(\xi+q)^2-m^2\big]\big(q^2-m^2\big)}\nonumber\\
&=&\frac{i}{16\pi}\bigg[\frac{\Lambda^2}{2}+\frac{\xi^2}{6}\bigg(\frac{5}{3}-8\frac{m^2}{\xi^2}+\big(1-6\frac{m^2}{\xi^2}\big)\log\frac{\Lambda^2}{m^2}+\big(1-4\frac{m^2}{\xi^2}\big)^{3/2}\log\Big(\frac{\sqrt{1-\frac{4m^2}{\xi^2}}-1}{\sqrt{1-\frac{4m^2}{\xi^2}}+1}\Big)\bigg)\bigg]\nonumber\\
&\times&C_{\mu\nu\rho\lambda}
+\frac{i}{8\pi}\bigg[-\frac{7}{18}+\frac{2}{3}\frac{m^2}{\xi^2}-\frac{1}{3}\big(1-\frac{m^2}{\xi^2}\big)\sqrt{1-\frac{4m^2}{\xi^2}} \log\bigg(\frac{\sqrt{1-\frac{4m^2}{\xi^2}}-1}{\sqrt{1-\frac{4m^2}{\xi^2}}+1}\bigg)
\bigg]\bigg[\eta_{\mu\nu}\xi_{\rho}\xi_{\lambda}\nonumber\\
&+&\eta_{\mu\rho}\xi_{\nu}\xi_{\lambda}+\eta_{\mu\lambda}\xi_{\nu}\xi_{\rho}+\eta_{\nu\rho}\xi_{\mu}\xi_{\lambda}+\eta_{\nu\lambda}\xi_{\rho}\xi_{\mu}+\eta_{\rho\lambda}\xi_{\mu}\xi_{\nu}\bigg]
+\frac{i}{4\pi \xi^2}\bigg[\frac{11}{6}-\frac{m^2}{\xi^2}\nonumber\\
&+&\frac{1-4\frac{m^2}{\xi^2}+2\frac{m^4}{\xi^4}}{\sqrt{1-\frac{4m^2}{\xi^2}}}\log\bigg(\frac{\sqrt{1-\frac{4m^2}{\xi^2}}-1}{\sqrt{1-\frac{4m^2}{\xi^2}}+1}\bigg)\bigg]\xi_{\mu}\xi_{\nu}\xi_{\rho}\xi_{\lambda}
\label{L4_int}
\eea
where $C_{\mu\nu\rho\lambda}$ is given in (\ref{C_mnrl}). 
By (\ref{eqn1}), (\ref{eqn2}) and (\ref{L4_int}),
\bea
L^{(s)}_{\mu\nu\rho\lambda}(\theta)&=&\frac{i}{16\pi}\bigg[\frac{\Lambda^2}{2}+\frac{2}{3}m^2\cosh^2\frac{\theta}{2}\bigg(\frac{5}{3}-\frac{2}{\cosh^2\frac{\theta}{2}}+\big(1-\frac{3}{2\cosh^2\frac{\theta}{2}}\big)\log\frac{\Lambda^2}{m^2}+\big(i\pi-\theta\big)\tanh^3\frac{\theta}{2}\bigg)\bigg]\nonumber\\
&\times&C_{\mu\nu\rho\lambda}+\frac{i}{8\pi}\bigg[-\frac{7}{18}+\frac{1}{6\cosh^2\frac{\theta}{2}}-\frac{1}{3}\big(i\pi-\theta\big)\tanh\frac{\theta}{2}\big(1-\frac{1}{4\cosh^2\frac{\theta}{2}}\big)-\frac{1}{3}\log\frac{\Lambda^2}{m^2}\bigg]\bigg[\eta_{\mu\nu}\xi_{\rho}\xi_{\lambda}\nonumber\\
&+&\eta_{\mu\rho}\xi_{\nu}\xi_{\lambda}+\eta_{\mu\lambda}\xi_{\nu}\xi_{\rho}+\eta_{\nu\rho}\xi_{\mu}\xi_{\lambda}+\eta_{\nu\lambda}\xi_{\rho}\xi_{\mu}+\eta_{\rho\lambda}\xi_{\mu}\xi_{\nu}\bigg]+\frac{i}{16\pi m^2\cosh^2\frac{\theta}{2}}\bigg[\frac{11}{6}-\frac{1}{4\cosh^2\frac{\theta}{2}}\nonumber\\
&+&\big(i\pi-\theta\big)\big(\coth\frac{\theta}{2}-\frac{2}{\sinh\theta}+\frac{1}{4\sinh\theta \cosh^2\frac{\theta}{2}}\big)\bigg]\xi_{\mu}\xi_{\nu}\xi_{\rho}\xi_{\lambda}.
\label{L4_theta}
\eea
\bea
\text{(vi)} \ \ \ \ L^{(s)}_{\mu\nu\rho\lambda\delta}(\xi)&=&\int  \frac{d^2 q}{(2\pi)^2} \frac{q_{\mu}q_{\nu}q_{\rho}q_{\lambda}q_{\delta}}{\big[(\xi+q)^2-m^2\big]\big(q^2-m^2\big)}\nonumber\\
&=&-\frac{i}{16\pi}\bigg[\frac{\Lambda^2}{4}+\xi^2\bigg(\frac{5}{36}-\frac{2}{3}\frac{m^2}{\xi^2}+\frac{1}{12}\big(1-6\frac{m^2}{\xi^2}\big)\log\frac{\Lambda^2}{m^2}\nonumber\\
&+&\frac{1}{12}\big(1-4\frac{m^2}{\xi^2}\big)^{3/2}\log\Big(\frac{\sqrt{1-\frac{4m^2}{\xi^2}}-1}{\sqrt{1-\frac{4m^2}{\xi^2}}+1}\Big)\bigg)\bigg]D_{\mu\nu\rho\lambda\delta}
-\frac{i}{8\pi}\bigg[-\frac{1}{3}+\frac{m^2}{\xi^2}-\frac{1}{4}\log\frac{\Lambda^2}{m^2}\nonumber\\
&-&\frac{1}{4}\big(1-2\frac{m^2}{\xi^2}\big)\sqrt{1-\frac{4m^2}{\xi^2}}\log\Big(\frac{\sqrt{1-\frac{4m^2}{\xi^2}}-1}{\sqrt{1-\frac{4m^2}{\xi^2}}+1}\Big)\bigg]B_{\mu\nu\rho\lambda\delta}
\nonumber
\eea\bea
&-&\frac{i}{4\pi \xi^2}\bigg[\frac{25}{12}-\frac{5}{2}\frac{m^2}{\xi^2}
+\frac{1-5\frac{m^2}{\xi^2}+5\frac{m^4}{\xi^4}}{\sqrt{1-\frac{4m^2}{\xi^2}}}\log\bigg(\frac{\sqrt{1-\frac{4m^2}{\xi^2}}-1}{\sqrt{1-\frac{4m^2}{\xi^2}}+1}\bigg)\bigg]\xi_{\mu}\xi_{\nu}\xi_{\rho}\xi_{\lambda}\xi_{\delta}\ ,
\label{L5_int}
\eea
where,
\bea
D_{\mu\nu\rho\lambda\delta}&=&C_{\mu\nu\rho\lambda}\xi_{\delta}+C_{\nu\rho\lambda\delta}\xi_{\mu}+C_{\rho\lambda\delta\mu}\xi_{\nu}+C_{\lambda\delta\mu\nu}\xi_{\rho}+C_{\delta\mu\nu\rho}\xi_{\lambda}\ ,
\label{B_mnrld}\\ \vspace{2mm}
B_{\mu\nu\rho\lambda\delta}&=&\eta_{\mu\nu}\xi_{\rho}\xi_{\lambda}\xi_{\delta}
+\eta_{\mu\rho}\xi_{\nu}\xi_{\lambda}\xi_{\delta}+\eta_{\mu\lambda}\xi_{\nu}\xi_{\rho}\xi_{\delta}
+\eta_{\mu\delta}\xi_{\nu}\xi_{\rho}\xi_{\lambda}+\eta_{\nu\rho}\xi_{\mu}\xi_{\lambda}\xi_{\delta}
\nonumber\\&&+\eta_{\nu\lambda}\xi_{\mu}\xi_{\rho}\xi_{\delta}+\eta_{\nu\delta}\xi_{\mu}\xi_{\rho}\xi_{\lambda}
+\eta_{\rho\lambda}\xi_{\mu}\xi_{\nu}\xi_{\delta}+\eta_{\rho\delta}\xi_{\mu}\xi_{\nu}\xi_{\lambda}
+\eta_{\lambda\delta}\xi_{\mu}\xi_{\nu}\xi_{\rho}\ .
\label{D_mnrld}
\eea
By (\ref{eqn1}), (\ref{eqn2}) and (\ref{L5_int}),
\bea
L^{(s)}_{\mu\nu\rho\lambda\delta}(\theta)&=&-\frac{i}{16\pi}\bigg[\frac{\Lambda^2}{4}+4m^2\cosh^2\frac{\theta}{2}\bigg(\frac{5}{36}-\frac{1}{6\cosh^2\frac{\theta}{2}}+\frac{1}{12}\big(1-\frac{3}{2\cosh^2\frac{\theta}{2}}\big)\log\frac{\Lambda^2}{m^2}\nonumber\\
&+&\frac{1}{12}\big(i\pi-\theta\big)\tanh^3\frac{\theta}{2}\bigg)\bigg]D_{\mu\nu\rho\lambda\delta}
-\frac{i}{8\pi}\bigg[-\frac{1}{3}+\frac{1}{4\cosh^2\frac{\theta}{2}}-\frac{\cosh\theta}{8\cosh^2\frac{\theta}{2}}\big(i\pi-\theta\big)\tanh\frac{\theta}{2}\nonumber\\
&-&\frac{1}{4}\log\frac{\Lambda^2}{m^2}\bigg]B_{\mu\nu\rho\lambda\delta}
-\frac{i}{16\pi m^2\cosh^2\frac{\theta}{2}}\bigg[\frac{25}{12}-\frac{5}{8\cosh^2\frac{\theta}{2}}
+\big(i\pi-\theta\big)\big(\coth\frac{\theta}{2}-\frac{5}{2\sinh\theta}\nonumber\\
&+&\frac{5}{8\sinh\theta \cosh^2\frac{\theta}{2}}\big)\bigg]\xi_{\mu}\xi_{\nu}\xi_{\rho}\xi_{\lambda}\xi_{\delta}.
\label{L5_theta}
\eea
\bea
\text{(vii)}   &&\ L^{(s)}_{\mu\nu\rho\lambda\delta\omega}(\xi)=\int  \frac{d^2 q}{(2\pi)^2} \frac{q_{\mu}q_{\nu}q_{\rho}q_{\lambda}q_{\delta}q_{\omega}}{\big[(\xi+q)^2-m^2\big]\big(q^2-m^2\big)}\nonumber\\
&=&-\frac{i}{96\pi}\bigg[\frac{\Lambda^4}{4}+\frac{1}{20}\xi^2\Lambda^2+\frac{1}{10}\xi^4\big(1-\frac{4m^2}{\xi^2}\big)^2\bigg(1+\frac{1}{2}\sqrt{1-\frac{4m^2}{\xi^2}}\log\Big(\frac{\sqrt{1-\frac{4m^2}{\xi^2}}-1}{\sqrt{1-\frac{4m^2}{\xi^2}}+1}\Big)\bigg)\nonumber\\
&+&\xi^4\big(\frac{1}{20}-\frac{1}{2}\frac{m^2}{\xi^2}+\frac{3}{2}\frac{m^4}{\xi^4}\big)\log\frac{\Lambda^2}{m^2}\bigg]F_{\mu\nu\rho\lambda\delta\omega}
+\frac{i\xi^2}{96\pi}\bigg[\frac{\Lambda^2}{\xi^2}+\frac{13}{25}-\frac{44}{15}\frac{m^2}{\xi^2}+\frac{8}{5}\frac{m^4}{\xi^4}\nonumber\\
&+&\frac{1}{10}\big(3-14\frac{m^2}{\xi^2}+8\frac{m^4}{\xi^4}\big)\sqrt{1-\frac{4m^2}{\xi^2}}\log\Big(\frac{\sqrt{1-\frac{4m^2}{\xi^2}}-1}{\sqrt{1-\frac{4m^2}{\xi^2}}+1}\Big)+\frac{1}{10}\big(3-20\frac{m^2}{\xi^2}\big)\log\frac{\Lambda^2}{m^2}\bigg]H_{\mu\nu\rho\lambda\delta\omega}\nonumber\\
&+&\frac{i}{8\pi}\bigg[-\frac{89}{300}+\frac{37}{30}\frac{m^2}{\xi^2}-\frac{2}{5}\frac{m^4}{\xi^4}-\frac{1}{5}\big(1-3\frac{m^2}{\xi^2}+\frac{m^4}{\xi^4}\big)\sqrt{1-\frac{4m^2}{\xi^2}}\log\Big(\frac{\sqrt{1-\frac{4m^2}{\xi^2}}-1}{\sqrt{1-\frac{4m^2}{\xi^2}}+1}\Big)\nonumber\\
&-&\frac{1}{5}\log\frac{\Lambda^2}{m^2}\bigg]M_{\mu\nu\rho\lambda\delta\omega}
+\frac{i}{4\pi\xi^2}\bigg[\frac{137}{60}-\frac{13}{3}\frac{m^2}{\xi^2}+\frac{m^4}{\xi^4}+\frac{1}{\sqrt{1-\frac{4m^2}{\xi^2}}}\Big(1-6\frac{m^2}{\xi^2}+9\frac{m^4}{\xi^4}\nonumber\\
&-&2\frac{m^6}{\xi^6}\Big)\log\Big(\frac{\sqrt{1-\frac{4m^2}{\xi^2}}-1}{\sqrt{1-\frac{4m^2}{\xi^2}}+1}\Big)\bigg]\xi_{\mu}\xi_{\nu}\xi_{\rho}\xi_{\lambda}\xi_{\delta}\xi_{\omega}
\label{L6_int}
\eea
where,
\bea
H_{\mu\nu\rho\lambda\delta\omega}&=&C_{\mu\nu\rho\lambda}\xi_{\delta}\xi_{\omega}+C_{\mu\nu\rho\delta}\xi_{\lambda}\xi_{\omega}+C_{\mu\nu\rho\omega}\xi_{\lambda}\xi_{\delta}+C_{\mu\nu\lambda\delta}\xi_{\rho}\xi_{\omega}+C_{\mu\nu\lambda\omega}\xi_{\rho}\xi_{\delta}+C_{\mu\rho\lambda\delta}\xi_{\nu}\xi_{\omega}\nonumber\\
&+&C_{\mu\rho\lambda\omega}\xi_{\nu}\xi_{\delta}+C_{\nu\rho\lambda\delta}\xi_{\mu}\xi_{\omega}+C_{\nu\rho\lambda\omega}\xi_{\mu}\xi_{\delta}+C_{\mu\rho\delta\omega}\xi_{\nu}\xi_{\lambda}+C_{\mu\lambda\delta\omega}\xi_{\nu}\xi_{\rho}+C_{\lambda\rho\delta\omega}\xi_{\nu}\xi_{\mu}\nonumber\\ 
&+&C_{\nu\rho\delta\omega}\xi_{\mu}\xi_{\lambda}+C_{\mu\nu\delta\omega}\xi_{\rho}\xi_{\lambda}+C_{\nu\lambda\delta\omega}\xi_{\mu}\xi_{\rho}\ , 
\label{H_mnrldo}\\ \vspace{1mm}
M_{\mu\nu\rho\lambda\delta\omega}&=&\xi_{\mu}\xi_{\nu}\xi_{\rho}\xi_{\lambda}\eta_{\delta\omega}+\xi_{\mu}\xi_{\nu}\xi_{\rho}\xi_{\delta}\eta_{\lambda\omega}+\xi_{\mu}\xi_{\nu}\xi_{\rho}\xi_{\omega}\eta_{\lambda\delta}+\xi_{\mu}\xi_{\nu}\xi_{\lambda}\xi_{\delta}\eta_{\rho\omega}+\xi_{\mu}\xi_{\nu}\xi_{\lambda}\xi_{\omega}\eta_{\rho\delta}\nonumber\\
&+&\xi_{\mu}\xi_{\rho}\xi_{\lambda}\xi_{\delta}\eta_{\nu\omega}+\xi_{\mu}\xi_{\rho}\xi_{\lambda}\xi_{\omega}\eta_{\nu\delta}+\xi_{\nu}\xi_{\rho}\xi_{\lambda}\xi_{\delta}\eta_{\mu\omega}+\xi_{\nu}\xi_{\rho}\xi_{\lambda}\xi_{\omega}\eta_{\mu\delta}+\xi_{\mu}\xi_{\rho}\xi_{\delta}\xi_{\omega}\eta_{\nu\lambda}\nonumber\\
&+&\xi_{\mu}\xi_{\lambda}\xi_{\delta}\xi_{\omega}\eta_{\nu\rho}+\xi_{\lambda}\xi_{\rho}\xi_{\delta}\xi_{\omega}\eta_{\nu\mu}+\xi_{\nu}\xi_{\rho}\xi_{\delta}\xi_{\omega}\eta_{\mu\lambda}+\xi_{\mu}\xi_{\nu}\xi_{\delta}\xi_{\omega}\eta_{\rho\lambda}+\xi_{\nu}\xi_{\lambda}\xi_{\delta}\xi_{\omega}\eta_{\mu\rho}
\label{M_mnrldo}
\eea
and $F_{\mu\nu\rho\lambda\delta\omega}$ is given in (\ref{F_mnrldo}). By (\ref{eqn1}), (\ref{eqn2}) and (\ref{L6_int}),
\bea
L^{(s)}_{\mu\nu\rho\lambda\delta\omega}(\theta)&=&-\frac{i}{96\pi}\bigg[\frac{\Lambda^4}{4}+\frac{\Lambda^2m^2}{5}\cosh^2\frac{\theta}{2}+4m^4\cosh^4\frac{\theta}{2}\big(\frac{1}{5}-\frac{1}{2\cosh^2\frac{\theta}{2}}+\frac{3}{8\cosh^4\frac{\theta}{2}}\big)\log\frac{\Lambda^2}{m^2}\nonumber\\
&+&\frac{8}{5}m^4\sinh^4\frac{\theta}{2}\big(1+\frac{1}{2}(i\pi-\theta)\tanh\frac{\theta}{2}\big)\bigg]F_{\mu\nu\rho\lambda\delta\omega}
+\frac{im^2\cosh^2\frac{\theta}{2}}{24\pi}\bigg[\frac{13}{25}-\dfrac{11}{15\cosh^2\frac{\theta}{2}}\nonumber\\
&+&\frac{1}{10\cosh^4\frac{\theta}{2}}+\frac{\Lambda^2}{4m^2\cosh^2\frac{\theta}{2}}+\frac{1}{10}\big(i\pi-\theta\big)\big(3-\frac{7}{2\cosh^2\frac{\theta}{2}}+\frac{1}{2\cosh^4\frac{\theta}{2}}\big)\tanh\frac{\theta}{2}\nonumber\\
&+&\frac{1}{10}\big(3-\frac{5}{\cosh^2\frac{\theta}{2}}\big)\log\frac{\Lambda^2}{m^2}\bigg]H_{\mu\nu\rho\lambda\delta\omega}
+\frac{i}{8\pi}\bigg[-\frac{89}{300}+\frac{37}{120\cosh^2\frac{\theta}{2}}-\frac{1}{40\cosh^4\frac{\theta}{2}}\nonumber\\
&-&\frac{1}{5}\log\frac{\Lambda^2}{m^2}-\frac{1}{5}(i\pi-\theta)\tanh\frac{\theta}{2}\big(1-\frac{3}{4\cosh^2\frac{\theta}{2}}+\frac{1}{16\cosh^4\frac{\theta}{2}}\big)\bigg]M_{\mu\nu\rho\lambda\delta\omega}\nonumber\\
&+&\frac{i}{16\pi m^2\cosh^2\frac{\theta}{2}}\bigg[\frac{137}{60}-\frac{13}{12\cosh^2\frac{\theta}{2}}+\frac{1}{16\cosh^4\frac{\theta}{2}}
+(i\pi-\theta)\coth\frac{\theta}{2}\big(1-\frac{3}{2\cosh^2\frac{\theta}{2}}\nonumber\\
&+&\frac{9}{16\cosh^4\frac{\theta}{2}}-\frac{1}{32\cosh^6\frac{\theta}{2}}\big)\bigg]\xi_{\mu}\xi_{\nu}\xi_{\rho}\xi_{\lambda}\xi_{\delta}\xi_{\omega}.
\label{L6_theta}
\eea
The above integrals (\ref{L_int}) - (\ref{L6_int}) can be used to evaluate the one-loop integrals needed in the computation of the $s$-channel amplitudes in section \ref{Bubble_Contribution}. When no derivatives are present the loop integral looks like,
\bea
I^{(s)}_{abcd}(\xi)&=&\int  \frac{d^2 q}{(2\pi)^2}G_{ab}(\xi+q)G_{cd}(q)\nonumber\\
&=&\int  \frac{d^2 q}{(2\pi)^2} \frac{i\big(\gamma\cdot (\xi+q)+m\big)_{ab}}{(\xi+q)^2-m^2} \frac{i\big(\gamma\cdot q+m\big)_{cd}}{q^2-m^2} \nonumber\\
&=&-\int  \frac{d^2 q}{(2\pi)^2}\frac{\gamma^{\mu}_{ab}\gamma^{\nu}_{cd}(\xi+q)_{\mu}q_{\nu}+m\delta_{ab}\gamma^{\mu}_{cd}q_{\mu}+m\gamma^{\mu}_{ab}(\xi+q)_{\mu}\delta_{cd}+m^2\delta_{ab}\delta_{cd}}
{\big[(\xi+q)^2-m^2\big](q^2-m^2)}\nonumber\\
&=&-\Big(\gamma^{\mu}_{ab}\gamma^{\nu}_{cd}\xi_{\mu}L^{(s)}_{\nu}(\xi)+\gamma^{\mu}_{ab}\gamma^{\nu}_{cd}L^{(s)}_{\mu\nu}(\xi)+m\delta_{ab}\gamma^{\mu}_{cd}L^{(s)}_{\mu}(\xi)+m\gamma^{\mu}_{ab}\xi_{\mu}\delta_{cd}L^{(s)}(\xi)\nonumber\\
&&+m\gamma^{\mu}_{ab}L^{(s)}_{\mu}(\xi)\delta_{cd}+m^2\delta_{ab}\delta_{cd}L^{(s)}(\xi)\Big)\nonumber\eea\bea
&=&-L^{(s)}(\xi)\Big(m\gamma^{\mu}_{ab}\xi_{\mu}\delta_{cd}+m^2\delta_{ab}\delta_{cd}\Big)
-\Big(\gamma^{\mu}_{ab}\xi_{\mu}\gamma^{\nu}_{cd}+m\delta_{ab}\gamma^{\nu}_{cd}+m\gamma^{\nu}_{ab}\delta_{cd}\Big)L^{(s)}_{\nu}(\xi)\nonumber\\
&&-\gamma^{\mu}_{ab}\gamma^{\nu}_{cd}L^{(s)}_{\mu\nu}(\xi)\nonumber\\
&=&-L^{(s)}(\xi)\Big(m\slashed{\xi}_{ab}\delta_{cd}+m^2\delta_{ab}\delta_{cd}\Big)-\Big(\slashed{\xi}_{ab}\slashed{L}^{(s)}_{cd}(\xi)+m\delta_{ab}\slashed{L}^{(s)}_{cd}(\xi)+m\slashed{L}^{(s)}_{ab}(\xi)\delta_{cd}\Big)\nonumber\\
&&-\gamma^{\mu}_{ab}\gamma^{\nu}_{cd}L^{(s)}_{\mu\nu}(\xi)\ ,
\label{I_abcd}
\eea
where $ \ \slashed{\xi}_{ab}=\gamma^{\mu}_{ab}\xi_{\mu}\ $ and $ \ \slashed{L}^{(s)}_{ab}=\gamma^{\mu}_{ab}L^{(s)}_{\mu} $ with $L^{(s)}$, $L^{(s)}_{\mu}$ and $L^{(s)}_{\mu\nu}$ are given by (\ref{L_theta}), (\ref{L1_theta}) and (\ref{L2_theta}), respectively.
\vspace{2mm}
\\
When it involves derivatives, a similar procedure can be performed to evaluate the one-loop integrals in the s-channel case,
\bea
(I_{\mu})^{(s)}_{abcd}(\xi)&=&\int  \frac{d^2 q}{(2\pi)^2} q_{\mu}G_{ab}(\xi+q)G_{cd}(q)\nonumber\\
&=&\int  \frac{d^2 q}{(2\pi)^2} q_{\mu}\frac{i\big(\gamma\cdot (\xi+q)+m\big)_{ab}}{(\xi+q)^2-m^2} \frac{i\big(\gamma\cdot q+m\big)_{cd}}{q^2-m^2}\nonumber\\
&=&-\Big(m\slashed{\xi}_{ab}\delta_{cd}+m^2\delta_{ab}\delta_{cd}\Big)L^{(s)}_{\mu}(\xi)-\Big(\slashed{\xi}_{ab}\gamma^{\nu}_{cd}+m\delta_{ab}\gamma^{\nu}_{cd}+m\gamma^{\nu}_{ab}\delta_{cd}\Big)L^{(s)}_{\mu\nu}(\xi)\nonumber\\
&&-\gamma^{\nu}_{ab}\gamma^{\rho}_{cd}L^{(s)}_{\mu\nu\rho}(\xi)
\label{I1_abcd}
\eea
\bea
(I_{\mu\nu})^{(s)}_{abcd}(\xi)&=&\int  \frac{d^2 q}{(2\pi)^2} q_{\mu}q_{\nu}G_{ab}(\xi+q)G_{cd}(q)\nonumber\\
&=&\int  \frac{d^2 q}{(2\pi)^2} q_{\mu}q_{\nu}\frac{i\big(\gamma\cdot (\xi+q)+m\big)_{ab}}{(\xi+q)^2-m^2} \frac{i\big(\gamma\cdot q+m\big)_{cd}}{q^2-m^2}\nonumber\\
&=&-\Big(m\slashed{\xi}_{ab}\delta_{cd}+m^2\delta_{ab}\delta_{cd}\Big)L^{(s)}_{\mu\nu}(\xi)-\Big(\slashed{\xi}_{ab}\gamma^{\rho}_{cd}+m\delta_{ab}\gamma^{\rho}_{cd}+m\gamma^{\rho}_{ab}\delta_{cd}\Big)L^{(s)}_{\mu\nu\rho}(\xi\nonumber\\
&&-\gamma^{\rho}_{ab}\gamma^{\lambda}_{cd}L^{(s)}_{\mu\nu\rho\lambda}(\xi)
\label{I2_abcd}
\eea
\bea
(I_{\mu\nu\rho})^{(s)}_{abcd}(\xi)&=&\int  \frac{d^2 q}{(2\pi)^2} q_{\mu}q_{\nu}q_{\rho}G_{ab}(\xi+q)G_{cd}(q)\nonumber\\
&=&\int  \frac{d^2 q}{(2\pi)^2} q_{\mu}q_{\nu}q_{\rho}\frac{i\big(\gamma\cdot (\xi+q)+m\big)_{ab}}{(\xi+q)^2-m^2} \frac{i\big(\gamma\cdot q+m\big)_{cd}}{q^2-m^2}\nonumber\\
&=&-\Big(m\slashed{\xi}_{ab}\delta_{cd}+m^2\delta_{ab}\delta_{cd}\Big)L^{(s)}_{\mu\nu\rho}(\xi)-\Big(\slashed{\xi}_{ab}\gamma^{\lambda}_{cd}+m\delta_{ab}\gamma^{\lambda}_{cd}+m\gamma^{\lambda}_{ab}\delta_{cd}\Big)L^{(s)}_{\mu\nu\rho\lambda}(\xi)\nonumber\\
&-&\gamma^{\lambda}_{ab}\gamma^{\delta}_{cd}L^{(s)}_{\mu\nu\rho\lambda\delta}(\xi)
\label{I3_abcd}
\eea
\bea
(I_{\mu\nu\rho\lambda})^{(s)}_{abcd}(\xi)&=&\int  \frac{d^2 q}{(2\pi)^2} q_{\mu}q_{\nu}q_{\rho}q_{\lambda}G_{ab}(\xi+q)G_{cd}(q)\nonumber\\
&=&\int  \frac{d^2 q}{(2\pi)^2} q_{\mu}q_{\nu}q_{\rho}q_{\lambda}\frac{i\big(\gamma\cdot (\xi+q)+m\big)_{ab}}{(\xi+q)^2-m^2} \frac{i\big(\gamma\cdot q+m\big)_{cd}}{q^2-m^2}\nonumber\\
&=&-\Big(m\slashed{\xi}_{ab}\delta_{cd}+m^2\delta_{ab}\delta_{cd}\Big)L^{(s)}_{\mu\nu\rho\lambda}(\xi)-\Big(\slashed{\xi}_{ab}\gamma^{\delta}_{cd}+m\delta_{ab}\gamma^{\delta}_{cd}+m\gamma^{\delta}_{ab}\delta_{cd}\Big)L^{(s)}_{\mu\nu\rho\lambda\delta}(\xi)\nonumber\\
&-&\gamma^{\delta}_{ab}\gamma^{\omega}_{cd}L^{(s)}_{\mu\nu\rho\lambda\delta\omega}(\xi)
\label{I4_abcd}
\eea
Where $L^{(s)}_{\mu\nu}(\xi)$, $L^{(s)}_{\mu\nu\rho}(\xi)$,  $L^{(s)}_{\mu\nu\rho\lambda}(\xi)$, $L^{(s)}_{\mu\nu\rho\lambda\delta}(\xi)$ and $L^{(s)}_{\mu\nu\rho\lambda\delta\omega}(\xi)$ are respectively given by (\ref{L2_theta}), (\ref{L3_theta}), (\ref{L4_theta}), (\ref{L5_theta}) and (\ref{L6_theta}). 

\subsubsection*{\underline{$t$-channel one-loop integrals}}
Next, evaluate the one-loop integrals that appear while evaluating the $t$-channel amplitudes.
 These integrals are much simpler as $t=(p_1-p_3)^2=0$,
\bea
I_{abcd}^{(t)}&&=\int  \frac{d^2 q}{(2\pi)^2}G_{ab}(q)G_{cd}(q)
=-m^2\delta_{ab}\delta_{cd}L^{(t)}-m\big(\delta_{ab}\slashed{L}^{(t)}_{cd}+\slashed{L}^{(t)}_{ab}\delta_{cd}\big)-\gamma^{\mu}_{ab}\gamma^{\nu}_{cd}L^{(t)}_{\mu\nu}\nonumber\\
(I_{\mu})^{(t)}_{abcd}&&=\int  \frac{d^2 q}{(2\pi)^2} \ q_{\mu} \ G_{ab}(q)G_{cd}(q)
=-m^2\delta_{ab}\delta_{cd}L^{(t)}_{\mu}-m\big(\delta_{ab}\gamma^{\nu}_{cd}+\gamma^{\nu}_{ab}\delta_{cd}\big)L^{(t)}_{\mu\nu}-\gamma^{\nu}_{ab}\gamma^{\rho}_{cd}L^{(t)}_{\mu\nu\rho}\nonumber\\
(I_{\mu\nu})^{(t)}_{abcd}&&=\int  \frac{d^2 q}{(2\pi)^2} \ q_{\mu} q_{\nu}\ G_{ab}(q)G_{cd}(q)
=-m^2\delta_{ab}\delta_{cd}L^{(t)}_{\mu\nu}-m\big(\delta_{ab}\gamma^{\rho}_{cd}+\gamma^{\rho}_{ab}\delta_{cd}\big)L^{(t)}_{\mu\nu\rho}\nonumber\\
&&\hspace{5.6cm}-\gamma^{\rho}_{ab}\gamma^{\lambda}_{cd}L^{(t)}_{\mu\nu\rho\lambda}\nonumber\\
(I_{\mu\nu\rho})^{(t)}_{abcd}&&=\int  \frac{d^2 q}{(2\pi)^2} \ q_{\mu} q_{\nu} q_{\rho}\ G_{ab}(q)G_{cd}(q)
=-m^2\delta_{ab}\delta_{cd}L^{(t)}_{\mu\nu\rho}-m\big(\delta_{ab}\gamma^{\lambda}_{cd}+\gamma^{\lambda}_{ab}\delta_{cd}\big)L^{(t)}_{\mu\nu\rho\lambda}\nonumber\\
&&\hspace{5.8cm}-\gamma^{\lambda}_{ab}\gamma^{\delta}_{cd}L^{(t)}_{\mu\nu\rho\lambda\delta}\nonumber\\
(I_{\mu\nu\rho\lambda})^{(t)}_{abcd}&&=\int  \frac{d^2 q}{(2\pi)^2} q_{\mu} q_{\nu} q_{\rho} q_{\lambda}G_{ab}(q)G_{cd}(q)
=-m^2\delta_{ab}\delta_{cd}L^{(t)}_{\mu\nu\rho\lambda}-m\big(\delta_{ab}\gamma^{\delta}_{cd}+\gamma^{\delta}_{ab}\delta_{cd}\big)L^{(t)}_{\mu\nu\rho\lambda\delta}\nonumber\\
&&\hspace{6cm}-\gamma^{\delta}_{ab}\gamma^{\omega}_{cd}L^{(t)}_{\mu\nu\rho\lambda\delta\omega}\ ,
\label{I_t_abcd}
\eea
where
\bea
L^{(t)}&=&\int  \frac{d^2 q}{(2\pi)^2} \frac{1}{\big(q^2-m^2\big)^2}=\frac{i}{4\pi m^2}\nonumber\\
L^{(t)}_{\mu}&=&\int  \frac{d^2 q}{(2\pi)^2} \frac{q_{\mu}}{\big(q^2-m^2\big)^2}=0\nonumber\\
L^{(t)}_{\mu\nu}&=&\int  \frac{d^2 q}{(2\pi)^2}\frac{q_{\mu}q_{\nu}}{\big(q^2-m^2\big)^2}=\frac{i}{8\pi}\Big(1-\log\frac{\Lambda^2}{m^2}\Big)\eta_{\mu\nu}\nonumber\\
L^{(t)}_{\mu\nu\rho}&=&\int  \frac{d^2 q}{(2\pi)^2}\frac{q_{\mu}q_{\nu}q_{\rho}}{\big(q^2-m^2\big)^2}=0\nonumber\\
L^{(t)}_{\mu\nu\rho\lambda}&=&\int  \frac{d^2 q}{(2\pi)^2}\frac{q_{\mu}q_{\nu}q_{\rho}q_{\lambda}}{\big(q^2-m^2\big)^2}=\frac{i m^2}{32\pi}\Big(\frac{\Lambda^2}{m^2}-2\log\frac{\Lambda^2}{m^2}\Big)\big(\eta_{\mu\nu}\eta_{\rho\lambda}+\eta_{\mu\rho}\eta_{\nu\lambda}+\eta_{\mu\lambda}\eta_{\nu\rho}\big)\nonumber\\
L^{(t)}_{\mu\nu\rho\lambda\delta}&=&\int  \frac{d^2 q}{(2\pi)^2}\frac{q_{\mu}q_{\nu}q_{\rho}q_{\lambda}q_{\delta}}{\big(q^2-m^2\big)^2}=0\nonumber\\
L^{(t)}_{\mu\nu\rho\lambda\delta\omega}&=&\int  \frac{d^2 q}{(2\pi)^2}\frac{q_{\mu}q_{\nu}q_{\rho}q_{\lambda}q_{\delta}q_{\omega}}{\big(q^2-m^2\big)^2}=-\frac{i m^4}{384\pi}\Big(\frac{\Lambda^4}{m^4}+6\log\frac{\Lambda^2}{m^2}\Big)F_{\mu\nu\rho\lambda\delta\omega},
\label{L_t_integrals}
\eea
and $F_{\mu\nu\rho\lambda\delta\omega}$ is given by (\ref{F_mnrldo}). The above integrals have been derived from the corresponding $s$-channel integrals (\ref{L_theta}, \ref{L1_theta}, \ref{L2_theta}, \ref{L3_theta}, \ref{L4_theta}, \ref{L5_theta} and \ref{L6_theta}) by taking the limit $\theta\rightarrow i\pi$ and setting $\xi\rightarrow 0$.

\subsubsection*{\underline{$u$-channel one-loop integrals}}
Finally, evaluate the one-loop integrals that appear while evaluating the $u$-channel amplitudes. When there are no derivatives the one-loop integral can be written as, 
\bea
I_{abcd}^{(u)}(\zeta)&=&\int  \frac{d^2 q}{(2\pi)^2}G_{ab}(\zeta-q)G_{cd}(q)=\int  \frac{d^2 q}{(2\pi)^2} \frac{i\big(\gamma\cdot (\zeta-q)+m\big)_{ab}}{(\zeta-q)^2-m^2} \frac{i\big(\gamma\cdot q+m\big)_{cd}}{q^2-m^2}\nonumber\\
 &=&-\int  \frac{d^2 q}{(2\pi)^2}\frac{\gamma^{\mu}_{ab}\gamma^{\nu}_{cd}(\zeta-q)_{\mu}q_{\nu}+m\delta_{ab}\gamma^{\mu}_{cd}q_{\mu}+m\gamma^{\mu}_{ab}(\zeta-q)_{\mu}\delta_{cd}+m^2\delta_{ab}\delta_{cd}}
{\big[(\zeta-q)^2-m^2\big](q^2-m^2)}\nonumber\\
&=&-L^{(u)}(\zeta)\Big(m\slashed{\zeta}_{ab}\delta_{cd}+m^2\delta_{ab}\delta_{cd}\Big)-\Big(\slashed{\zeta}_{ab}\slashed{L}_{cd}^{(u)}(\zeta)+m\delta_{ab}\slashed{L}_{cd}^{(u)}(\zeta)-m\slashed{L}_{ab}^{(u)}(\zeta)\delta_{cd}\Big)\nonumber\\
&&+\gamma^{\mu}_{ab}\gamma^{\nu}_{cd}L_{\mu\nu}^{(u)}(\zeta)\ ,
\label{I_abcd_u_final}
\eea
 where $\zeta_{\mu}=(p_1-p_4)_{\mu}=(p_1-p_2)_{\mu}$, \hspace{2mm}$ \slashed{\zeta}_{ab}=\gamma^{\mu}_{ab}\zeta_{\mu} $  \hspace{2mm}and
 \bea
 L^{(u)}(\zeta)&=&\int  \frac{d^2 q}{(2\pi)^2} \frac{1}{\big[(\zeta-q)^2-m^2\big]\big(q^2-m^2\big)}\ ,\nonumber\\
 L_{\mu}^{(u)}(\zeta)&=&\int  \frac{d^2 q}{(2\pi)^2} \frac{q_{\mu}}{\big[(\zeta-q)^2-m^2\big]\big(q^2-m^2\big)}\ ,\nonumber\\
 L_{\mu\nu}^{(u)}(\zeta)&=&\int  \frac{d^2 q}{(2\pi)^2} \frac{q_{\mu}q_{\nu}}{\big[(\zeta-q)^2-m^2\big]\big(q^2-m^2\big)}\ .
 \eea
 Comparing the above integrals with their $s$-channel counterparts (\ref{L_int}, \ref{L1_int} and \ref{L2_int}) one sees that the structure of the integrals are identical apart from the fact that the $u$-channel integrands depends on $\zeta-q$ while the $s$-channel integrands depend on $\xi+q$.
 \vspace{2mm}
 \\
 Since the $u$-channel corresponds to $\theta\rightarrow i\pi-\theta$, the above integrals can be derived directly from the results of their $s$-channel counterparts (\ref{L_theta}, \ref{L1_theta} and \ref{L2_theta}) by replacing $\theta\rightarrow i\pi-\theta$. Notice that replacing $\theta$ by $i\pi-\theta$ is equivalent to replacing $\xi^2$ by $\zeta^2$. However, since the functional dependence of the integrands on $\zeta$ is $\zeta-q$, an extra minus sign must be included whenever the integrand is odd in $q$. Thus,
 \bea
 L^{(u)}(\theta)&=&i\frac{\theta}{4\pi m^2\sinh\theta}\nonumber\\
 L^{(u)}_{\mu}(\theta)&=&i\frac{\theta}{8\pi m^2\sinh\theta}\zeta_{\mu}\nonumber\\
 L^{(u)}_{\mu\nu}(\theta)&=&-\frac{i}{8\pi}\bigg[1+\log \frac{\Lambda^2}{m^2}-\theta\coth\frac{\theta}{2}\bigg]\eta_{\mu\nu}-\frac{i}{16\pi m^2\sinh^2\frac{\theta}{2}}\big(1-\theta\coth\theta\big)\zeta_{\mu}\zeta_{\nu} 
 \eea
 In a similar manner, one can derive the one-loop integrals in the $u$-channel case when there are derivatives,
 \bea
 (I_{\mu})_{abcd}^{(u)}(\zeta)&=&\int  \frac{d^2 q}{(2\pi)^2} \ q_{\mu}G_{ab}(\zeta-q)G_{cd}(q)\nonumber\\
 &=&-\Big(m\slashed{\zeta}_{ab}\delta_{cd}+m^2\delta_{ab}\delta_{cd}\Big)L_{\mu}^{(u)}(\zeta)-\Big(\slashed{\zeta}_{ab}\gamma^{\nu}_{cd}+m\delta_{ab}\gamma^{\nu}_{cd}+m\gamma^{\nu}_{ab}\delta_{cd}\Big)L_{\mu\nu}^{(u)}(\zeta)\nonumber\\
 &&-\gamma^{\nu}_{ab}\gamma^{\rho}_{cd}L_{\mu\nu\rho}^{(u)}(\zeta)
 \label{I1_abcd_u}
 \eea
 \bea
 (I_{\mu\nu})_{abcd}^{(u)}(\zeta)&=&\int  \frac{d^2 q}{(2\pi)^2} q_{\mu}q_{\nu}G_{ab}(\zeta-q)G_{cd}(q)\nonumber\\
 &=&-\Big(m\slashed{\zeta}_{ab}\delta_{cd}+m^2\delta_{ab}\delta_{cd}\Big)L_{\mu\nu}^{(u)}(\zeta)-\Big(\slashed{\zeta}_{ab}\gamma^{\rho}_{cd}+m\delta_{ab}\gamma^{\rho}_{cd}+m\gamma^{\rho}_{ab}\delta_{cd}\Big)L_{\mu\nu\rho}^{(u)}(\zeta)\nonumber\\
 &&-\gamma^{\rho}_{ab}\gamma^{\lambda}_{cd}L_{\mu\nu\rho\lambda}^{(u)}(\zeta)
 \label{I2_abcd_u}
 \eea
 \bea
 (I_{\mu\nu\rho})_{abcd}^{(u)}(\zeta)&=&\int  \frac{d^2 q}{(2\pi)^2} q_{\mu}q_{\nu}q_{\rho}G_{ab}(\zeta-q)G_{cd}(q)\nonumber\\
 &=&-\Big(m\slashed{\zeta}_{ab}\delta_{cd}+m^2\delta_{ab}\delta_{cd}\Big)L_{\mu\nu\rho}^{(u)}(\zeta)-\Big(\slashed{\zeta}_{ab}\gamma^{\lambda}_{cd}+m\delta_{ab}\gamma^{\lambda}_{cd}+m\gamma^{\lambda}_{ab}\delta_{cd}\Big)L_{\mu\nu\rho\lambda}^{(u)}(\zeta)\nonumber\\
 &&-\gamma^{\lambda}_{ab}\gamma^{\delta}_{cd}L_{\mu\nu\rho\lambda\delta}^{(u)}(\zeta)
 \label{I3_abcd_u}
 \eea
 \bea
 (I_{\mu\nu\rho\lambda})_{abcd}^{(u)}(\zeta)&=&\int  \frac{d^2 q}{(2\pi)^2} q_{\mu}q_{\nu}q_{\rho}q_{\lambda}G_{ab}(\zeta-q)G_{cd}(q)\nonumber\\
 &=&-\Big(m\slashed{\zeta}_{ab}\delta_{cd}+m^2\delta_{ab}\delta_{cd}\Big)L_{\mu\nu\rho\lambda}^{(u)}(\zeta)-\Big(\slashed{\zeta}_{ab}\gamma^{\delta}_{cd}+m\delta_{ab}\gamma^{\delta}_{cd}+m\gamma^{\delta}_{ab}\delta_{cd}\Big)L_{\mu\nu\rho\lambda\delta}^{(u)}(\zeta)\nonumber\\
 &&-\gamma^{\delta}_{ab}\gamma^{\omega}_{cd}L_{\mu\nu\rho\lambda\delta\omega}^{(u)}(\zeta)\ ,
 \label{I4_abcd_u}
 \eea
where,
\bea
L_{\mu\nu\rho}^{(u)}(\theta)&=&\int  \frac{d^2 q}{(2\pi)^2} \frac{q_{\mu}q_{\nu}q_{\rho}}{\big[(\zeta-q)^2-m^2\big]\big(q^2-m^2\big)}\nonumber\\
&=&-\frac{i}{16\pi}\big(1+\log\frac{\Lambda^2}{m^2}-\theta\coth\frac{\theta}{2}\big)\big(\eta_{\mu\nu}\zeta_{\rho}+\eta_{\nu\rho}\zeta_{\mu}+\eta_{\rho\mu}\zeta_{\nu}\big)\nonumber\\
&&-\frac{i}{32\pi m^2\sinh^2\frac{\theta}{2}}\big(3-2\theta\coth\theta
-\theta \csch\theta\big)\zeta_{\mu}\zeta_{\nu}\zeta_{\rho}
\eea
\bea
L_{\mu\nu\rho\lambda}^{(u)}(\theta)&=&\int  \frac{d^2 q}{(2\pi)^2} \frac{q_{\mu}q_{\nu}q_{\rho}q_{\lambda}}{\big[(\zeta-q)^2-m^2\big]\big(q^2-m^2\big)}\nonumber\\
&=&-\frac{i m^2}{288\pi}\bigg[14-9\frac{\Lambda^2}{m^2}+12\log\frac{\Lambda^2}{m^2}+2\cosh\theta\big(5+3\log\frac{\Lambda^2}{m^2}\big)-6\theta\big(2\coth\frac{\theta}{2}+\sinh\theta\big)\bigg] C_{\mu\nu\rho\lambda}\nonumber\\
&+&\frac{i}{288\pi\sinh^2\frac{\theta}{2}}\bigg[1+6\log\frac{\Lambda^2}{m^2}-\cosh\theta\big(7+6\log\frac{\Lambda^2}{m^2}\big)+3\theta\cosh\frac{3\theta}{2}\csch\frac{\theta}{2}\bigg]\bigg[\eta_{\mu\nu}\zeta_{\rho}\zeta_{\lambda}\nonumber\\
&+&\eta_{\mu\rho}\zeta_{\nu}\zeta_{\lambda}+\eta_{\mu\lambda}\zeta_{\nu}\zeta_{\rho}+\eta_{\nu\rho}\zeta_{\mu}\zeta_{\lambda}+\eta_{\nu\lambda}\zeta_{\rho}\zeta_{\mu}+\eta_{\rho\lambda}\zeta_{\mu}\zeta_{\nu}\bigg]+\frac{i}{192\pi m^2\sinh^4\frac{\theta}{2}}\bigg[8-11\cosh\theta\nonumber\\
&+&3\theta\cosh2\theta\csch\theta\bigg]\zeta_{\mu}\zeta_{\nu}\zeta_{\rho}\zeta_{\lambda}
\eea
\bea
L_{\mu\nu\rho\lambda\delta}^{(u)}(\theta)&=&\int  \frac{d^2 q}{(2\pi)^2} \frac{q_{\mu}q_{\nu}q_{\rho}q_{\lambda}q_{\delta}}{\big[(\zeta-q)^2-m^2\big]\big(q^2-m^2\big)}\nonumber\\
&=&-\frac{i m^2}{576\pi}\bigg[14-9\frac{\Lambda^2}{m^2}+12\log\frac{\Lambda^2}{m^2}+2\cosh\theta\big(5+3\log\frac{\Lambda^2}{m^2}\big)-6\theta\big(2\coth\frac{\theta}{2}+\sinh\theta\big)\bigg] D'_{\mu\nu\rho\lambda\delta}\nonumber\\
&-&\frac{i}{192\pi}\bigg[8+6\log\frac{\Lambda^2}{m^2}+6\csch^2\frac{\theta}{2}-\frac{3}{4}\theta\sinh2\theta\csch^4\frac{\theta}{2}\bigg]B'_{\mu\nu\rho\lambda\delta}\nonumber\eea\bea
&+&\frac{i}{384\pi m^2\sinh^4\frac{\theta}{2}}\bigg[10
-25\cosh\theta+3\theta\big(\csch\theta+2\coth\theta+2\cosh2\theta\csch\theta\big)\bigg]\zeta_{\mu}\zeta_{\nu}\zeta_{\rho}\zeta_{\lambda}\zeta_{\omega}\nonumber\\
\eea
\bea
&&L_{\mu\nu\rho\lambda\delta\omega}^{(u)}(\theta)=\int  \frac{d^2 q}{(2\pi)^2} \frac{q_{\mu}q_{\nu}q_{\rho}q_{\lambda}q_{\delta}q_{\omega}}{\big[(\zeta-q)^2-m^2\big]\big(q^2-m^2\big)}\nonumber\\
&=&-\frac{i m^4}{3840\pi}\bigg[4\frac{\Lambda^2}{m^2}+10\frac{\Lambda^4}{m^4}+4(8+\cosh2\theta)\log\frac{\Lambda^2}{m^2}+64\cosh^4\frac{\theta}{2}-4\cosh\theta\big(\frac{\Lambda^2}{m^2}-6\log\frac{\Lambda^2}{m^2}\big)\nonumber\\
&-&\theta\sinh^5\theta\csch^6\frac{\theta}{2}\bigg]F_{\mu\nu\rho\lambda\delta\omega}
-\frac{im^2}{2400\pi}\bigg[\frac{142}{3}-25\frac{\Lambda^2}{m^2}+5\log\frac{\Lambda^2}{m^2}(7+3\cosh\theta)+26\cosh\theta\nonumber\\
&+&10\csch^2\frac{\theta}{2}+5\theta(2-3\cosh\theta)\coth^3\frac{\theta}{2}\bigg]H'_{\mu\nu\rho\lambda\delta\omega}
+\frac{i}{19200\pi\sinh^4\frac{\theta}{2}}\bigg[43-180\frac{\Lambda^2}{m^2}\nonumber\\
&-&\cosh2\theta\big(89+60\log\frac{\Lambda^2}{m^2}\big)-2\cosh\theta\big(7-120\log\frac{\Lambda^2}{m^2}\big)+30\theta\cosh\frac{5\theta}{2}\csch\frac{\theta}{2}\bigg]M'_{\mu\nu\rho\lambda\delta\omega}\nonumber\\
&+&\frac{i}{15360\pi m^2\sinh^6\frac{\theta}{2}\sinh\theta}\bigg[60\theta\cosh3\theta-225\sinh\theta+288\sinh2\theta-137\sinh3\theta\bigg]\zeta_{\mu}\zeta_{\nu}\zeta_{\rho}\zeta_{\lambda}\zeta_{\delta}\zeta_{\omega},\nonumber\\
\eea
 $C_{\mu\nu\rho\lambda}$ and $F_{\mu\nu\rho\lambda\delta\omega}$ are given by (\ref{C_mnrl}) and (\ref{F_mnrldo}) respectively and $D'_{\mu\nu\rho\lambda\delta}$, $B'_{\mu\nu\rho\lambda\delta}$, $H'_{\mu\nu\rho\lambda\delta\omega}$ and $M'_{\mu\nu\rho\lambda\delta\omega}$ are given by (\ref{B_mnrld}), (\ref{D_mnrld}), (\ref{H_mnrldo}) and (\ref{M_mnrldo}) with $\zeta_{\mu}$ instead of $\xi_{\mu}$.

\bigskip
\bibliographystyle{JHEP}

\begin{thebibliography}{99}
\bibitem{Wilson}
K. G. Wilson, ``The renormalization group and critical phenomena'',
Rev. Mod. Phys. 55 (1983) 583-600.
\bibitem{Polchinski}
J. Polchinski, ``Renormalization and Effective Lagrangians'',
Nucl. Phys. B231 (1984) 269-295.

\bibitem{Zamolodchikov}
A. B. Zamolodchikov, ``Expectation value of composite field $T\bar{T}$ in two-dimensional
quantum field theory'', [arXiv:hep-th/0401146].

\bibitem{Smirnov}
F. A. Smirnov and A. B. Zamolodchikov, ``On space of integrable quantum field theories",
Nucl. Phys. B915 (2017) 363-383, [arXiv:1608.05499 [hep-th]].

\bibitem{Andrea}
C. Andrea, S. Negro, I. M. Szecsenyi and R. Tateo, ``$T\bar{T}$-deformed $2D$ Quantum Field
Theories'', JHEP 10 (2016) 112, [arXiv:1608.05534 [hep-th]].

\bibitem{Bonelli_2018}
G. Bonelli, N. Doroud, Nima and M. Zhu, ``$ T\bar{T}$-deformations in closed form'',
JHEP 06 (2018) 149, [arXiv:1804.10967 [hep-th]].

\bibitem{McGough} 
L. McGough, M. Mezei and H. Verlinde, ``Moving the CFT into the bulk with $T\bar{T}$'',
JHEP 04 (2018) 010, [arXiv:1611.03470 [hep-th]].

\bibitem{Kraus}
P. Kraus, J. Liu and D. Marolf, ``Cutoff AdS$_3$ versus the \ttb-deformation'', 
JHEP 07 (2018) 027, [arXiv:1801.02714 [hep-th]].

\bibitem{Taylor}
M. Taylor, ``\ttb-deformations in general dimensions'', [arXiv:1805.10287 [hep-th]].

\bibitem{Hartman}
T. Hartman, J. Kruthoff, E. Shaghoulian and A. Tajdini, ``Holography at finite cutoff with a $T^2$ deformation'', JHEP 03  (2019) 004 [arXiv:1807.11401 [hep-th]].

\bibitem{Caputa}
P. Caputa, S. Datta and V. Shyam, ``Sphere partition functions \& cut-off AdS'',
JHEP 05 ( 2019) 112, [arXiv:1902.10893 [hep-th]].

\bibitem{Giveon1}
A. Giveon, N. Itzhaki and D. Kutasov, ``$T\bar{T}$ and LST'', 
JHEP 07 (2017) 122, [arXiv:1701.05576 [hep-th]].

\bibitem{Giveon2}
A. Giveon, N. Itzhaki and D. Kutasov, ``A solvable irrelevant deformation of
AdS$_3$/CFT$_2$'', JHEP 12 (2017) 155, [arXiv:1707.05800 [hep-th]].

\bibitem{Asrat}
M. Asrat, A. Giveon, N. Itzhaki and D. Kutasov, ``Holography Beyond AdS'', 
Nucl. Phys. B 932 (2018) 241-253, [arXiv:1711.02690 [hep-th]].

\bibitem{Chakraborty1}
S. Chakraborty, A. Giveon, N. Itzhaki and D. Kutasov, ``Entanglement Beyond AdS'', 
Nucl. Phys. B 935 (2018) 290-309, [arXiv:1805.06286 [hep-th]].

\bibitem{Chakraborty2}
S. Chakraborty, ``Wilson loop in a $T\bar{T}$ like deformed CFT$_2$'', Nucl. Phys. B 938 (2019) 605-620,  [arXiv:1809.01915 [hep-th]].

\bibitem{Giveon3}
A. Giveon, ``Comments on $T\bar{T}$, $J\bar{T}$ and string theory'', [arXiv:1903.06883 [hep-th]].

\bibitem{Giribet}
 G. Giribet, ``\ttb--deformations, AdS/CFT and correlation functions'',
JHEP 02 (2018) 114, [arXiv:1711.02716 [hep-th]].

\bibitem{Barbon}
J. L. F. Barbon and E. Rabinovici, ``Remarks On The Thermodynamic Stability Of \ttb-Deformations'',
 J. Phys. A: Math. Theor. 53 (2020)  424001, [arXiv:2004.10138 [hep-th]].

\bibitem{Cardy1}
J. Cardy, ``The $T\bar{T}$ deformation of quantum field theory as random geometry'',
JHEP 10 (2018) 186, [arXiv:1801.06895 [hep-th]].

\bibitem{Datta}
S. Datta and Y. Jiang, ``$T\bar{T}$ deformed partition functions'', JHEP 08 (2018) 106,
[arXiv:1806.07426 [hep-th]].

\bibitem{Aharony}
O. Aharony, S. Datta, A. Giveon, Y. Jiang and D. Kutasov, ``Modular invariance and uniqueness of $T\bar{T}$ deformed CFT'', JHEP 01(2019) 086 [arXiv:1808.02492 [hep-th]].

\bibitem{Jiang}
Y. Jiang, ``A pedagogical review on solvable irrelevant deformations of $2d$ quantum field theory",
Commun. Theor. Phys. 73 (2021) 057201, [arXiv:1904.13376 [hep-th]].

\bibitem{Delfino}
G. Delfino and G. Niccoli, ``Matrix elements of the operator $T\bar{T}$ in integrable
quantum field theory'', Nucl. Phys. B 707 [FS] (2005) 381–404, [arXiv:hep-th/0407142].

\bibitem{Dubovsky1}
S. Dubovsky, R. Flauger and V. Gorbenko, ``Solving the Simplest Theory of Quantum
Gravity'', JHEP 09 (2012) 133, [arXiv:1205.6805 [hep-th]].

\bibitem{Dubovsky2}
 S. Dubovsky, R. Flauger and V. Gorbenko, ``Evidence from Lattice Data for a New
Particle on the Worldsheet of the QCD Flux Tube'',
Phys. Rev. Lett. 111 (2013) 062006, [arXiv:1301.2325 [hep-th]].

\bibitem{Caselle}
M. Caselle, D. Fioravanti, F. Gliozzi and R. Tateo, ``Quantisation of the effective
string with TBA'', JHEP 07 (2013) 071, [arXiv:1305.1278 [hep-th]].

\bibitem{Rosenhaus}
V. Rosenhaus and M. Smolkin, ``Integrability and renormalization under $T\bar T$'',  
Phys. Rev. D. 102 (2020) 065009 [arXiv:1909.02640[hep-th]].
	
\bibitem{ZZ}
A. B. Zamolodchikov and A. B. Zamolodchikov, ``Factorized $S$-matrices in two-dimensions as the exact solutions of certain relativistic quantum field models'', Annals of Physics 120 (1979) 253-291.
	
\bibitem{Polyakov}
A. M. Polyakov, ``Hidden symmetry of the two-dimensional chiral fields'', Phys. Lett. B 72 (1977) 224.
	
\bibitem{Shankar}
 R. Shankar and E. Witten, ``$S$ matrix of the supersymmetric nonlinear $\sigma$ model'', 
 Phys. Rev. D 17 (1978) 2134.
	
\bibitem{Parke} 
S. J. Parke, ``Absence of particle production and factorization of the $S$-matrix in $(1+1)$-dimensional models'', Nucl. Phys. B 174 (1980) 166-182.
	
\bibitem{Iagolnitzer} 
D. Iagolnitzer, ``Factorization of the multiparticle $S$ matrix in two-dimensional space-time models'', 
Phys. Rev. D 18 (1978) 1275.
	
\bibitem{Dorey}	
 P. Dorey, ``Exact $S$-matrices'', [arXiv:hep-th/9810026].
 
\bibitem{Bombardelli} 
 D. Bombardelli, ``$S$-matrices and integrability'', 
 J. Phys. A: Math. Theor. 49 (2016) 323003, [arXiv:1606.02949 [hep-th]].
 
\bibitem{Gorbenko}
 S. Dubovsky, V. Gorbenko and M. Mirbabayi, ``Asymptotic fragility, near AdS$_2$ holography and $T\bar{T}$'', 
 JHEP 09 (2017) 136, [arXiv:1706.06604 [hep-th]].
 
\bibitem{DGH} 
S. Dubovsky, V. Gorbenko and G. Hernandez-Chifflet, ``$T\bar{T}$-partition function from topological gravity'', 
JHEP 09 (2018) 158, [arXiv:1805.07386 [hep-th]].
  
 \bibitem{Conti1}
 R. Conti, S. Negro and R. Tateo, ``The $T\bar{T}$ perturbation and its geometric interpretation'', 
 JHEP 02 (2019) 085, [arXiv:1809.09593 [hep-th]].

\bibitem{Conti2} 
R. Conti, L. Iannella, S. Negro and R. Tateo, ``Generalised Born-Infeld models, Lax operators and the $T\bar{T}$ perturbation'', JHEP 11 (2018) 007, [arXiv:1806.11515 [hep-th]].

\bibitem{He1}
 S. He and Y. Sun, ``Correlation functions of CFTs on a torus with a \ttb-deformation'',  
 Phys. Rev. D 102 (2020) 026023, [arXiv:2004.07486 [hep-th]].
 
 \bibitem{He2}
S. He, Y. Sun and Y. X. Zhang, ``\ttb-flow effects on torus partition functions'', [arXiv:2011.02902 [hep-th]].

\bibitem{He3}
S. He, ``Note on higher-point correlation functions of the $T\bar{T}$ or $J\bar{T}$ deformed CFTs'', [arXiv:2012.06202 [hep-th]].

\bibitem{Dey}
A. Dey, M. Goykhman and M. Smolkin, 
``Composite operators in \ttb-deformed free QFTs'', 
JHEP 06 (2021) 006, [arXiv:2012.15605 [hep-th]].

\bibitem{Cardy2}
J. Cardy, ``$T\bar{T}$ deformation of correlation functions'', 
JHEP12(2019)160, [arXiv:1907.03394 [hep-th]].

\bibitem{Haruna}
J. Haruna, T. Ishii, H. Kawai, K. Sakai and K. Yoshida, ``Large N Analysis of \ttb-deformation and Unavoidable Negative-norm States'', JHEP 04 (2020) 127, [arXiv:2002.01414 [hep-th]].

\end{thebibliography}

\end{document}